\documentclass[11pt,aps,prd,amsmath,amssymb,superscriptaddress,nofootinbib,preprintnumbers]{revtex4-2}

\usepackage{soul}

\usepackage[hmargin=2cm,vmargin=2cm]{geometry}
\usepackage[indent,skip=0.1\baselineskip]{parskip}
\usepackage{xcolor}
\usepackage{enumitem}
\usepackage{mathrsfs}
\usepackage[mathscr]{eucal}
\usepackage[normalem]{ulem}
\usepackage{braket}
\usepackage[range-phrase=\text{--}]{siunitx}
\usepackage{multirow}
\usepackage{graphicx}
\usepackage{float}
\usepackage[hidelinks]{hyperref}
\usepackage[all]{hypcap}
\usepackage{orcidlink}

\hypersetup{colorlinks, linkcolor={red!50!black}, citecolor={blue!50!black}, urlcolor={blue!80!black}}

\makeatletter
\g@addto@macro\bfseries{\boldmath}
\makeatother

\DeclareMathOperator{\tr}{tr}
\DeclareMathOperator{\SU}{\mathrm{SU}}
\DeclareMathOperator{\order}{\mathcal{O}}
\renewcommand{\Re}{\operatorname{Re}}
\renewcommand{\Im}{\operatorname{Im}}
\newcommand{\var}{\qty[parse-numbers=false]}

\begin{document}
\preprint{MIT-CTP/6148}

\title{Emergence and polymer physics of the center-vortex network:\texorpdfstring{\\}{} Confronting \texorpdfstring{\(\SU(4)\)}{SU(4)} lattice Yang-Mills theory}

\author{Luis E. Oxman \orcidlink{0000-0002-4670-5597}}
\affiliation{Instituto de F{\'i}sica, Universidade Federal Fluminense, 24210-346 Niter{\'o}i, RJ, Brasil}

\author{Jackson A. Mickley \orcidlink{0000-0001-5294-2823}}
\affiliation{Centre for the Subatomic Structure of Matter, Department of Physics, The University of Adelaide, South Australia 5005, Australia}
\affiliation{Center for Theoretical Physics --- a Leinweber Institute, Massachusetts Institute of Technology, Cambridge, MA 02139, USA}

\author{Derek B. Leinweber \orcidlink{0000-0002-4745-6027}}
\affiliation{Centre for the Subatomic Structure of Matter, Department of Physics, The University of Adelaide, South Australia 5005, Australia}

\begin{abstract}
The geometric organization and statistical properties of center-vortex networks provide a window into the infrared dynamics underlying confinement in Yang-Mills theory. Motivated by this connection, we investigate new consequences of a previously proposed confinement mechanism based on physically equivalent elementary Cartan fluxes. We show that, in the percolating phase, these fluxes dynamically reorganize into center-charge sectors whose dominant vortex tensions satisfy a Casimir scaling law, similar to that observed for confining string tensions. We then confront this result with the first nontrivial Yang-Mills case:\ the doubly charged vortex chains recently observed in \(\SU(4)\) lattice simulations. To extract the parameters needed to connect the lattice ensemble with the effective description, we characterize how center-vortex chains grow and bend in \(\SU(3)\) and \(\SU(4)\) using polymer-inspired techniques not previously applied to center vortices.  We find good quantitative agreement for the \(\SU(4)\) doubly charged sector, providing further evidence for the proposed mechanism.
\end{abstract}

\maketitle

\section{Introduction} \label{sec:intro}
Lattice studies of Yang-Mills theory broadly fall into two groups. The first group concerns the confining properties themselves:\ Wilson loops, string tensions at intermediate and asymptotic distances, and confining flux-tube profiles for different quark representations. These can be mapped as the gauge group, the spacetime dimensionality, or the temperature are varied~\cite{Wilson:1974sk, Bali:1994de, DElia:1997sdk, Bali:2000gf, Lucini:2001nv, Kratochvila:2003zj, Lucini:2004my, Stokes:2013oaa, Trewartha:2015ida, Cea:2017ocq, Athenodorou:2021qvs}. The second group concerns the structure of the infrared configurations believed to dominate the Monte Carlo ensemble and the evidence linking them to confinement. Candidate infrared degrees of freedom include monopoles~\cite{Maedan:1988yi, Chernodub:1997dr, Ambjorn:1999ym, Stack:2002ysv, Sakumichi:2014xpa} and center vortices~\cite{tHooft:1977nqb, tHooft:1979rtg, Nielsen:1979xu, DelDebbio:1996lih, Langfeld:1997jx, DelDebbio:1998luz, Engelhardt:1998wu, Langfeld:1998cz, deForcrand:1999our, Engelhardt:1999fd, Faber:1999sq, deForcrand:2000pg, Langfeld:2001cz, Greensite:2003bk, Langfeld:2003ev, OMalley:2011aa, Greensite:2016pfc, Biddle:2022zgw, Mickley:2024zyg, Mickley:2024vkm} identified in Abelian-projected and center-projected variables, respectively. A proposed understanding of the infrared regime of \(\SU(N)\) Yang-Mills theory provides a compelling explanation of confinement only if it simultaneously accounts for both groups of observables, and does so economically. That is, the fewer independent parameters required, and the more numerous and detailed the connections established between the two sets of observables, the stronger the case that the mechanism captures the underlying physics.

The geometry and statistical properties of center vortices have likewise been probed as physical conditions are varied~\cite{Bertle:1999tw, Spengler:2018dxt, Biddle:2019gke, Biddle:2023lod, Mickley:2024zyg, Mickley:2024vkm, Mickley:2025ksp, Mickley:2025qgk}. In particular, \textit{ab initio} Yang-Mills simulations combined with center-projection techniques have determined the structure of the center-vortex network in \(\SU(2)\)~\cite{Bertle:1999tw} and \(\SU(3)\)~\cite{Biddle:2023lod}. In \(\SU(3)\), a percolating network of charge-one center-vortex lines on fixed-time slices containing three-way center monopoles was characterized. More recently, calculations revealed an \(\SU(4)\) network formed by larger charge-one lines attached in pairs to typically shorter charge-two lines~\cite{Mickley:2025ksp}. These objects have no analog in \(\SU(2)\) or \(\SU(3)\), since the distinction between elementary and doubly charged vortices only becomes physical for \(N \geq 4\).

In addition, phenomenological random-surface models have been used to describe \(\SU(2)\) and \(\SU(3)\) deconfinement~\cite{Engelhardt:1999xw, Engelhardt:1999wr, Engelhardt:2003wm} and were later extended to \(\SU(4)\) by introducing independent curvature couplings for the charge-one and charge-two sectors, along with an explicit branching term~\cite{Engelhardt:2005qu}. In general \(\SU(N)\), noting \(\pm k\) center charges are simply opposite orientations in spacetime, this construction would require an independently tuned parameter for each sector \(k = 1,\ldots,\lfloor N/2 \rfloor\), together with additional couplings describing the various branching processes.

In a series of works~\cite{Oxman:2018dzp, Junior:2019fty, Junior:2021vpd, Junior:2022bol, Junior:2024urr, Junior:2023tjg}, a confinement mechanism based on a mixed ensemble of oriented and nonoriented center vortices was developed. At the level provided by Abelian projection, the mechanism initially considers elementary center vortices, which carry the \(N\) physically equivalent Cartan fluxes \(\beta_i\), \(i = 1,\ldots,N\), associated with unit center charge. Their equivalence implies they share the same physical properties, such as tension and stiffness. The mechanism then possesses a natural hierarchy. Its first stage consists of the ``oriented'' center-vortex condensate, where these elementary Cartan fluxes interact through a single \(N\)-matching interaction:\ \(N\)-way center monopoles. The second stage introduces a milder component associated with Cartan monopoles, which mediate transitions between different Cartan fluxes on ``nonoriented'' vortex components. This gives rise to confining flux tubes, directly connecting the condensation of collimated Cartan fluxes in the center-vortex ensemble to the formation of the confining string. Moreover, their asymptotic tensions depend only on the \(N\)-ality of the quark representation and can display Casimir scaling. In addition, the flux-tube profiles can exhibit Abelian-like behavior.

This raises the question as to whether this mechanism can also account for the charge-two objects observed in \textit{ab initio} \(\SU(4)\) Yang-Mills simulations. In this work, we address this question by analyzing the first stage of the hierarchy, i.e.\ the oriented vortex component, in the percolating phase. We use it as a simple reference model and confront its first nontrivial predictions with the \(\SU(4)\) Yang-Mills center-vortex network. To this end, we introduce novel observables inspired by the mechanics of polymer chains, including the persistence length and mean-square end-to-end distance, and analyze how center-vortex chains grow and bend over short, intermediate, and long length scales. We analyze whether the network exhibits the expected polymer-like behavior in both \(\SU(3)\) and \(\SU(4)\), mapping separately the properties of the charge-one and charge-two sectors in the latter case. We then use these measured properties to determine the effective lattice spacing entering the model, allowing for a quantitative comparison with the Yang-Mills network.

The paper is organized as follows. In Sec.~\ref{sec:collimatedfluxes}, we review the effective description of collimated elementary Cartan fluxes and its lattice representation. The center-vortex network emerging from these elementary constituents is obtained in Sec.~\ref{sec:vortexnetworks}. Subsequently, Sec~\ref{sec:latticeobservables} introduces new observables inspired by polymer physics to characterize the bending and growth properties of center-vortex chains. These are applied to numerical simulations in Sec.~\ref{sec:simulations}, where we present an extensive study of the \(\SU(3)\) and \(\SU(4)\) lattice Yang-Mills ensembles. In Sec.~\ref{sec:discussion}, we use our numerical findings to confront the predictions of the effective model. Finally, our conclusions are presented in Sec.~\ref{sec:conclusion}.

\section{Ensembles of collimated fluxes} \label{sec:collimatedfluxes}
\subsection{4D partition function} \label{subsec:4Dpartitionfunction}
The mixed ensemble of oriented and nonoriented center vortices was originally proposed in Ref.~\cite{Oxman:2018dzp}. An ensemble measure describing percolating center-vortex worldsurfaces coupled in pairs to monopole worldlines was introduced, with the effective description written directly in terms of the corresponding Goldstone modes. A general phase of the ensemble was parametrized on the lattice in Ref.~\cite{Junior:2024urr}, starting from a sum over closed center-vortex worldsurfaces \(\{\mathcal{S}\}\),
\begin{equation}
	Z = \sum_{\{\mathcal{S}\}} \psi_{\{\mathcal{S}\}} \,.
\end{equation}
The weight factor \(\psi_{\{\mathcal{S}\}}\) encodes the physical properties of center vortices, such as their tension, stabilizing repulsive interactions and natural correlations. Using a generalized Weingarten lattice representation~\cite{Weingarten:1979gn}, the partition function for the oriented center-vortex sector was represented by
\begin{gather}
	Z_\mathrm{c.v.} = \frac{1}{\mathcal{N}} \int D\Phi \; e^{-W_\mathrm{c.v.}[\Phi]} \,, \qquad D\Phi \equiv \prod_{\{x,y\}} d\Phi(x,y) \,, \qquad \mathcal{N} = \int D\Phi \; e^{-\sum_{\{x,y\}} \mathcal{Q}(\Phi(x,y))} \,, \label{eq:4DZcv} \\[0.4em]
	W_\mathrm{c.v.}[\Phi] = -\sum_{p} \tr\, \bigl(\Phi(p)\bigr) + \sum_{\{x,y\}} \left[\mathcal{Q} \bigl(\Phi(x,y)\bigr) - \xi \, \bigl(\det \Phi(x,y) + \det \Phi(y,x) \bigr) \right] \,, \label{eq:4DWcv} \\[0.4em]
	\mathcal{Q}(\Phi) = \tr \,\bigl(\eta \, \Phi^\dagger \Phi
	+ \lambda \, (\Phi^\dagger \Phi)^2 \bigr) \,, \label{eq:4DQ}
\end{gather}
where \(\Phi(p)\) is a product of the link variables \(\Phi(x,y)\) around the plaquette \(p\). The first sum is over oriented plaquettes, while \(\sum_{\{x,y\}}\) runs over nonoriented links. 

In the general non-Abelian formulation, \(\Phi(x,y)\) is an \(N \times N\) complex matrix and
\begin{equation}
	d\Phi = \prod_{i,j} \pi^{-1} \, d\!\Re(\Phi_{ij}) \, d\!\Im(\Phi_{ij}) \,.
\end{equation}
Here, we shall focus on the Abelian-projected formulation, naturally embedded in the general description. In this setting, \(\Phi(x,y)\) is restricted to be diagonal with entries \(\phi_i(x,y) \in \mathbb{C}\)~\cite{Junior:2024urr}. The diagonal link variables \(\phi_i(x,y)\) are the fundamental degrees of freedom in the model, and correspond to elementary center vortices carrying center charge \(\pm 1\). The index \(i\) identifies one of the \(N\) physically equivalent Cartan fluxes \(\beta_i\), \(i = 1,\ldots,N\), which are also the magnetic weights of the defining representation of \(\mathfrak{su}(N)\). Each \(\beta_i\) is a tuple of \(N-1\) real components, defined to satisfy
\begin{align} \label{eq:magneticweights}
	e^{i2\pi\beta_i\cdot T} &= e^{-i2\pi/N} I \,, & \beta\cdot T &= \beta|_q \, T_q \,,
\end{align}
where \(T_q\), \(q = 1, \ldots, N-1\), are the Cartan generators of \(\SU(N)\). Our conventions are summarized in Appendix~\ref{app:suNweights}. The ensemble is invariant under permutations of the \(\beta_i\).

With this characterization, \(N\) elementary center vortices each carrying a different Cartan flux \(\beta_i\) can be naturally matched because of the ``\(N\)-matching'' property
\begin{equation}
	\beta_1 + \cdots + \beta_N = 0 \,.
\end{equation}
This corresponds to an \(N\)-way center monopole, in which \(N\) vortex sheets join at a single link, and is implemented in the partition function by the determinant term. Accordingly, \(Z_\mathrm{c.v.}\) generates a sum over closed elementary center-vortex surfaces carrying one of the \(N\) Cartan fluxes, which can interact via the elementary \(N\)-matching process.

The complete partition function \(Z\) is then obtained by including nonoriented center vortices, along which the flux can change from \(\beta_i\) to \(\beta_j\) due to the presence of a Cartan monopole carrying flux \(\beta_i-\beta_j\). This contrasts with the \(N\)-way center junctions discussed above, at which there is zero net Cartan flux. To describe this mixed ensemble of oriented and nonoriented center vortices, complex fields \(\zeta_\alpha\) were introduced, where \(\alpha\) collectively represents a pair of Cartan-flux indices \(\{i,j\}\) with \(i<j\)~\cite{Oxman:2018dzp,Junior:2024urr}. The resulting partition function is
\begin{equation} \label{eq:4Dnonoriented}
\begin{gathered}[b]
	Z = \frac{1}{\mathcal{N}} \int D\Phi\,  D\zeta \; e^{-W[\Phi,\zeta]} \;, \\[0.4em]
	W[\Phi,\zeta] = W_{\rm c.v.}[\Phi] - \sum_l\sum_{\alpha} \bar{\zeta}_{\alpha}(x) \, \bar{\phi}_j(x,y)\phi_i(x,y) \, \zeta_{\alpha}(y) + \sum_x \sum_{\alpha} \left(\tilde{\eta} \, |\zeta_\alpha(x)|^2 + \tilde{\lambda} \, |\zeta_\alpha(x)|^4 \right) \,.
\end{gathered}
\end{equation}
In this case, \(l = (x,y)\) runs over oriented links. Since \(\phi_i(y,x) = \bar{\phi}_i(x,y)\), reversing the link generates the complex-conjugate hopping term.

\subsection{Partition function on fixed-time slices} \label{subsec:fixedtimepartitionfunction}
In order to study whether the \(\SU(4)\) Yang-Mills lattice simulations support the mixed ensemble, we are interested in describing the center-vortex network on fixed-time slices, where the configurations \(\{\mathcal{S}\}\) produce a network \(\{\gamma\}\) of lines and Cartan fluxes governed by the partition function
\begin{align}
	\mathscr{Z} = \sum_{\{\gamma\}} \psi_{\{\gamma\}} \,.
\end{align}
The statistical properties and the effective description in this case are expected to correspond with those already parametrized in Refs.~\cite{Junior:2019fty, Junior:2022bol} to describe the partition function in 3D Euclidean spacetime and the vacuum wave functional in four dimensions, which must be specified at a given time. On the lattice and in the absence of Cartan monopoles, this description is given by the lower-dimensional version of the model in Eqs.~\eqref{eq:4DZcv} and \eqref{eq:4DWcv}. That is, the partition function \(\mathscr{Z}_\mathrm{c.v.}\) for oriented center-vortex lines is
\begin{gather}
	\mathscr{Z}_\mathrm{c.v.} = \frac{1}{\mathcal{N}} \int D\Phi \; e^{-W_\mathrm{c.v.}[\Phi]} \,, \qquad D\Phi \equiv \prod_x d\Phi(x) \,, \qquad \mathcal{N} = \int D\Phi \; e^{-\sum_x  \mathcal{Q}(\Phi(x))} \,, \label{eq:3DZcv} \\[0.4em]
	W_\mathrm{c.v.}[\Phi] = -\sum_{l} \tr \, \bigl(\Phi^\dagger(x) \Phi(y)\bigr) + \sum_x \left[\mathcal{Q}\bigl(\Phi(x)\bigr) - \xi \, \bigl(\det\Phi(x) + \det\Phi^\dagger(x) \bigr) \right] \,, \label{eq:3DWcv}
\end{gather}
where \(\sum_l\) sums over the oriented links \(l = (x,y)\) and the local potential \(\mathcal{Q}\) is given by Eq.~\eqref{eq:4DQ}. Although the parameters in 3D and 4D  have different dimensions, we use the same names whenever they have a similar physical meaning. Namely, \(\Phi(x)\) is an \(N \times N\) diagonal matrix defined at the sites \(x \in \mathbb{Z}^3\), whose entries are \(N\) complex scalar fields \(\phi_i(x)\) carrying Cartan flux \(\beta_i\). These generate elementary center-vortex lines, with the possibility of \(N\)-matching. 

The three-dimensional counterpart of the nonoriented vortex component in Eq.~\eqref{eq:4Dnonoriented} is obtained by simply adding the term
\begin{align} \label{eq:3Dnonoriented}
	- \zeta \sum_x \sum_{i \neq j} \bar{\phi}_j(x) \phi_i(x) \,,
\end{align}
to \(W_\mathrm{c.v.}[\Phi]\). This produces Cartan monopoles where a Cartan flux of type \(i\) changes into one of type \(j\), quantified by the mixing coefficient \(\zeta\).

Obtaining an effective representation of surfaces or lines in the continuum always requires stiffness. Introducing stiffness as a local curvature cost is essential to obtain an effective field representation of the ensemble, quadratic in derivatives at large distances. This property was explicitly implemented in Refs.~\cite{Junior:2019fty, Junior:2022bol} using polymer techniques developed in Ref.~\cite{Fredrickson:2005} (see also Ref.~\cite{deLemos:2011ww}). Similarly, in lattice models, the relevance of stiffness in preventing surface “crumpling”~\cite{Ambjorn:1997di} is well known. 

Alternatively, rather than taking the continuum limit with explicit stiffness terms, one may introduce a physical lattice spacing \(\tilde{a}\), which plays the role of a persistence length. To see this, we rewrite \(W_\mathrm{c.v.}[\Phi]\) in terms of the finite difference \(\Delta_\mu \Phi = \Phi(x + \hat{\mu}) - \Phi(x)\),
\begin{gather}
	W_\mathrm{c.v.}[\Phi] = \sum_{x,\,\mu} \tr \,\bigl((\Delta_\mu \Phi)^\dagger \Delta_\mu \Phi\bigr) + V(\Phi,\Phi^\dagger) - \sum_x \xi \,\bigl(\det\Phi(x) + \det\Phi^\dagger(x) \bigr) \,, \\
  V(\Phi,\Phi^\dagger) = \sum_x \tr \,\bigl(\rho \, \Phi^\dagger \Phi + \lambda \, (\Phi^\dagger \Phi )^2 \bigr) \,. \label{eq:potential}
\end{gather} 
This recovers Eq.~\eqref{eq:3DWcv} with \(\eta = 2d + \rho\), where \(d = 3\) is the number of dimensions. We can then introduce dimensionful variables
\begin{align}
	\Phi_\mathrm{c}(\tilde{a}x) &= \tilde{a}^{-1/2} \, \Phi(x) \,, & x &\in \mathbb{Z}^3
\end{align}
to obtain a ``discretized'' action quadratic in the derivatives \(\Delta_\mu \Phi_\mathrm{c}/\tilde{a}\), with squared mass \(m^2 = \rho \tilde{a}^{-2}\) and coupling constant \(\lambda \tilde{a}^{-1}\). At large distances compared with \(\tilde{a}\), this is expected to reproduce continuum field behavior, applicable only in that regime.

\section{The structure of center-vortex networks} \label{sec:vortexnetworks}
As we will see, the morphology of the center-vortex ensemble changes qualitatively depending on the sign of \(\rho\) in Eq.~\eqref{eq:potential}. In previous works, the confining properties of the center-vortex ensemble emerge by considering the following hierarchy~\cite{Oxman:2018dzp, Junior:2019fty, Junior:2022bol, Junior:2024urr}. First, a center-vortex condensate is formed for \(\rho < 0\). Within this condensed phase, a strong \(N\)-matching effect (controlled by \(\xi\)) is subsequently considered. Finally, a milder nonoriented component due to Cartan monopoles is incorporated. Because of this hierarchy, it is interesting to discuss the geometry of the center-vortex network while disregarding the Cartan monopole component. That said, it is important to bear in mind that this component is essential to generate the appropriate mass scales and the spontaneous-symmetry-breaking pattern required for the formation of confining flux tubes.

We start by considering the ``normal phase'', with \(\rho > 0\). For \(\xi=0\), the hopping expansion generates a dilute gas of closed center-vortex loops carrying the elementary Cartan fluxes \(\beta_i\). Since \(\Phi\) is diagonal, vortices carrying different fluxes \(\beta_i\), \(\beta_j\), \(i\neq j\), do not interact at this level, while the quartic interaction produces a local repulsion when vortex lines of the same type \(i = j\) meet. For a given elementary type, the contribution of a closed loop \(\mathcal C\) to the partition function is
\begin{align} \label{eq:singleoccupation}
	e^{-\mu\,l(\mathcal C)} \,, && e^{-\mu} &= \frac{1}{z} \int d\phi \; |\phi|^2 \, e^{-\eta|\phi|^2-\lambda|\phi|^4} \,, & z &= \int d\phi \; e^{-\eta|\phi|^2-\lambda|\phi|^4} \,,
\end{align}
where \(l(\mathcal C)\) is the number of links in \(\mathcal C\) and \(d\phi = \pi^{-1} \, d\!\Re(\phi) \, d\!\Im(\phi)\). This expansion is briefly reviewed in Appendix~\ref{app:tensiondecayrate}. For small \(\xi \neq 0\), the perturbative expansion, together with the integration over the field variables, will then select configurations in which one chain of hopping terms of each type \(i\), ending or starting at \(x\), combines with \(\det\Phi(x)\) or \(\det\Phi^\dagger(x)\), respectively, to form a singlet at \(x\). This corresponds to the \(N\) different elementary center-vortex types matching at an \(N\)-way center monopole or antimonopole. In this case, the parameter \(\xi\) plays the role of a fugacity for these pointlike matching events.

While these center monopoles do appear in Monte Carlo simulations of both \(\SU(3)\) and \(\SU(4)\) Yang-Mills theory, the percolating \(\SU(4)\) ensemble exhibits a considerably richer morphology than suggested by the dilute expansion in terms of the elementary processes. In particular, extended vortex matching (VM) chains are far more frequent than pointlike four-way center monopoles~\cite{Mickley:2025ksp}. Vortex convergence (VC), in which two elementary vortex components combine to form a doubly charged chain, is also observed with a frequency comparable to that of VM.

Here, based on the three-dimensional Villain approximation, we show that the condensate naturally gives rise to this structure, with charge-one chains matched in pairs to typically shorter charge-two chains, as observed in Yang-Mills simulations. The distinction between noninteracting (\(i \neq j\)) and interacting (\(i = j\)) elementary vortex components provides a simple mechanism for this morphology: repulsive interactions can stabilize the network, while the superposition of elementary vortices with \(\beta_i\neq\beta_j\) allows the formation of doubly charged chains.

When \(\mu\) goes below a critical value, \(\rho\) becomes negative. Consequently, the potential in Eq.~\eqref{eq:potential} can be written as
\begin{align}
	V(\Phi,\Phi^\dagger) &= \lambda \, \sum_x \tr \,\bigl(\Phi^\dagger \Phi - \vartheta^2 \bigr)^2 + \mathrm{const.} \,, & \vartheta^2 &= -\frac{\rho}{2\lambda} \,,
\end{align}
with \(\vartheta^2\) positive. The potential therefore develops minima at
\begin{align}
	\Phi &= \vartheta U \,, & U^\dagger U &= I \implies U \in U(1)^N \,.
\end{align}
In the thermodynamic limit, the system undergoes spontaneous symmetry breaking, so it becomes dominated by the modes 
\begin{align}
	\Phi(x) &= \vartheta U(x) \,, & U(x) &\in U(1)^N \,,
\end{align}
which minimize \(V(\Phi, \Phi^\dagger)\). This is the center-vortex condensate, or percolating, phase.

The condensate is thus governed by the action
\begin{gather}
	W_\mathrm{c.v.}[\Phi] \approx \vartheta^2 \sum_{x,\,\mu} \tr \, \bigl((\Delta_\mu U)^\dagger \Delta_\mu U \bigr) - \xi \, \vartheta^N \sum_x \bigl(\det U(x) + \det U^\dagger(x)\bigr) \,.
\end{gather}
A sufficiently strong \(\xi\) forces \(\det U(x) = 1\). That is, because of \(N\)-matching, \(U(x)\) is not only approximated as a diagonal unitary matrix, but it is also close to the Cartan subgroup of \(\SU(N)\),
\begin{align}
	U(x) &= e^{i\theta(x)\cdot T} \in U(1)^{N-1} \,, & \theta \cdot T &= \theta|_q T_q \,,
\end{align}
where \(T_q\) are the Cartan generators. Therefore, defining \(\beta = 2\vartheta^2\), the effective description becomes
\begin{align} \label{eq:percolatingWcv}
	W_\mathrm{c.v.}[\theta] &\approx \beta \sum_{x,\,\mu} \tr \left(I -\cos(\Delta_\mu(x) \cdot T)  \right) \,, & \Delta_\mu(x) &= \theta(x + \hat{\mu}) - \theta(x) \,.
\end{align} 
This takes a similar form to the classical XY model, but with a Cartan angle \(\Delta_\mu(x) \cdot T\) in place of a real angle. In this case, excluding Cartan monopoles, there are large phase fluctuations. To expose the resulting vortex-network structure, we introduce a Villain approximation that implements the correct periodicity in the Cartan subalgebra of \(\mathfrak{su}(N)\).

\subsection{Villain representation} \label{subsec:villainapprox}
In the usual 3D XY model, neighboring sites interact via \(\cos\Delta_\mu(x)\), where \(\Delta_\mu(x)=\theta(x+\hat{\mu})-\theta(x)\) is the difference of neighboring angles. The Villain approximation~\cite{Villain:1974ir} implements the \(2\pi\)-periodicity of the real angle \(\Delta\),
\begin{align} \label{eq:villain}
	e^{\beta\cos\Delta} &\approx R_\mathrm{V} \sum_{n\in\mathbb Z} e^{-\frac{\beta_\mathrm{V}}{2}(\Delta-2\pi n)^2} \,, & R_\mathrm{V} &= I_0(\beta)\sqrt{2\pi\beta_\mathrm{V}} \,, & e^{-\frac{1}{2\beta_\mathrm{V}}} &= \frac{I_1(\beta)}{I_0(\beta)} \,,
\end{align}
where \(I_n\) are modified Bessel functions of the first kind. 

In our case, the weight factor involves a Cartan ``angle'' 
\begin{align} \label{eq:expcos}
	& e^{\beta \tr \cos(\Delta \cdot T)} \,, &
	\cos(\Delta \cdot T) &= \frac{e^{i \Delta \cdot T} + e^{-i \Delta \cdot T}}{2} \,,
\end{align} 
As each diagonal entry of the Cartan interaction has the same form as the ordinary XY interaction, the Villain approximation can be applied entry by entry with the same function \(\beta_\mathrm{V}(\beta)\). It is clear that Eq.~\eqref{eq:expcos} is periodic under shifts of the form \(\Delta \to \Delta + 2\pi\nu\) for any \(\nu\) satisfying
\begin{equation}
	e^{i2\pi\nu\cdot T} = I \,.
\end{equation}
The diagonal entries of the Cartan generators are precisely the components of the weights \(\omega_i\) of the defining representation [cf.\ Eq.~\eqref{eq:weights}], so that
\begin{equation}
	\left. e^{i2\pi\nu\cdot T}\right|_{ii} = e^{i2\pi\nu\cdot\omega_i} \,.
\end{equation}
Therefore, the periodicity condition is satisfied when \(\nu\cdot\omega_i \in \mathbb Z\) for every \(i\). This defines the coroot lattice of \(\mathfrak{su}(N)\). Accordingly, the periodic integer sums associated with the diagonal entries are organized as a sum over the coroot lattice. The derivation, presented in Appendix~\ref{app:villainapprox}, leads to
\begin{equation} \label{eq:generalizedvillain}
	e^{\beta \tr\cos(\Delta \cdot T)} \approx \mathrm{const.} \sum_{\{\nu\}} e^{-\frac{\beta_\mathrm{V}}{4N} (\Delta-2\pi\nu)^2} \,.
\end{equation}
Using Poisson summation over the coroot lattice, this can be rewritten as
\begin{gather}
	e^{\beta \tr\cos(\Delta \cdot T)} \approx \mathrm{const.} \sum_{w} e^{-\frac{N}{\beta_\mathrm{V}} w^2} e^{i w \cdot \Delta} \,,
\end{gather}
where \(w\) belongs to the weight lattice generated by integer combinations of the defining weights \(\omega_i\).

Implementing this approximation in the fixed time-slice partition function, we have 
\begin{equation} \label{eq:villainZcv}
\begin{aligned}[b]
	\mathscr{Z}_\mathrm{c.v.} & \approx \frac{1}{\mathcal{N}} \int D\theta \; e^{-\beta \sum_{x,\mu} \tr \, \left(I - \cos(\Delta_\mu(x) \cdot T)\right)} \\[0.3em]
	& \approx \frac{1}{\mathcal{N}} \int D\theta \; \sum_{\{w\}} e^{-\frac{N}{\beta_\mathrm{V}} \sum_{x,\mu} \left(w_\mu(x) \right) ^2} e^{i \sum_{x,\mu} w_\mu(x) \cdot \Delta_\mu(x)} \,,
\end{aligned}
\end{equation}
where any constant factors have been absorbed into \(\mathcal{N}\). Now, note that
\begin{align}
	\sum_{x,\,\mu} w_\mu(x) \cdot \Delta_\mu(x) &= -\sum_x (\operatorname{div} w)(x) \cdot \theta(x) \,, & (\operatorname{div} w)(x) = \sum_{\mu} \left(w_\mu(x) - w_\mu(x-\mu)\right) \,.
\end{align}
Substituting this into Eq.~\eqref{eq:villainZcv}, the integral over \(\theta(x)\) enforces that \(w_\mu(x)\) must be divergenceless in the absence of Cartan monopoles. Thus, the partition function for the oriented center-vortex condensate simplifies to the following sum over divergenceless weight-lattice fluxes,
\begin{align} \label{eq:divZcv}
	\mathscr{Z}_\mathrm{c.v.} &\propto \sum_{\{\lambda,\,\operatorname{div}\lambda=0\}} e^{-W[\lambda]} \,, & W[\lambda] &= \frac{1}{4N\beta_\mathrm{V}} \sum_{x,\,\mu} \left(\lambda_\mu(x) \right)^2 \,,
\end{align} 
where \(\lambda_\mu(x) = 2Nw_\mu(x)\) takes values in the lattice of magnetic weights, generated by \(\beta_i = 2N\omega_i\).

This is the Cartan counterpart of the conserved integer-current representation of the 3D XY model, where integration over the angular variables imposes the corresponding zero-divergence constraint~\cite{Kleinert:1989}. There, the integer-valued currents admit a representation in terms of closed non-backtracking paths carrying unit flux. Immediate forward-backward retracing is then removed as a redundant description of the same current configuration. The present construction admits a natural Cartan generalization of this. A convenient independent basis of the lattice of magnetic weights is provided by
\begin{equation}
\begin{gathered}
	\lambda = \sum_{q=1}^{N-1} m^q \lambda_q \,, \quad\quad m^q \in \mathbb{Z} \,, \\
	\lambda_1 = \beta_1 \,, \quad\quad \lambda_2 = \beta_1+\beta_2 \,, \quad\quad \ldots \,, \quad\quad \lambda_{N-1} = \beta_1+\cdots+\beta_{N-1} \,,
\end{gathered}
\end{equation}
where the \(N-1\) fundamental magnetic weights \(\lambda_q\) generate the lattice without overcounting. Unlike the \(N\) physically equivalent elementary Cartan fluxes \(\beta_i\), which satisfy the constraint \(\beta_1+\cdots+\beta_N=0\), the \(N-1\) magnetic weights \(\lambda_q\) constitute an independent basis and provide convenient representatives naturally organized according to center charge. Namely, each \(\lambda_q\) carries center charge \(q\). This follows from its action on Wilson loops:\ \(e^{i 2\pi \lambda_q \cdot T} = e^{-i 2\pi q/N}I\) [cf.\ Eq.~\eqref{eq:magneticweights}].

Accordingly, the sum over \(\lambda\), \(\operatorname{div}\lambda = 0\), can be written without overcounting as a sum over networks
\begin{align}
	\{\mathcal{C}\} = \bigcup_q \, \{\mathcal{C}_q\} \,,
\end{align}
which define the divergenceless fluxes
\begin{equation} \label{eq:divergencelessfluxes}
	\lambda_\mu(x) = \sum_{q=1}^{N-1} \lambda_q \, j_\mu^q(x) \,.
\end{equation}
Here, \(\{\mathcal{C}_q\}\) denotes the collection of non-backtracking loops associated with the magnetic weight \(\lambda_q\) in a given network configuration, and \(j_\mu^q(x)\) counts the signed total number of times the loops in \(\{\mathcal{C}_q\}\) pass a given link in the positive (\(+\)) or negative (\(-\)) sense.

\subsection{Energetics of center-charge sectors} \label{subsec:energetics}
The effective model is intended to provide a continuum description of the infrared Yang-Mills ensemble and should therefore be considered in the scaling regime, where the correlation length is much larger than the lattice spacing. In the 3D XY model, this regime is associated with the proliferation of large loops as the critical point \(\beta_{\rm c}\simeq0.454\) is approached. The analogue of the weight-lattice-valued current \(w_\mu(x)\) introduced above is an integer-valued link current, which counts the signed number of passages through a given link. Although a given link may be traversed more than once, leading to a current with magnitude larger than one, the hierarchy of the modified Bessel coefficients in the exact current expansion suppresses such higher-current values~\cite{Kleinert:1989}. Indeed, within a range of smaller \(\beta\) extending up to the phase transition, link currents of unit magnitude dominate over those with larger magnitude~\cite{Janke:1986ej}. This provides a natural basis for a single-occupation approximation near criticality. Consistently, an explicit truncation to singly occupied links, \(w_\mu(x)=0,\pm1\), is known to preserve the XY universality class~\cite{Winter:2008}.

This behavior has a natural counterpart in the Cartan network. Recall that the Villain approximation was applied independently to each diagonal entry of the Cartan interaction. Since \(T_q|_{ii}=\omega_i|_q\), the \(i\)th diagonal entry depends on \(\Delta\cdot\omega_i\), so that the associated integer measures the multiplicity of the elementary Cartan flux \(\beta_i=2N\omega_i\) on a given link. Thus, multiple occupation by the same elementary Cartan flux \(\beta_i\) is expected to be suppressed. This is consistent with the original center-vortex model, where elementary vortex lines carrying the same Cartan flux repel, whereas vortices carrying different elementary fluxes may coexist.

Since each \(\lambda_q = \beta_1 + \cdots + \beta_q\) is itself a sum of distinct elementary fluxes, multiple occupation by the same \(\lambda_q\) would necessarily repeat these elementary components. We therefore restrict the individual currents to
\begin{equation} \label{eq:jmusingleoccupation}
	j_\mu^q(x) = 0, \pm 1 \,.
\end{equation}
In principle, this restriction does not prevent currents of different types from occupying the same link. Let us therefore consider the possibility that loops of type \(q\) share some links with loops of a different type \(p\neq q\). The collection of \(q\)-type paths can then be decomposed as the disjoint union
\begin{equation} \label{eq:Cqunion}
	\{C_q\} = \{\gamma_q\} \,\bigcup_{\substack{p,\\p \neq q}} \{\gamma_{qp}^+\} \, \bigcup_{\substack{p,\\p \neq q}} \{\gamma_{qp}^-\} \,,
\end{equation}
where \(\{\gamma_q\}\) contains links occupied only by loops of type \(q\), while \(\{\gamma_{qp}^\pm\}\) contains links simultaneously occupied by loops of type \(q\) and \(p\) with equal (\(+\)) or opposite (\(-\)) orientations, which carry (absolute) center charge \(\lvert q\pm p\rvert\pmod N\). 

Substituting Eq.~\eqref{eq:divergencelessfluxes} into Eq.~\eqref{eq:divZcv}, the action can be written as
\begin{equation}
\begin{aligned}[b]
	W[\lambda] &= \frac{1}{4N\beta_\mathrm{V}} \sum_{x,\,\mu} \left(\sum_q \lambda_q^2 \, j_\mu^q(x)^2 + 2\sum_{q<p} \left(\lambda_q\cdot\lambda_p\right) j_\mu^q(x) j_\mu^p(x) \right) \\
	&= \frac{1}{4N\beta_\mathrm{V}} \left(\sum_q \lambda_q^2 \, l\{\mathcal C_q\} + 2\sum_{q<p} \left(\lambda_q\cdot\lambda_p\right) \Bigl(l\{\gamma_{qp}^+\} - l\{\gamma_{qp}^-\} \Bigr) \right) \,,
\end{aligned} 
\end{equation}
where we have used the restriction in Eq.~\eqref{eq:jmusingleoccupation}. Here, \(l\{\gamma\}\) denotes the total length (number of links) of the paths in \(\{\gamma\}\). In addition, the total length of \(\{\mathcal C_q\}\) can be decomposed according to Eq.~\eqref{eq:Cqunion} in terms of \(\{\gamma_q\}\) and \(\{\gamma_{qp}^\pm\}\),
\begin{equation}
\begin{aligned}[b]
	\sum_q \lambda_q^2 \, l\{\mathcal C_q\} &= \sum_q \lambda_q^2\,l\{\gamma_q\} + \sum_{p\neq q}\lambda_q^2 \Bigl(l\{\gamma_{qp}^+\} + l\{\gamma_{qp}^-\}\Bigr) \\
	&= \sum_q \lambda_q^2 \, l\{\gamma_q\} + \sum_{q<p} \left(\lambda_q^2 + \lambda_p^2\right) \Bigl(l\{\gamma_{qp}^+\} + l\{\gamma_{qp}^-\} \Bigr) \,.
\end{aligned}
\end{equation}
The action in the percolating phase can thus be expressed as
\begin{equation} \label{eq:networktensions}
\begin{gathered}
	W[\lambda] = \sum_q \mu_q \, l\{\gamma_q\} + \sum_{q<p} \Bigl(\mu_{qp}^+ \, l\{\gamma_{qp}^+\} + \mu_{qp}^- \, l\{\gamma_{qp}^-\}\Bigr) \,, \\
	\mu_q = \frac{\lambda_q^2}{4N\beta_\mathrm{V}} \,, \qquad\qquad \mu_{qp}^\pm = \frac{(\lambda_q\pm\lambda_p)^2}{4N\beta_\mathrm{V}} \,,
\end{gathered}
\end{equation}
where \(\mu_q\) and \(\mu_{qp}^\pm\) are the tensions associated with the corresponding types of paths.

For a given center charge \(q\), the minimum tension is attained by sums of \(q\) distinct elementary Cartan fluxes (see Appendix~\ref{app:suNweights}). The fundamental weight \(\lambda_q = \beta_1 + \cdots + \beta_q\) is precisely one such representative, and as such the tension \(\mu_q\) is the smallest in that center-charge sector. Since each contribution to the partition function is weighted by \(e^{-\mu l}\), it is these minimum-tension realizations that dominate the long-distance statistics. On the other hand, the tensions \(\mu_{qp}^\pm\) associated with center charge \(q \pm p\) (mod \(N\)) do not generally coincide with the minimum tension in the corresponding center-charge sector. 

The percolating phase is therefore described by a network whose center-charge sectors emerge from the simple dynamics of the physically equivalent elementary Cartan fluxes, without needing to introduce independent phenomenological parameters for each sector. The dominant tensions satisfy
\begin{align} \label{eq:dominanttensions}
	\frac{\mu_q}{\mu_1} &= \frac{q(N-q)}{N-1} \,, & \mu_1 &= \frac{N-1}{2N\beta_\mathrm{V}} \,.
\end{align}
This has the same form as the asymptotic Casimir scaling law for \(k\)-string tensions, which we will return to in Sec.~\ref{sec:discussion}. We now proceed to discuss the specific application to \(\SU(3)\) and \(\SU(4)\).

\subsubsection{\texorpdfstring{\(\SU(3)\)}{SU(3)}} \label{subsubsec:SU3chains}
In \(\SU(3)\), there are four types of center-vortex chains:\ \(\gamma_1\), \(\gamma_2\), \(\gamma_{12}^-\) and \(\gamma_{12}^+\). The first three of these all realize the elementary center-charge sector \(|q| = 1\), as their associated magnetic weights \(\lambda_1\), \(\lambda_2\), and \(\lambda_1-\lambda_2\) simply correspond to the elementary Cartan fluxes,
\begin{align}
	\lambda_1 &= \beta_1 \,, & \lambda_2 &= \beta_1 + \beta_2 = -\beta_3 \,, & \lambda_1 - \lambda_2 &= -\beta_2 \,,
\end{align}
with the minus sign indicating the opposite spacetime orientation. As such, these chains are all described by the minimum tension \(\mu_1\).

The remaining chains of type \(\gamma_{12}^+\) instead carry Cartan flux 
\begin{gather}
	\lambda_1 + \lambda_2 = \beta_1 + \beta_1 + \beta_2 = \beta_1 - \beta_3 \,.
\end{gather}
Using the explicit form of the \(\beta_i\) for \(\mathfrak{su}(3)\) in Appendix~\ref{app:suNweights}, the associated tension is \(\mu_{12}^+ = 3\mu_1\) such that these chains are suppressed relative to the elementary ones. However, because their center charge is \(3 \equiv 0 \pmod{3}\), chains of type \(\gamma_{12}^+\) are invisible to center projection in Yang-Mills simulations.

\subsubsection{\texorpdfstring{\(\SU(4)\)}{SU(4)}} \label{subsubsec:SU4chains}
For \(\SU(4)\), there are two different groups of chains with \(|q| = 1\). The first group corresponds to the types \(\gamma_1\), \(\gamma_3\), \(\gamma_{12}^-\), \(\gamma_{23}^-\), which are elementary,
\begin{align}
	\lambda_1 &= \beta_1 \,, & \lambda_3 &= \beta_1 + \beta_2 + \beta_3 = -\beta_4 \,, & \lambda_1 - \lambda_2 &= -\beta_2 \,, & \lambda_2 - \lambda_3 &= -\beta_3 \,,
\end{align}
and all realize the minimal tension \(\mu_1\). The chains in the second group, \(\gamma_{12}^+\) and \(\gamma_{23}^+\), carry the same center charge but have the larger tension \(\mu^+_{12} = \mu^+_{23} = \frac{11}{3}\mu_1\). These lines are hence strongly suppressed with respect to the first group, though we note that the two groups are indistinguishable within center projection.

The ``doubly charged'' chains with center charge \(|q| = 2\), novel to \(\SU(4)\), are of type \(\gamma_2\) and \(\gamma_{13}^-\). Their magnetic weights are
\begin{align}
    \lambda_2 &= \beta_1+\beta_2=-(\beta_3+\beta_4) \,, &
    \lambda_1-\lambda_3 &= -(\beta_2+\beta_3)=\beta_1+\beta_4 \,.
\end{align}
These are different minimum-norm realizations of the charge-two sector, related by permutations of the \(\beta_i\). Both correspond to the coalescence of two elementary center vortices carrying distinct Cartan fluxes into a single doubly charged chain, and have tension
\begin{equation}
	\mu_2 = \mu_{13}^- = \frac{4}{3} \mu_1 \,.
\end{equation}
Finally, the chain \(\gamma_{13}^+\) with tension \(\mu_{13}^+ = \frac{8}{3}\mu_1\) belongs to the trivial center-charge sector. This plays an analogous role to the \(\mu_{12}^+\) chain in \(\SU(3)\), and cannot be detected by center projection.

\begin{figure}
	\centering
	\includegraphics[width=0.53\linewidth]{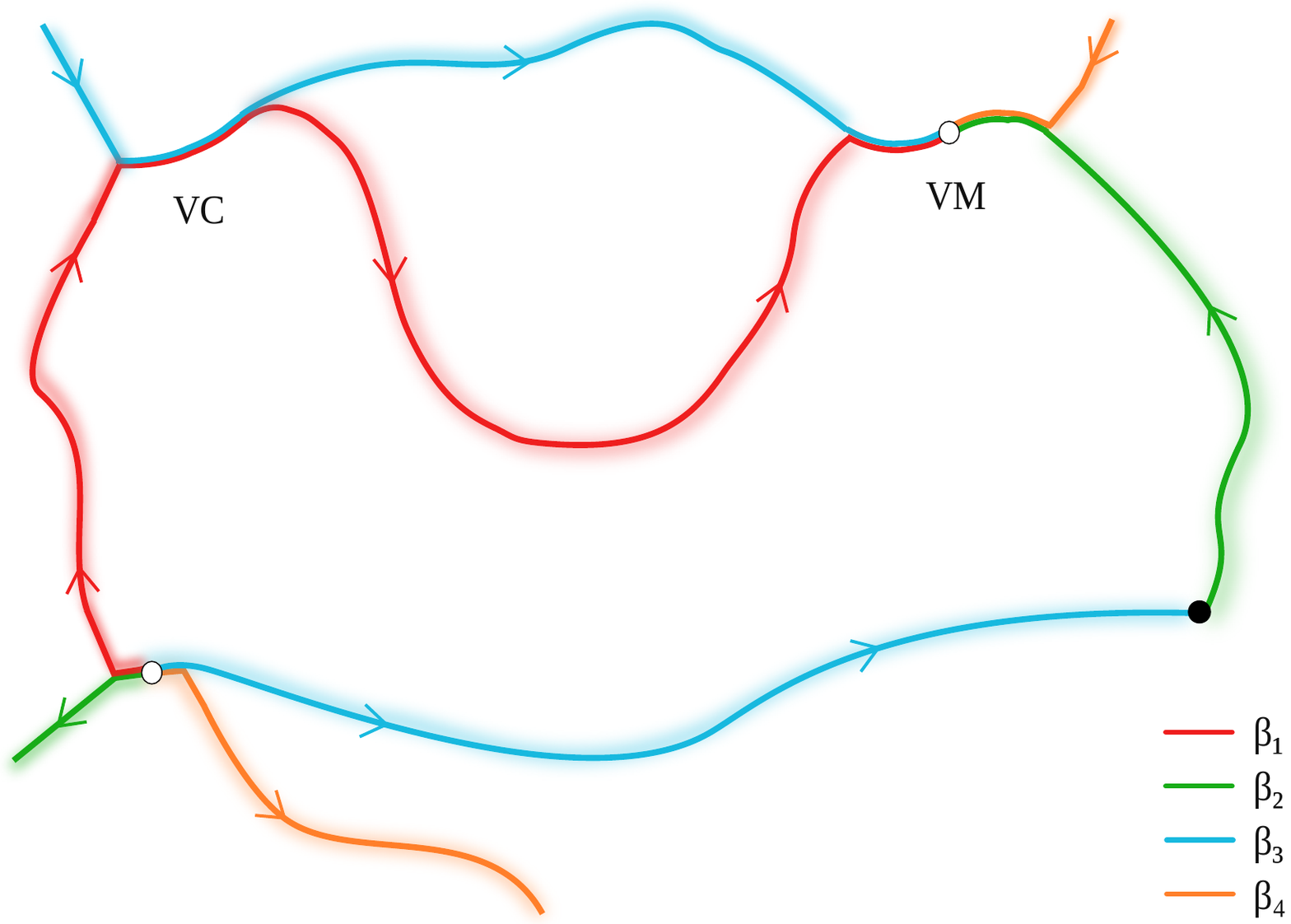}
	\caption{Illustration of the essential components of the \(\SU(4)\) network in the percolating phase. Four vortex lines can meet at a center monopole (white dots), while the Cartan flux can change its orientation in the Lie algebra through a Cartan monopole (black dot). In addition, novel to \(\SU(4)\), elementary center vortices carrying different Cartan fluxes can coalesce into doubly charged chains through vortex-convergence (VC) and vortex-matching (VM) processes. The former of these is formed by two elementary components that temporarily converge, while the latter looks like four vortices that match through a center monopole over an extended distance.}
	\label{fig:su4network}
\end{figure}

As explained above, the dominant contribution to the percolating phase is given by the minimum-tension realization in each observable center-charge sector, i.e.\ \(\mu_1\) and \(\mu_2 = \frac{4}{3}\mu_1\). The other realizations are exponentially suppressed by their larger tensions. Crucially, although \(\mu_2 > \mu_1\), it remains smaller than \(2\mu_1\). Thus, when two elementary center-vortex lines approach, their coalescence into a doubly charged chain is energetically favored over remaining separated. As such, while the normal phase is described by a gas of four-way center monopoles, the percolating phase is instead expected to develop a morphology consisting of long elementary center-vortex chains connected by typically shorter doubly charged chains. This is illustrated in Fig. \ref{fig:su4network}, and is in agreement with the \textit{ab initio} \(\SU(4)\) simulations of Ref.~\cite{Mickley:2025ksp}.

Accordingly, the simplest effective model, where nonoriented components are neglected, predicts that the tensions associated with the singly and doubly charged center-vortex chains identified in the center-projected Yang-Mills ensemble satisfy
\begin{equation} \label{eq:muratiomodel}
	\frac{\mu_2}{\mu_1} = \frac{4}{3} \,.
\end{equation}

\section{Effective model and lattice observables} \label{sec:latticeobservables}
For comparing to the model, we are interested in providing a determination of the tension ratio \(\mu_2/\mu_1\) using the center-vortex network observed in numerical simulations of Yang-Mills theory. To this end, in this section we introduce and motivate lattice observables that can be related to the effective model. For this, two ingredients are needed:\ an understanding of the entropic factors that govern how likely a chain is to reach a given length, and a physical criterion to fix the effective lattice spacing \(\tilde{a}\) from the geometry of the chains themselves.

\subsection{Entropy and the physical decay rates} \label{subsec:decayrates}
In the effective model, the center-vortex tensions \(\mu\) determine how energetically favored each configuration is according to the Boltzmann factor \(e^{-\mu k}\) for a chain of length \(k\). However, the probability of finding a chain of a given length is not controlled by the tension alone. The Boltzmann factor needs to be multiplied by the number of configurations with that length, an entropic factor that depends on how many ways a chain can continue at each step, known as the \emph{connectivity factor}. This effect is well known for growing paths in general, from Brownian paths to self-avoiding walks, and is discussed for the simple case of a noninteracting polymer in Appendix~\ref{app:tensiondecayrate}.

Denoting the connectivity factor by \(g\), the number of possible paths of length \(k\) is \(g^k\). Thus, the probability of a path formed by \(k\) monomers is expected to be proportional to
\begin{align} \label{eq:modelprobability}
	g^k e^{-\mu k} &= e^{-(\mu-\ln g) k } = \exp\left(-\frac{\mu -\ln g}{\tilde{a}} \, L \right) \,, & L &= k\tilde{a} \,,
\end{align}
where, as discussed in Sec.~\ref{subsec:fixedtimepartitionfunction}, \(\tilde{a}\) is the effective lattice spacing in the model. That is, the probability is governed by the combination \(\mu - \ln g\). For a simple random chain, \(g = 2d\) (backward and forward in each dimension). However, for the percolating center-vortex network described by the Villain approximation of Sec.~\ref{subsec:villainapprox}, the conserved-current representation naturally excludes an immediate reversal, since it does not contribute to the net current~\cite{Kleinert:2009}. The corresponding loops are therefore non-backtracking, with \(g=2d-1=5\) possible steps on a \(d=3\) time slice. The same type of correction has been considered in \(d = 4\) for monopole worldlines~\cite{Gubarev:2002kk, Zakharov:2002md, Zakharov:2003nm}. For example, in polymer networks the local connectivity factor remains unchanged when chains meet at junctions~\cite{Duplantier:1989, Schafer:1992vsw, Duplantier:2019ega}.

This motivates a study of the numerical probability distribution for the center-vortex chain lengths. If the lattice data were to follow an exponential distribution \(\sim e^{-\lambda L}\) with a decay rate \(\lambda\), then equating this with Eq.~\eqref{eq:modelprobability} leads to the identification
\begin{equation} \label{eq:decayratetension}
	\lambda = \frac{\mu -\ln g}{\tilde{a}} \,.
\end{equation}
This has previously been observed in \(\SU(3)\)~\cite{Biddle:2023lod, Mickley:2024zyg} and for doubly charged chains in \(\SU(4)\)~\cite{Mickley:2025ksp}, barring some short-distance effects. This will be elaborated on in Sec.~\ref{subsec:stoppingrates}, where we demonstrate that a finite physical decay rate for both charge-one and charge-two chains in \(\SU(4)\) is implied in the continuum. Therefore, we can translate these decay rates to center-vortex tension ratios in the model,
\begin{equation} \label{eq:muratiorates}
	\frac{\mu_2}{\mu_1} = \frac{\lambda_{(2)} \, \tilde{a} + \ln 5}{\lambda_{(1)} \, \tilde{a} + \ln 5} \,.
\end{equation}

\subsection{Effective lattice spacing and stiffness} \label{subsec:stiffness}
To evaluate Eq.~\eqref{eq:muratiorates}, the second ingredient needed to relate the effective model to Yang-Mills data is the physical lattice spacing \(\tilde{a}\). This represents stiffness and therefore is closely related with \emph{persistence length}, which is a local measure for the intrinsic tendency of a chain to stay aligned. Suppose we have a chain formed by \(n\) monomers \(\mathbf{s}_i\), \(i = 1, \ldots, n\), of length \(b\). Flory's definition of the persistence length \(\xi_\mathrm{p}(i)\) of a chain at site \(i\) corresponds to the average projection of the displacement vector between site \(i\) and the chain's endpoint onto the \(i\)th monomer~\cite{Flory:1969},
\begin{align} \label{eq:localpersistence}
	\xi_\mathrm{p}(i) &= \Braket{\mathbf{t}_i \cdot \sum_{j=i}^n \mathbf{s}_j} = b \sum_{j=i}^n \braket{\mathbf{t}_i \cdot \mathbf{t}_j} \,, & \mathbf{t}_i &= \frac{1}{b} \, \mathbf{s}_i \,.
\end{align}
This is a local definition, measuring the extent to which the chain deviates from its orientation at site \(i\) onward on average.

Since the tangent correlations \(\braket{\mathbf{t}_i \cdot \mathbf{t}_j}\) decay as \(j\) moves away from \(i\), Eq.~\eqref{eq:localpersistence} is expected to be \(i\)-independent provided \(i\) is far from the chain endpoints. Thus, up to a small boundary effect for long chains, the \emph{global} persistence length, \(\xi_\mathrm{p}(L)\), for a chain of length \(L = nb\) can be defined as an average over the local persistence on each segment,
\begin{equation} \label{eq:globalpersistence}
	\xi_\mathrm{p}(L) = \frac{1}{n} \sum_{i=1}^n \xi_\mathrm{p}(i) = b + \frac{b}{n} \sum_{i < j} \braket{\mathbf{t}_i \cdot \mathbf{t}_j} \,.
\end{equation}
The average relative tangent orientations \(\braket{\mathbf{t}_i \cdot \mathbf{t}_j}\) between sites \(i\) and \(j\) are usually assumed to only depend on the distance between monomers. Therefore, defining \(c_m = \braket{\mathbf{t}_i \cdot \mathbf{t}_{i+m}}\), we have
\begin{equation}
	\xi_\mathrm{p}(L) = b + \frac{b}{n} \sum_{m=1}^{n-1} \left(n - m\right) c_m \,.
\end{equation}
Now, a Markov wormlike chain (WLC) is characterized by \(c_m = c^m\), \(c = c_1 = \braket{\mathbf{t}_i \cdot \mathbf{t}_{i+1}}\). The above sum can then be evaluated, and written as
\begin{align} \label{eq:wlcpersistence}
	\xi_\mathrm{p}(L) &= \xi_0 \left[1 - \frac{\xi_0}{L} \left(1 - c^n\right) \right] \,, & \xi_0 &= \frac{b}{1-c} \,.
\end{align}
We see that the dominant contribution to the persistence length, \(\xi_0\), is \(L\)-independent.

For example, in the case of a simple random chain (SRC), after performing a step along \(\mathbf{t}_i\) the next six possible steps along \(\mathbf{t}_{i+1}\) have the same probability. Thus,
\begin{equation}
	c = \braket{\mathbf{t}_i \cdot \mathbf{t}_{i+1}} = 0 \implies \xi_0 = b \,.
\end{equation}
That is, the persistence length of a SRC is simply equal to the monomer size \(b\) for long chains. Nonetheless, we are interested in non-backtracking walks (NBW). In this case, the backtracking step has zero probability, while the other five steps (four orthogonal and one forward) have probability \(\frac{1}{5}\) such that
\begin{equation} \label{eq:xi0nbw}
	c = \braket{\mathbf{t}_i \cdot \mathbf{t}_{i+1}} = \frac{1}{5} \implies \xi_0 = \frac{5}{4} \, b \,.
\end{equation}
It is intuitive that the additional orientational growing property of the NBW leads to an increase of the persistence length compared to a SRC.

In the effective model, the monomer size is taken to be the effective lattice spacing \(\tilde{a}\). This opens up the possibility of fixing \(\tilde{a}\) based on the persistence length, motivating an exploration of how chains in the Yang-Mills center-vortex network actually grow. Previously, random surface models have incorporated stiffness via phenomenological curvature terms in the vortex action~\cite{Engelhardt:1999wr, Engelhardt:2003wm, Engelhardt:2005qu}. Subsequently, correlations between the Yang-Mills action and curvature of center-vortex surfaces were observed directly in lattice simulations~\cite{Buividovich:2007mm}. Polymer-inspired geometrical observables have also been introduced to characterize monopole trajectories~\cite{Bornyakov:2003kn}. 

In this work, we will instead characterize the persistence properties of center-vortex chains using the mean-square end-to-end distance measured directly on the observed network. As derived in Appendix~\ref{app:endtoend}, the average mean-square displacement between a chain's endpoints is generally related to its persistence length via
\begin{equation} \label{eq:R2xi}
	\braket{R^2} = 2 \left(\xi_\mathrm{p}(L) - \frac{b}{2}\right) L \,.
\end{equation}
Accordingly, by computing \(\braket{R^2}\) as a function of chain length \(L\), we will extract the persistence and relate this to the effective lattice spacing of the model. In the context of Yang-Mills simulations, the monomer size \(b\) is simply the lattice simulation spacing \(a\). As \(a\) is reduced to attain the continuum properties, Eq.~\eqref{eq:R2xi} simplifies to
\begin{equation} \label{eq:R2continuum}
	\braket{R^2} \underset{a \to 0}{=} 2 \, \xi_\mathrm{p}(L) \, L \,.
\end{equation}
This is valid for any possible correlations among the chain elements. For example, a linear relationship between \(\braket{R^2}\) and chain length, 
\begin{equation} \label{eq:linear}
	\braket{R^2} = 2 \, \xi_0 \, (L-L_0) \,,
\end{equation}
would be associated with the dominant persistence-length terms 
\begin{equation}
	\xi_\mathrm{p}(L) = \xi_0 + \order(L^{-1}) + \ldots \;.
\end{equation}
In this case, matching \(\xi_0\) with the model would allow the effective spacing to be identified according to Eq.~\eqref{eq:xi0nbw} as \(\tilde{a} = \frac{4}{5} \xi_0\). In the following section, we will see what Yang-Mills simulations reveal about the behavior of \(\braket{R^2}\) as a function of \(L\). To our knowledge, this is the first determination of the persistence properties from such polymer observables in Yang-Mills theory.

\section{Numerical simulations} \label{sec:simulations}
Having introduced the relevant aspects of polymer physics, we are now ready to detail our numerical analysis of center-vortex chains in lattice Yang-Mills theory. We consider both \(\SU(3)\) and \(\SU(4)\) to expose any differences that arise from the presence of doubly charged chains in the latter. To avoid confusion with the effective spacing \(\tilde{a}\) of the model, we will refer to the lattice spacing in our simulations, \(a\), as the ``lattice simulation'' spacing.

We perform our simulations using the Cabibbo-Marinari algorithm~\cite{Creutz:1980zw, Cabibbo:1982zn} with the standard Wilson action~\cite{Wilson:1974sk},
\begin{align}
	S_W &= \beta \sum_{x,\mu<\nu} \left[1 - \frac{1}{N} \Re\tr P_{\mu\nu}(x)\right] \,, & \beta = \frac{2N}{g^2} \,.
\end{align}
In \(\SU(3)\), we generate a single \(32^3 \times 64\) ensemble at \(\beta = 5.95\), which corresponds to a lattice simulation spacing of \(a\sqrt{\sigma} \simeq 0.236 \!\implies\! a \simeq \qty{0.10}{fm}\)~\cite{Lucini:2005vg}. For \(\SU(4)\), we consider four different \(\beta\) values to investigate the approach to the continuum limit, with the lattice extents varied to retain an approximately constant physical volume. The coarsest \(\SU(4)\) ensemble closely matches our \(\SU(3)\) ensemble, and the precise \(\SU(4)\) simulation details are provided in Table~\ref{tab:ensembledetails}. 100 configurations are generated for all ensembles.

\begin{table}
	\caption{Simulation details for our \(\SU(4)\) ensembles, including the \(\beta\) values, lattice simulation spacings (in units of the string tension and in \(\unit{fm}\)), and the spatial and temporal lattice extents, \(N_s\) and \(N_t\), respectively. The scale is set using string tension data from Refs.~\cite{Lucini:2005vg, Lucini:2008vi}, with the conversion to \(\unit{fm}\) performed using \(\sqrt{\sigma} \simeq \qty{465}{MeV}\)~\cite{Bali:1997am, Edwards:1997xf, Allton:2008ty, Athenodorou:2020ani}.}
	\label{tab:ensembledetails}
	\begin{ruledtabular}
		\begin{tabular}{@{}ccccS[table-format=3.0]@{}}
			\(\beta\) & \(a\sqrt{\sigma}\) & \(\var{a}{(fm)}\) & \(N_s\) & \(N_t\) \\
			\hline
			10.900    & 0.239              & 0.101             & 32      & 64      \\
			11.125    & 0.191              & 0.081             & 40      & 80      \\
			11.340    & 0.159              & 0.068             & 48      & 96      \\
			11.782    & 0.115              & 0.049             & 64      & 128     \\
		\end{tabular}
	\end{ruledtabular}
\end{table}

\subsection{\texorpdfstring{\(\SU(3)\)}{SU(3)}} \label{sec:su3}
\begin{figure}
	\includegraphics[width=0.5\textwidth]{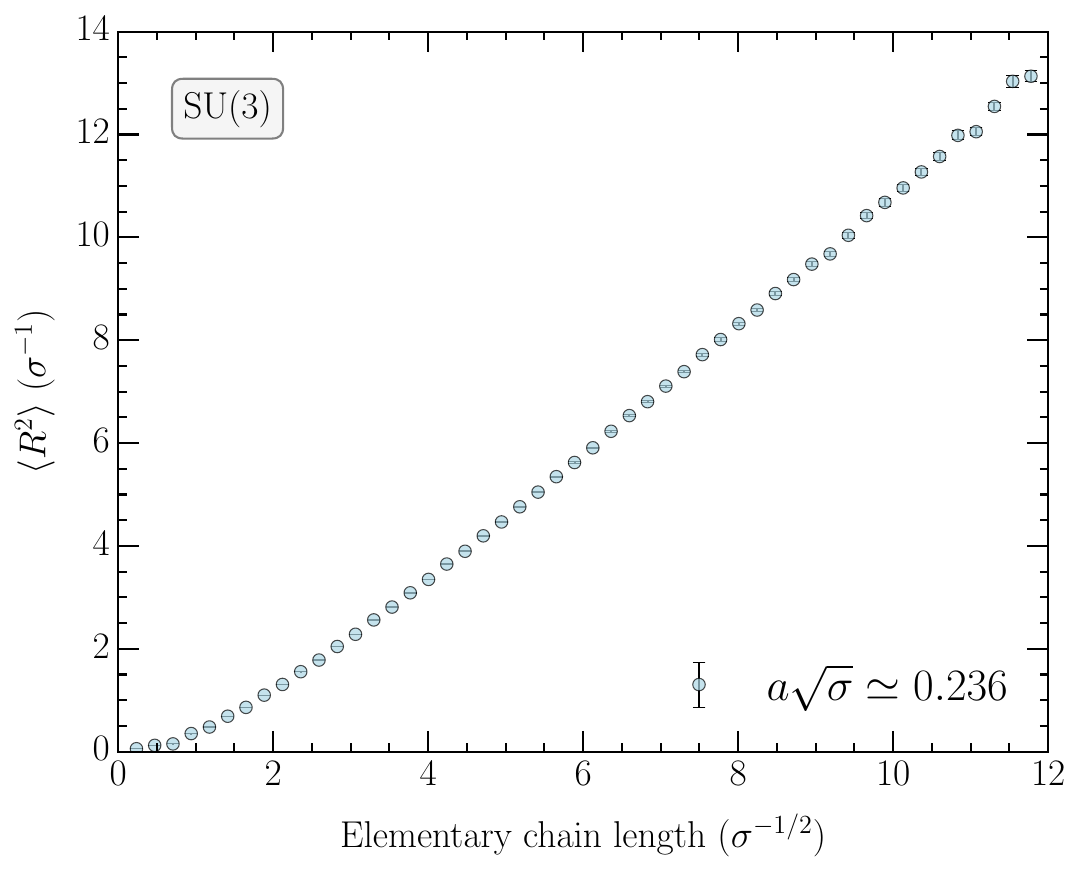}
	\caption{The mean-square end-to-end distance of center-vortex chains as a function of chain length for our \(\SU(3)\) ensemble. Following some nontrivial short distance behavior, the dependence is approximately linear at large chain lengths. Note that the uncertainty in the end-to-end distance is typically contained within the plot symbols.}
	\label{fig:R2su3}
\end{figure}

In \(\SU(3)\), the end points of center-vortex chains are three-way center monopoles. To compute the end-to-end distance \(R\) for a given chain, we explicitly keep track of the displacement from its start point as we follow the chain. This allows \(R\) to be determined while accounting for periodic boundary conditions, mitigating the effects of a finite volume without the need to simulate on excessively large lattices. For each chain length, we average over the set of obtained squared end-to-end distances and then plot \(\braket{R^2}\) as a function of chain length. The result for our \(\SU(3)\) ensemble is shown in Fig.~\ref{fig:R2su3}.

Following some nontrivial short-distance behavior, the data appears to obey an approximately linear relationship at large chain lengths. This is qualitatively similar to a continuum wormlike chain, whose properties are derived in Appendix~\ref{app:endtoend}. The relevant result is Eq.~\eqref{eq:R2wlc}, which provides the mean-square end-to-end distance as
\begin{equation} \label{eq:wlc}
	\braket{R^2} = 2 \xi_0 L \left[1 - \frac{\xi_0}{L} \left(1 - e^{-\frac{L}{\xi_0}}\right)\right] \,.
\end{equation}
This is asymptotically linear, \(\braket{R^2} = 2\xi_0\left(L - \xi_0\right)\) for \(L \gg \xi_0\), while for \(L \ll \xi_0\) it reduces to the rigid rod behavior \(\braket{R^2} = L^2\). Motivated by this, we consider fitting our numerical data to the ansatz
\begin{equation} \label{eq:wlcmodified}
	\braket{R^2} = 2\xi_0 L\left[1 - \frac{\xi_0}{\alpha L} \left(1 - e^{-\frac{\alpha L}{\xi_0}}\right)\right] \,,
\end{equation}
where the free parameter \(\alpha\) is introduced to allow for extra flexibility in the fit. This is still asymptotically linear, but reduces to the generalized short-distance behavior \(\braket{R^2} = \alpha L^2\). 

\begin{figure}
	\includegraphics[width=0.49\textwidth]{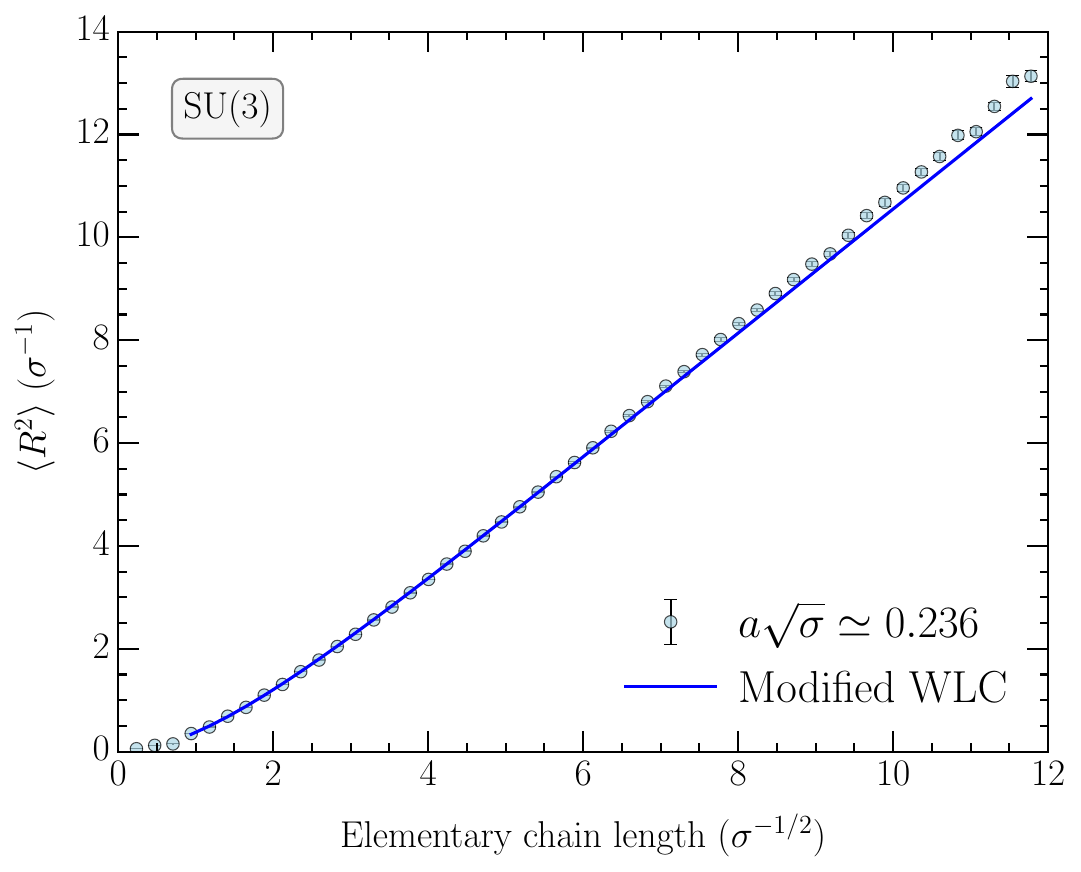}
	\includegraphics[width=0.49\textwidth]{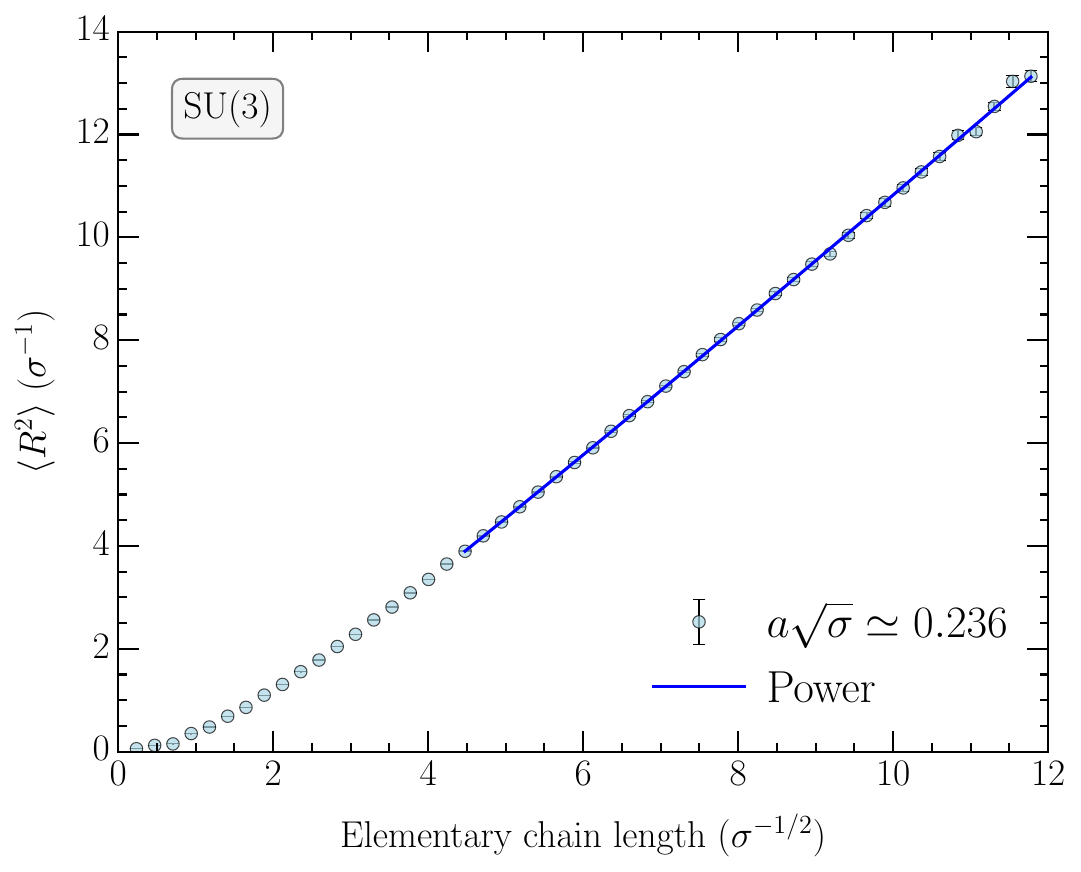}
	\caption{The mean-square end-to-end distances for vortex chains in \(\SU(3)\), fit to the generalized WLC model of Eq.~\eqref{eq:wlcmodified} (\textbf{left}) and power-law ansatz of Eq.~\eqref{eq:powerlaw} (\textbf{right}). The short-distance behavior is captured by the generalized WLC with a value \(\alpha \simeq 0.478\) for the parameter \(\alpha\) described in the text, suggesting short vortex chains grow roughly with \(\braket{R^2} \approx \frac{1}{2} L^2\). Meanwhile, the asymptotic long-distance data is accurately described by a power-law exponent slightly greater than one, with \(p \simeq 0.528\).}
	\label{fig:R2su3wlcpower}
\end{figure}

Care must be taken when performing these fits. Looking in Fig.~\ref{fig:R2su3}, there is a small but clear jump in value of \(\braket{R^2}\) moving from the third to fourth data point. This is fascinating, since it is known from prior \(\SU(3)\) work that center monopoles preferentially cluster within \(\numrange{1}{3}\) lattice units of each other~\cite{Biddle:2023lod, Mickley:2024zyg}. That is, chain lengths of \(\numrange{1}{3}\) are overwhelmingly the most probable. We have performed calculations that indicate this clustering occurs within \(\numrange{1}{3}\) lattice units regardless of the lattice simulation spacing, and we will see this phenomenon in action for \(\SU(4)\) in Sec.~\ref{subsec:stoppingrates}. Accordingly, the clustering appears to be a lattice artefact, and we perform the fit of Eq.~\eqref{eq:wlcmodified} only for chain lengths \(> 3\) lattice units. The result is shown in the left panel of Fig.~\ref{fig:R2su3wlcpower}.

We find that the modified WLC ansatz visibly describes the data well at short chain lengths. The obtained value for \(\alpha\) is \(\alpha \simeq 0.478\), which is in close proximity to \(0.5\). This indicates that the behavior at short chain lengths is closer to \(\braket{R^2} \approx \frac{1}{2} L^2\), contrasting the rodlike \(\braket{R^2} = L^2\) and demonstrating the importance of including the parameter \(\alpha\). However, the modified WLC ansatz is unable to simultaneously account for both the short- and long-distance behavior. Indeed, the asymptotic linear slope underestimates the end-to-end distance data for long chains, which is still subject to some gentle curvature.

Instead, we find that the end-to-end distances for long chains over our data range are better described by a power law with an exponent slightly greater than 1. To be precise, we consider the power-law ansatz
\begin{equation} \label{eq:powerlaw}
	\braket{R^2} = 2 A \left(L - L_0\right)^{2p} \,,
\end{equation}
where the fit is to be performed for chain lengths \(L > L_\mathrm{min}\) greater than some minimum length \(L_\mathrm{min}\). We make \(L_\mathrm{min}\) as small as possible while retaining a reasonable \(\chi^2/\mathrm{dof}\) value \(\simeq 1\) to determine the range over which a power-law description is valid. This is found to be \(L_\mathrm{min}\sqrt{\sigma} \simeq 4.5\), with \(\chi^2/\mathrm{dof} \simeq 1.06\). The corresponding fit is presented in the right panel of Fig.~\ref{fig:R2su3wlcpower}, with fit parameters of
\begin{align} \label{eq:su3powerfits}
	p &= 0.528(6) \,, & A \, \sigma^{1-p} &= 0.54(2) \,, & L_0 \sqrt{\sigma} &= 1.09(6) \,.
\end{align} 
Although \(p\) is only slightly greater than \(\frac{1}{2}\), the discrepancy is strongly statistically significant, reflecting a gentle curvature in the variation of end-to-end distances with chain length.

The fact that a power law provides a valid description of the data over a wide range of chain lengths is rather suggestive. In particular, end-to-end distances for a self-avoiding walk (i.e.\ a path that never visits a given point more than once) in three dimensions are governed by a power law with \(p \simeq 0.588\)~\cite{deGennes:1979, Clisby:2010}. The measured exponent is close to, but smaller than, this value, lying between the random-walk value \(p = 1/2\) and the self-avoiding-walk value. The possible physical mechanisms underlying this intermediate behavior will be discussed in Sec.~\ref{subsec:saw}.

\begin{figure}
	\centering
	\includegraphics[width=0.5\textwidth]{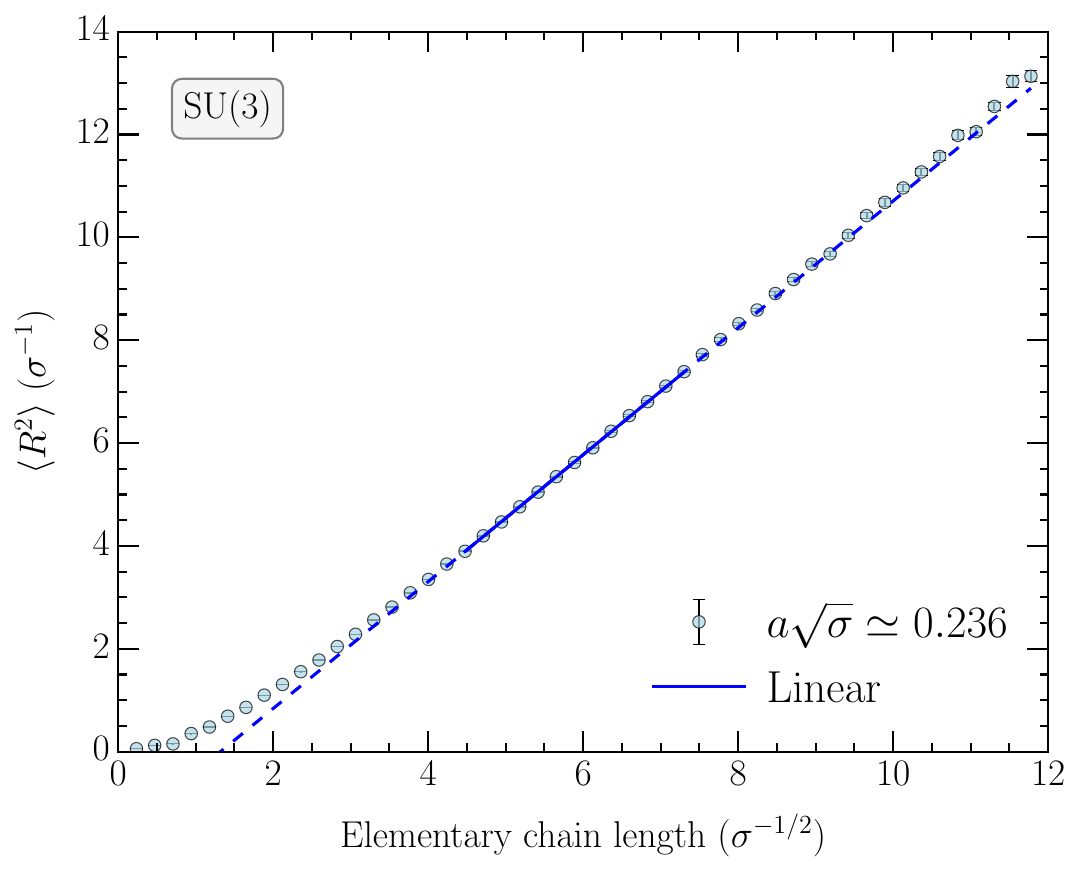}
	\caption{The linear ansatz of Eq.~\eqref{eq:wlcmodified} fit to the mean-square end-to-end distances for intermediate chain lengths \(L\sqrt{\sigma} \in [4.5,7.5]\) in \(\SU(3)\). The solid line represents the fit window, while the dashed lines show how the fit deviates from the data at short and long chain lengths.}
	\label{fig:R2su3linear}
\end{figure}

Now, our aim is to use the end-to-end distance data to determine the finite lattice spacing \(\tilde{a}\) of the effective model. To make contact with the model, we desire an estimate of the local stiffness before the asymptotic power-law scaling regime first emerges, just beyond the short-distance transient. As such, we seek to approximate the end-to-end distance behavior as locally linear in a finite window according to Eq.~\eqref{eq:linear}, whose slope gives access to the leading persistence length term \(\xi_0\). We start the linear fits at the same minimum chain length \(L_\mathrm{min}\sqrt{\sigma} \simeq 4.5\) as the power-law fits, and then vary the upper bound to fit over as large a window as possible while retaining a \(\chi^2/\mathrm{dof} \simeq 1\). This is determined to be approximately \(L\sqrt{\sigma} \in [4.5,7.5]\). The resulting linear fit is included in Fig.~\ref{fig:R2su3linear}. The extracted fit parameters, which will be relevant in Sec.~\ref{sec:discussion}, are
\begin{align} \label{eq:su3linearfits}
	\xi_0 \sqrt{\sigma} &= 0.633(3) \,, & L_0 \sqrt{\sigma} &= 1.46(4) \,.
\end{align}

\subsection{\texorpdfstring{\(\SU(4)\)}{SU(4)}} \label{sec:su4}
The analysis of end-to-end distances is complicated in \(\SU(4)\) by the presence of novel doubly charged chains, which connect the elementary chain components of the center-vortex network. Doubly charged chains have recently been explored numerically for the first time in Ref.~\cite{Mickley:2025ksp}. There, two primary types of doubly charged chains were classified:\ ``vortex convergences'' (VC), in which two elementary vortex lines converge to form a doubly charged chain and then split, and ``vortex matchings'' (VM), through which two pairs of elementary vortex lines match over an extended distance. Although there are minor differences between their properties, such as in their average lengths, these differences are rather subtle. This vindicates providing a unified analysis of all doubly charged chains here to determine their decay and persistence properties.

\subsubsection{Stopping rates} \label{subsec:stoppingrates}
As motivated in Sec.~\ref{subsec:decayrates}, we first desire an empirical determination of the chain-length probability distribution. Prior work in \(\SU(3)\) has demonstrated that, forgoing the clustering within \(\numrange{1}{3}\) lattice units, elementary center-vortex chain lengths are exponentially distributed \(\sim e^{-\lambda L}\)~\cite{Biddle:2023lod, Mickley:2024zyg}. This allows for the extraction of a decay rate \(\lambda\), which is related to the tension \(\mu\) of the effective model according to Eq.~\eqref{eq:decayratetension}.

In Ref.~\cite{Mickley:2025ksp}, a similar analysis was performed in \(\SU(4)\) separately for both VC and VM doubly charged chains. Here, we seek to obtain a single decay rate for doubly charged chains and accordingly will repeat this analysis on our updated ensembles with VC and VM chains combined, in addition to examining the chain-length distributions for elementary chains. There are three possibilities for how an elementary chain in \(\SU(4)\) can start or terminate, either at a four-way center monopole or at one of the two types of doubly charged chains. As a result, there are nine variations in the elementary vortex chain endpoints, and for this work we also group them into a single analysis.

\begin{figure}
	\includegraphics[width=0.49\textwidth]{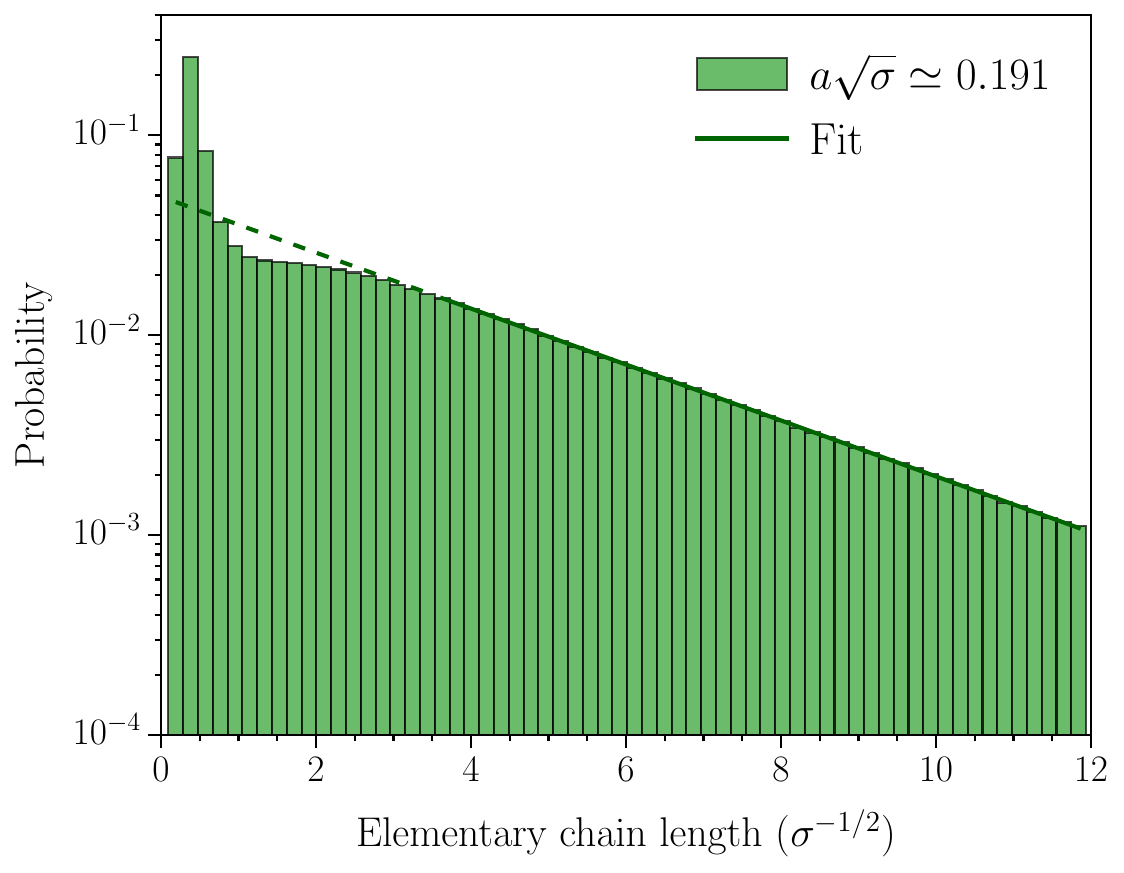}
	\hfill
	\includegraphics[width=0.49\textwidth]{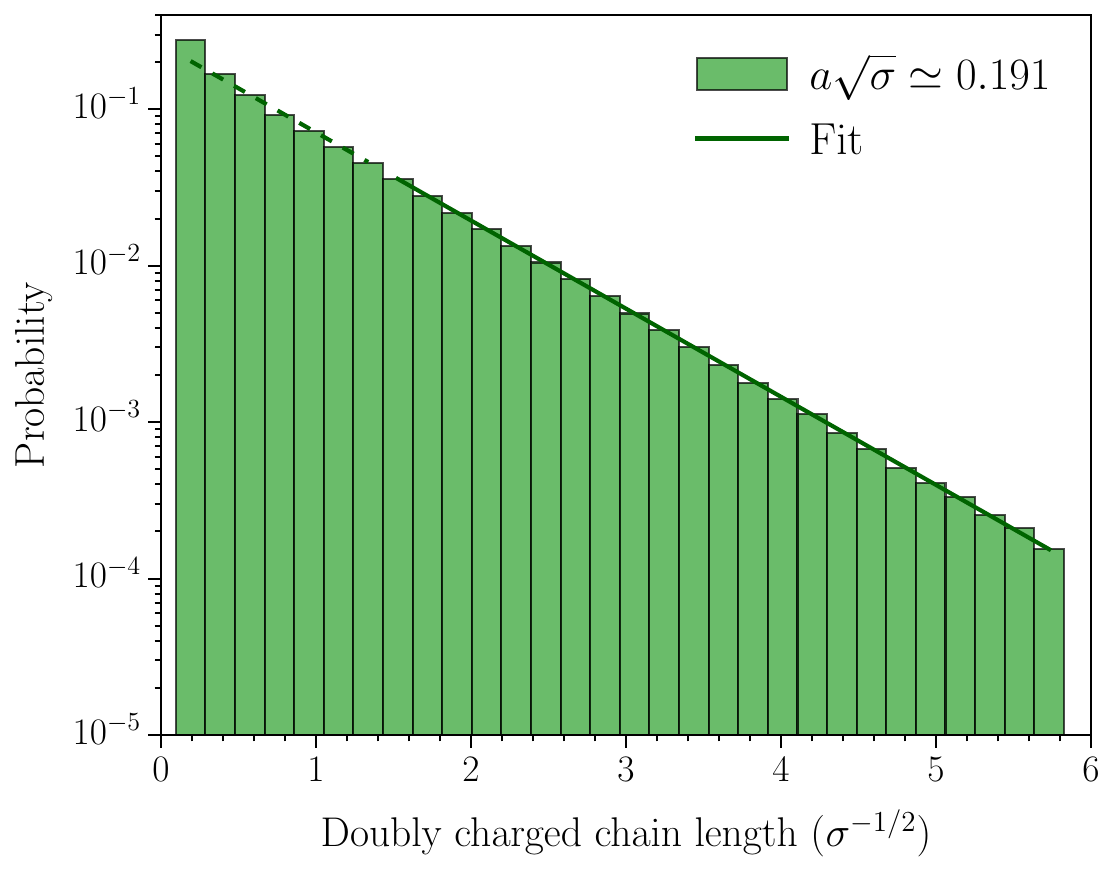}%
	\vspace{0.5em}
	\includegraphics[width=0.49\textwidth]{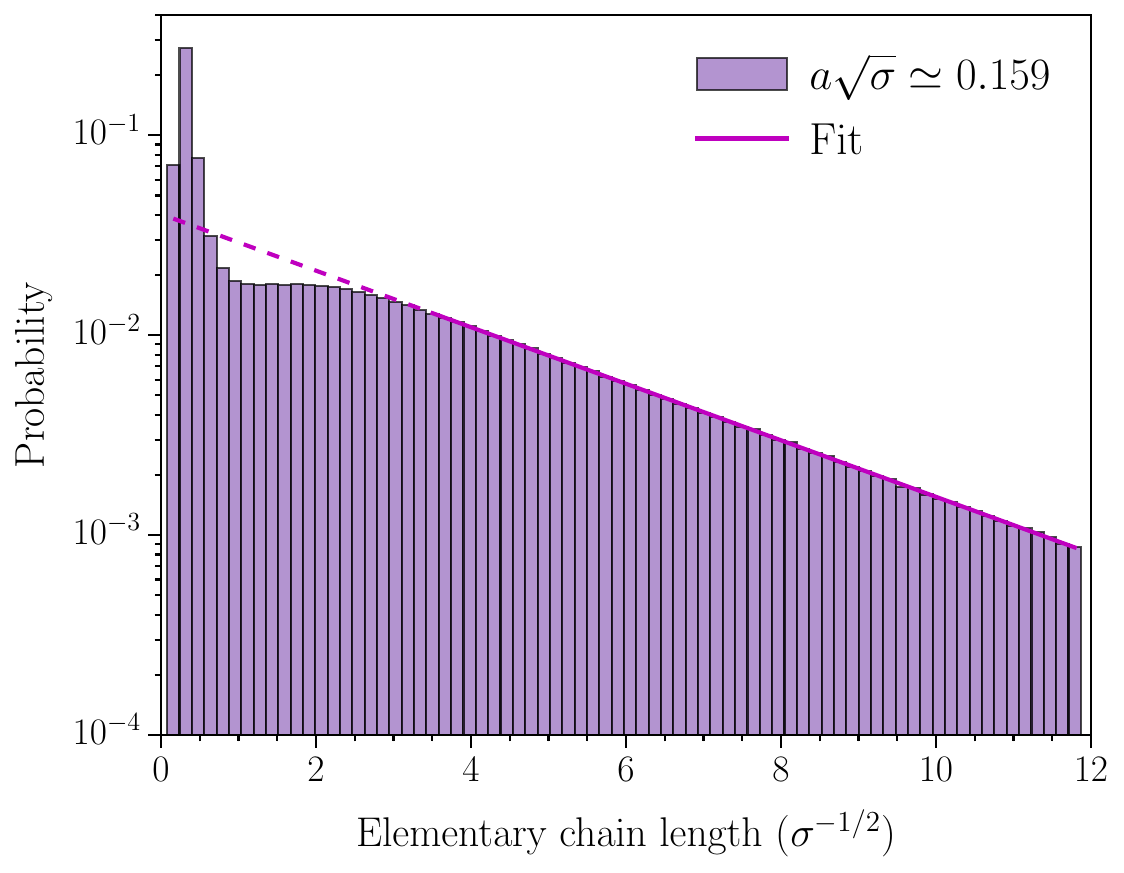}
	\hfill
	\includegraphics[width=0.49\textwidth]{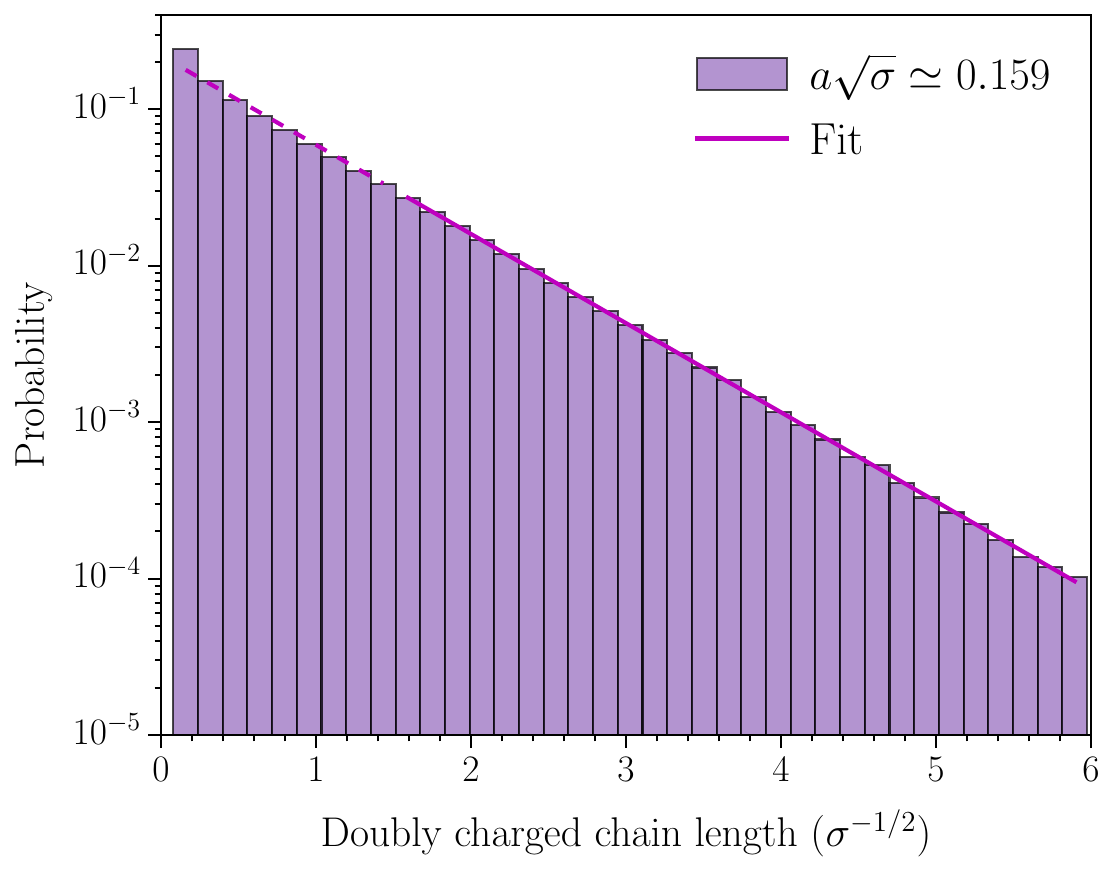}%
	\vspace{0.5em}
	\includegraphics[width=0.49\textwidth]{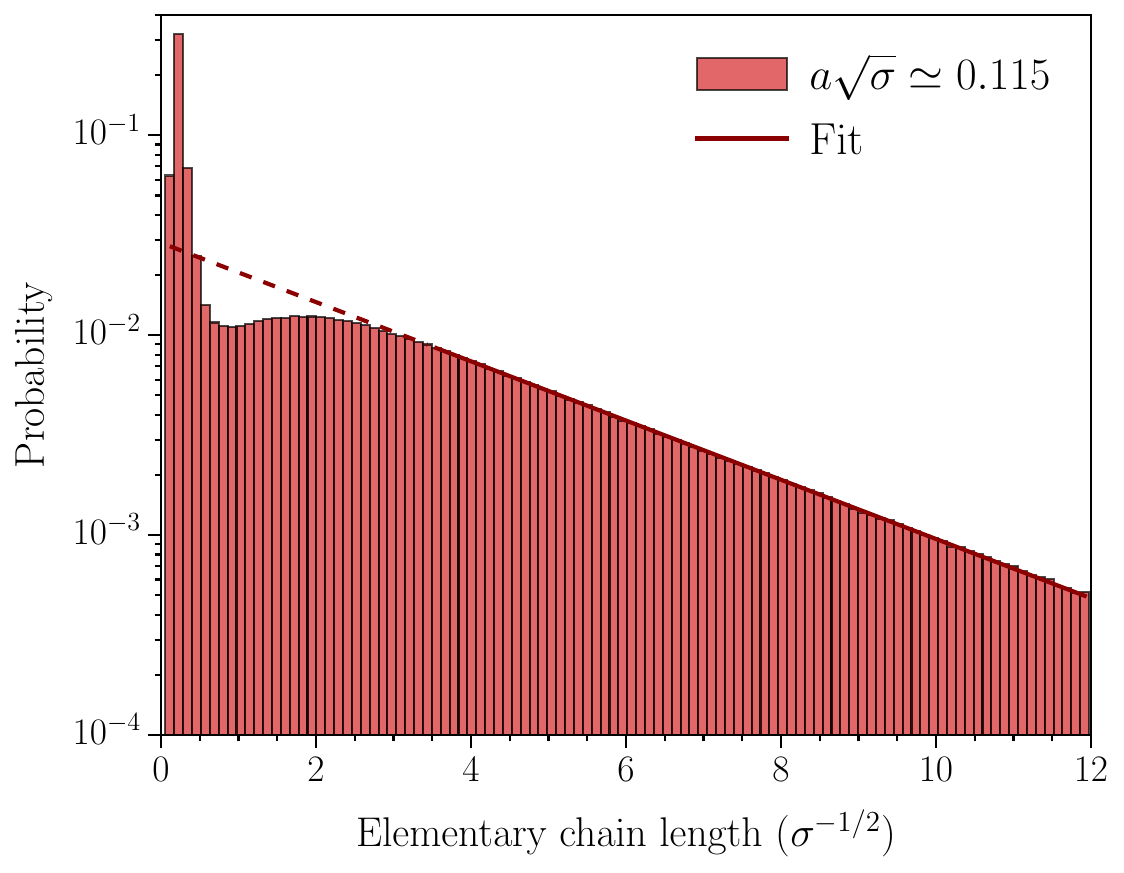}
	\hfill
	\includegraphics[width=0.49\textwidth]{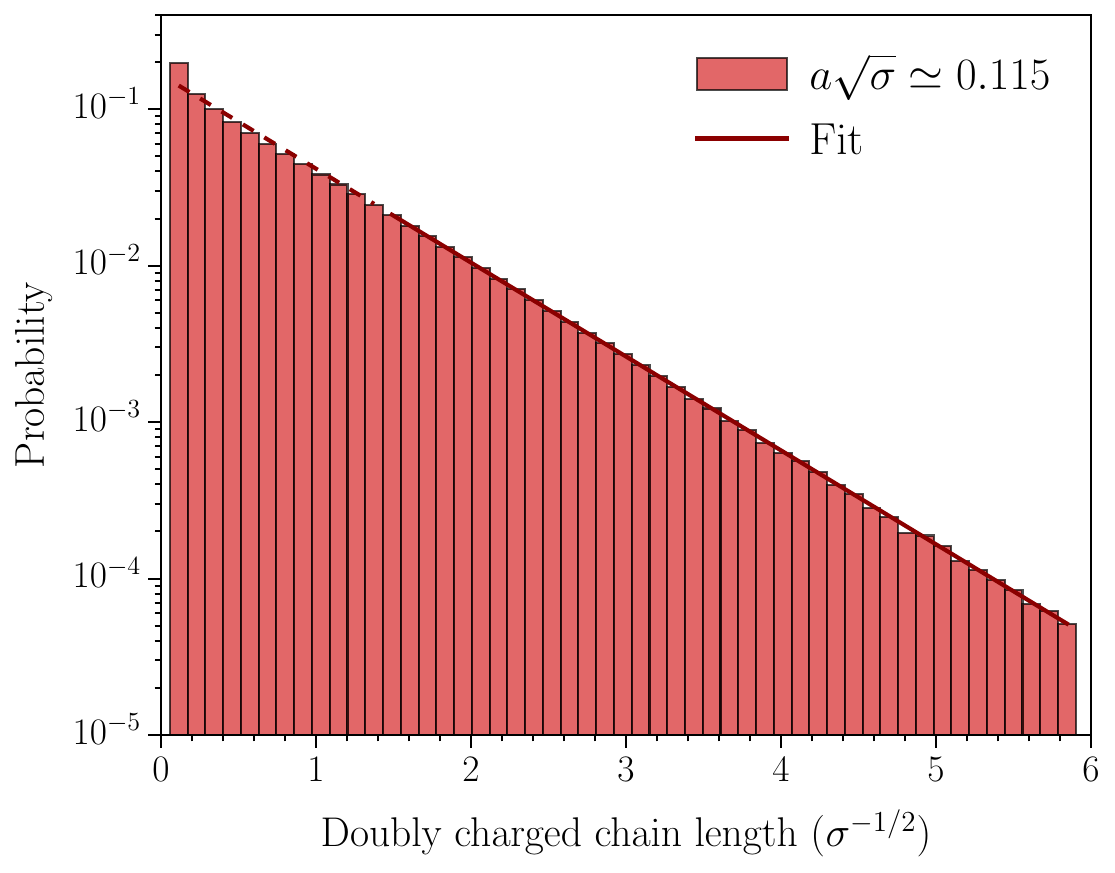}
	\caption{The distributions of center-vortex chain lengths for elementary (\textbf{left}) and doubly charged (\textbf{right}) chains on our three finest \(\SU(4)\) ensembles. Beyond some short-distance effects, which are more pronounced for elementary chains, both chain types are exponentially distributed. Exponential fits to elementary chain lengths \(L\sqrt{\sigma} \gtrsim 3.5\) and doubly charged chain lengths \(L\sqrt{\sigma} \gtrsim 1.5\) are overlaid. The solid lines represent the fit windows, while the dashed lines show how the fits deviate from the data at short chain lengths.}
	\label{fig:chainlengthdistributions}
\end{figure}

From the measured center-vortex chain lengths, we construct their probability distributions. These are shown in Fig.~\ref{fig:chainlengthdistributions} on a logarithmic scale for both elementary and doubly charged chains on our three finest \(\SU(4)\) ensembles. Their maximum length is fixed to be approximately constant in physical units across the three lattice simulation spacings. The distributions are seen to be exponential (linear on the logarithmic scale) beyond a certain length, but exhibit significant deviations from the exponential distribution at short chain lengths. This is especially pronounced for elementary chains where, as previously explained for \(\SU(3)\), a strong clustering within lengths of \(\numrange{1}{3}\) lattice units is manifest. This behavior persists as the lattice spacing is reduced, suggesting it is a lattice artefact.

Following the unphysical clustering, there is a range of elementary chain lengths that sit under the exponential trend line up to a chain length of \(L\sqrt{\sigma} \simeq 3.5\), which is approximately constant in physical units across our lattice spacings. This reflects a minimum length elementary chains tend to have, with shorter lengths (excluding the clustering) comparatively unlikely. An analogous but much softer effect is also present for doubly charged chains, for which there is a mild preference for chain lengths of 1 lattice unit after which the long-distance exponential trend slightly overestimates the data. This is most apparent on our finest ensemble, and was originally noted in Ref.~\cite{Mickley:2025ksp}.

Beyond this short-distance behavior, the chain lengths are exponentially distributed and we infer there is a constant long-distance ``stopping probability'' underlying the behavior of long vortex chains. Indeed, as done in Refs.~\cite{Biddle:2023lod, Mickley:2024zyg, Mickley:2025ksp}, the normalized exponential distribution
\begin{align} \label{eq:expdist}
	P(n) &= (e^\beta - 1) \, e^{-\beta n} \,, & \sum_{n=1}^\infty P(n) &= 1 \,,
\end{align}
can be rewritten as
\begin{align} \label{eq:geodist}
	P(n) &= (1-q)^{n-1} \, q \,, & q &= 1 - e^{-\beta} \,,
\end{align}
so that \(q\) and \(1-q\) are the probabilities to stop and continue growing, respectively, at each step along the chain. That is, the discrete exponential distribution is equivalent to a geometric distribution. These probabilities are extracted from exponential fits to the distributions of chain lengths \(L = na\), applied to elementary chains for \(L\sqrt{\sigma} \gtrsim 3.5\) and to doubly charged chains for \(L\sqrt{\sigma} \gtrsim 1.5\), over which the exponential description is valid. These fits are overlaid on the histograms in Fig.~\ref{fig:chainlengthdistributions}. Finally, the probabilities can be converted into physical ``stopping rates'' through dividing by the lattice simulation spacing, \(\lambda = q/a\). These are provided in Table~\ref{tab:stoppingrates} as \(\lambda_{(1)}\) for elementary chains and \(\lambda_{(2)}\) for doubly charged chains.

\begin{table}
	\caption{The physical long-distance stopping rates \(\lambda_{(1)}\) and \(\lambda_{(2)}\) for elementary and doubly charged vortex chains in \(\SU(4)\), respectively, obtained via exponential fits to the distributions of chain lengths for long chains. The rate remains finite in the continuum limit, which can be identified with the exponential decay rate.}
	\label{tab:stoppingrates}
	\begin{ruledtabular}
		\begin{tabular}{@{}cS[table-format=1.4(2)]c@{}}
			\(a\sqrt{\sigma}\) & \multicolumn{1}{c}{\(\lambda_{(1)}/\sqrt{\sigma}\)} & \(\lambda_{(2)}/\sqrt{\sigma}\) \\
			\hline
			0.239 & 0.3037(12) & 1.101(4) \\
			0.191 & 0.3108(7)  & 1.155(3) \\
			0.159 & 0.3158(5)  & 1.185(2) \\
			0.115 & 0.3207(4)  & 1.223(2) \\
			\hline
			0     & 0.3259(6)  & 1.260(2) \\
		\end{tabular}
	\end{ruledtabular}
\end{table}

The stopping rates for elementary chains are nearly a factor of four smaller than for doubly charged chains. Given the latter are typically considerably shorter than the former, this is the expected outcome. For both elementary and doubly charged chains, there is a soft increase in the stopping rates as the lattice simulation spacing \(a\) is reduced. This motivates performing an extrapolation to \(a = 0\). If the stopping rate remains finite in the continuum limit, then this signifies that the raw probability, \(q\), scales with the lattice spacing and tends to zero as \(a \to 0\). Since \(q = 1 - e^{-\beta} \approx \beta\) for small \(\beta\), it follows that \(e^{-\beta n} \approx e^{-\lambda L}\), such that the stopping rate coincides with the exponential decay rate in the continuum. That is, the local stopping rate measured in lattice simulations extrapolated to the continuum limit provides an estimate of the exponential decay rate.

\begin{figure}
	\includegraphics[width=0.49\textwidth]{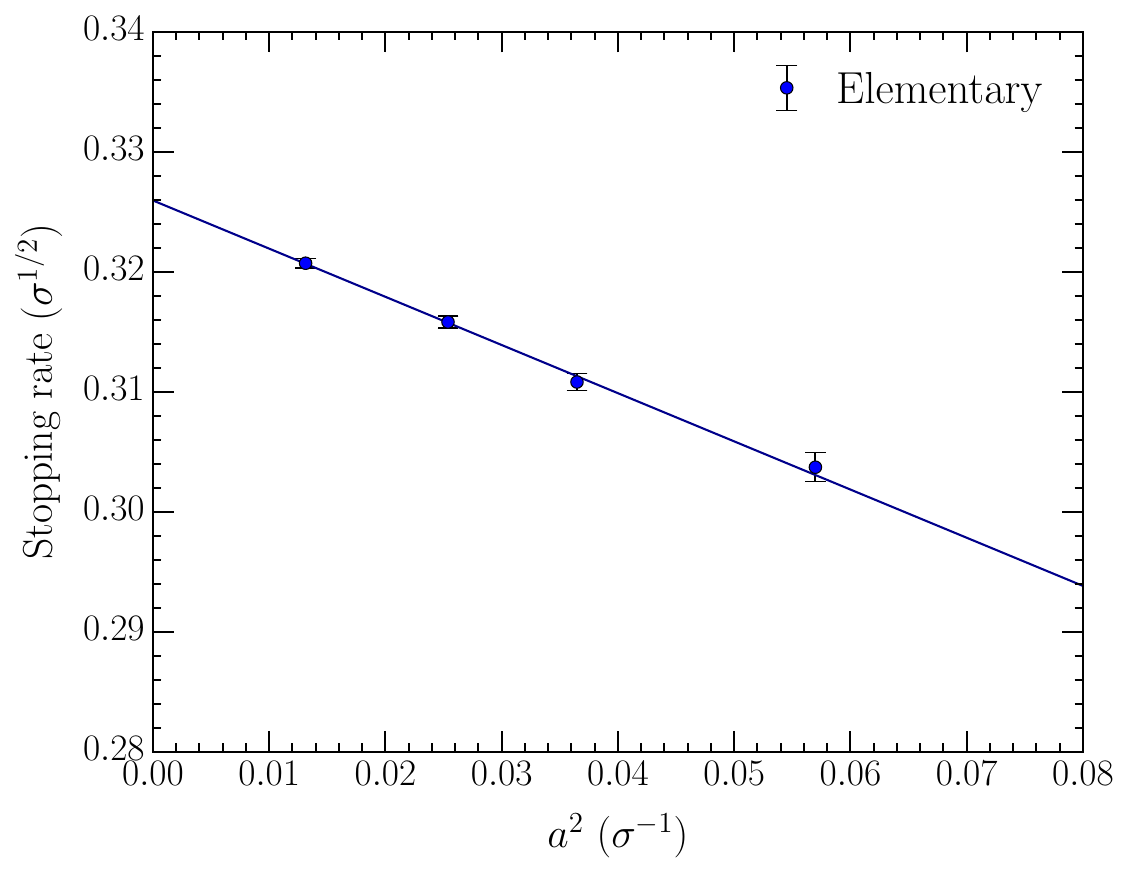}
	\hfill
	\includegraphics[width=0.49\textwidth]{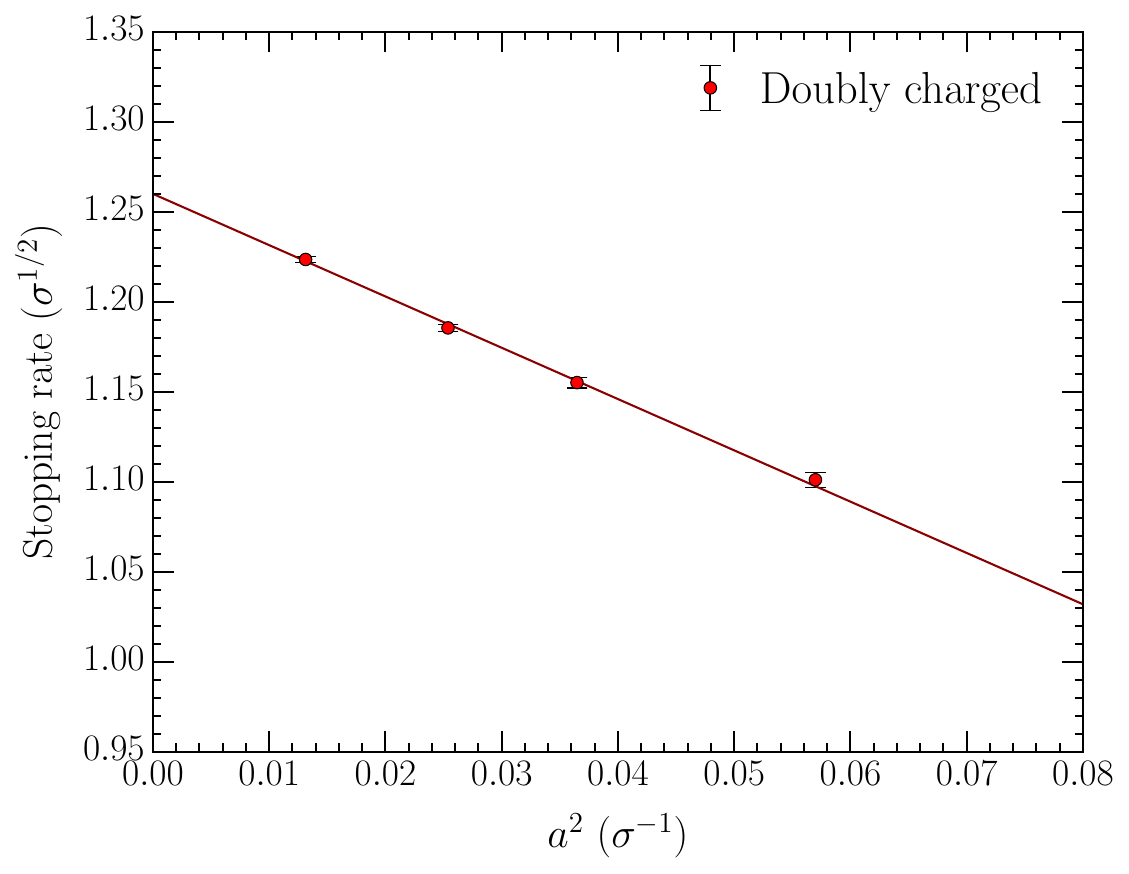}
	\caption{The stopping rates for elementary (\textbf{left}) and doubly charged (\textbf{right}) vortex chains in \(\SU(4)\) as a function of \(a^2\). Linear fits (in \(a^2\)) are overlaid, which describe the data well for both chain types and allow an extrapolation to the continuum limit.}
	\label{fig:stoppingrates}
\end{figure}

On account that the Wilson gauge action has leading errors of \(\order(a^2)\), we anticipate physical quantities to depend linearly on \(a^2\) provided higher-order terms are negligible. Accordingly, we plot \(\lambda_{(1)}\) and \(\lambda_{(2)}\) as a function of \(a^2\) in Fig.~\ref{fig:stoppingrates}. Both rates are found to be linear in \(a^2\), and fits of this form are overlaid and used to extrapolate to the continuum limit. These describe the data well, with \(\chi^2/\mathrm{dof}\) values \(\simeq 0.38\) for elementary chains and \(\simeq 1.01\) for doubly charged chains. The continuum-extrapolated (\(a = 0\)) stopping rates are listed in Table~\ref{tab:stoppingrates}, yielding the decay rates that enter in Eq.~\eqref{eq:muratiorates} for the ratio of center-vortex tensions.

\subsubsection{End-to-end distances} \label{subsec:endtoenddistances}
\begin{figure}
	\includegraphics[width=0.49\textwidth]{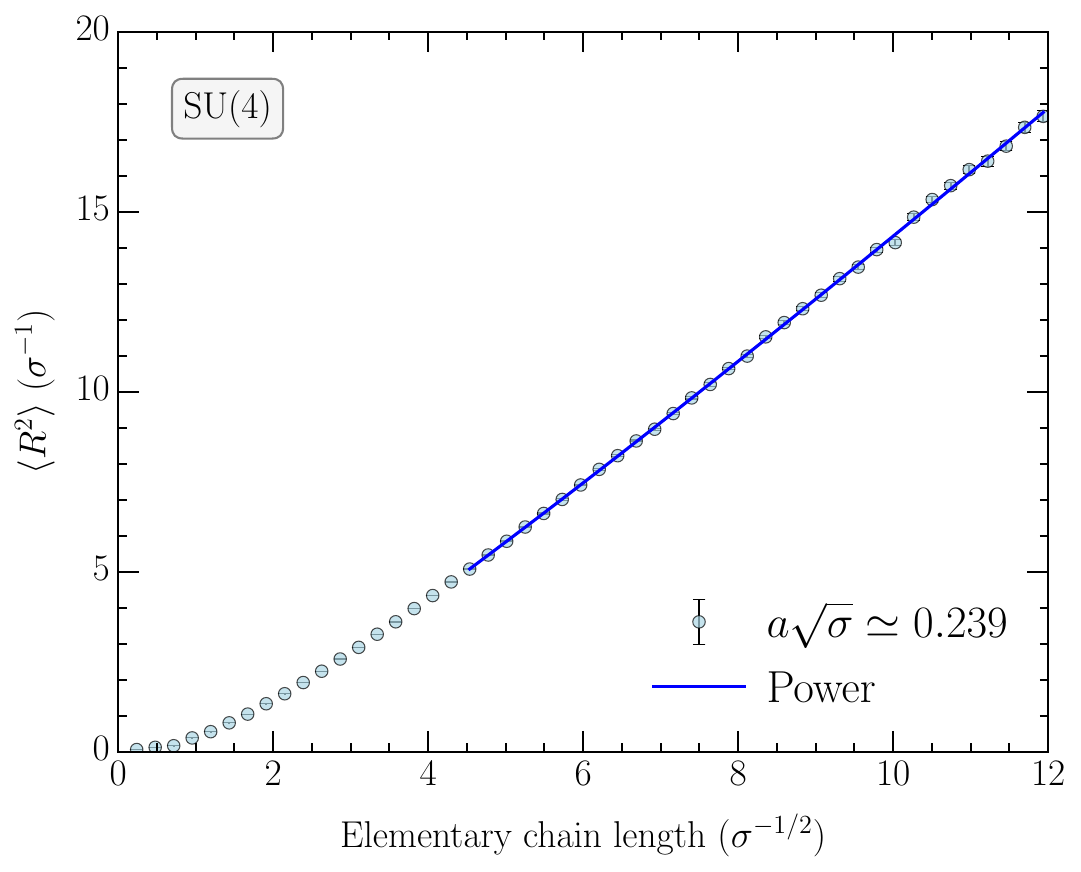}
	\hfill
	\includegraphics[width=0.49\textwidth]{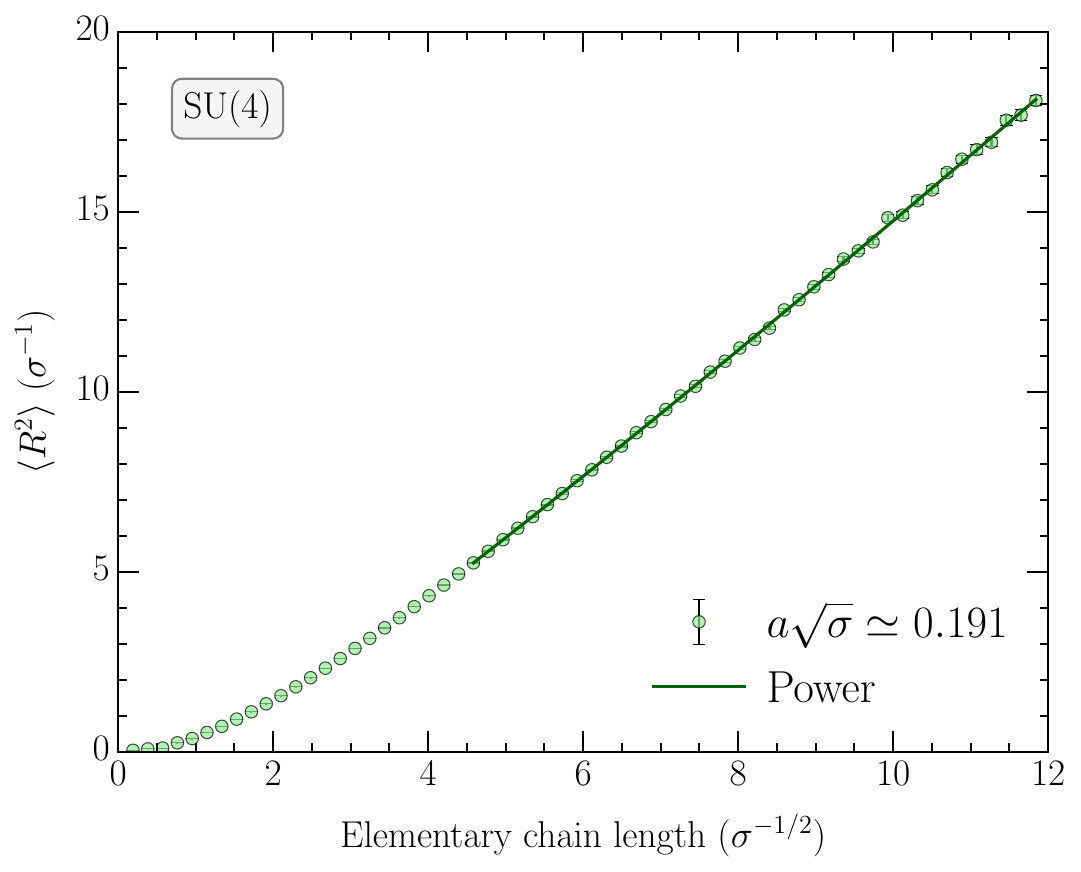}%
	\vspace{0.5em}
	\includegraphics[width=0.49\textwidth]{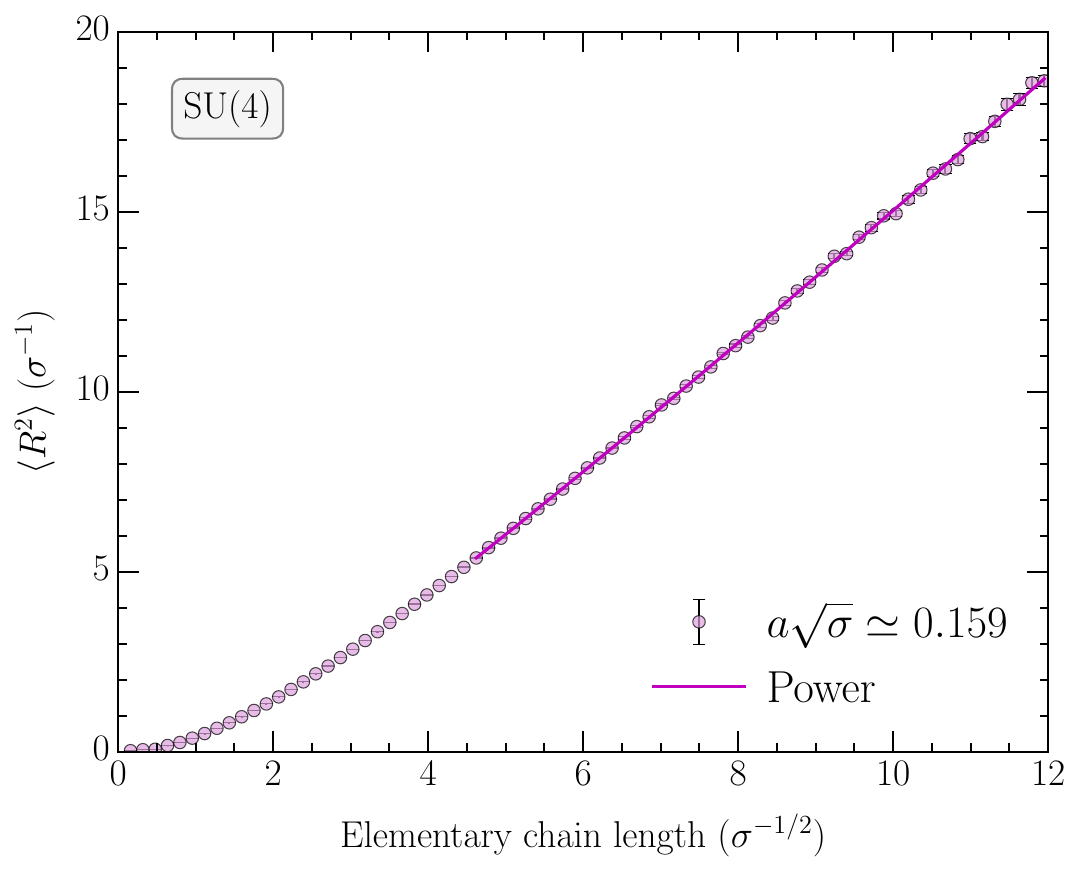}
	\hfill
	\includegraphics[width=0.49\textwidth]{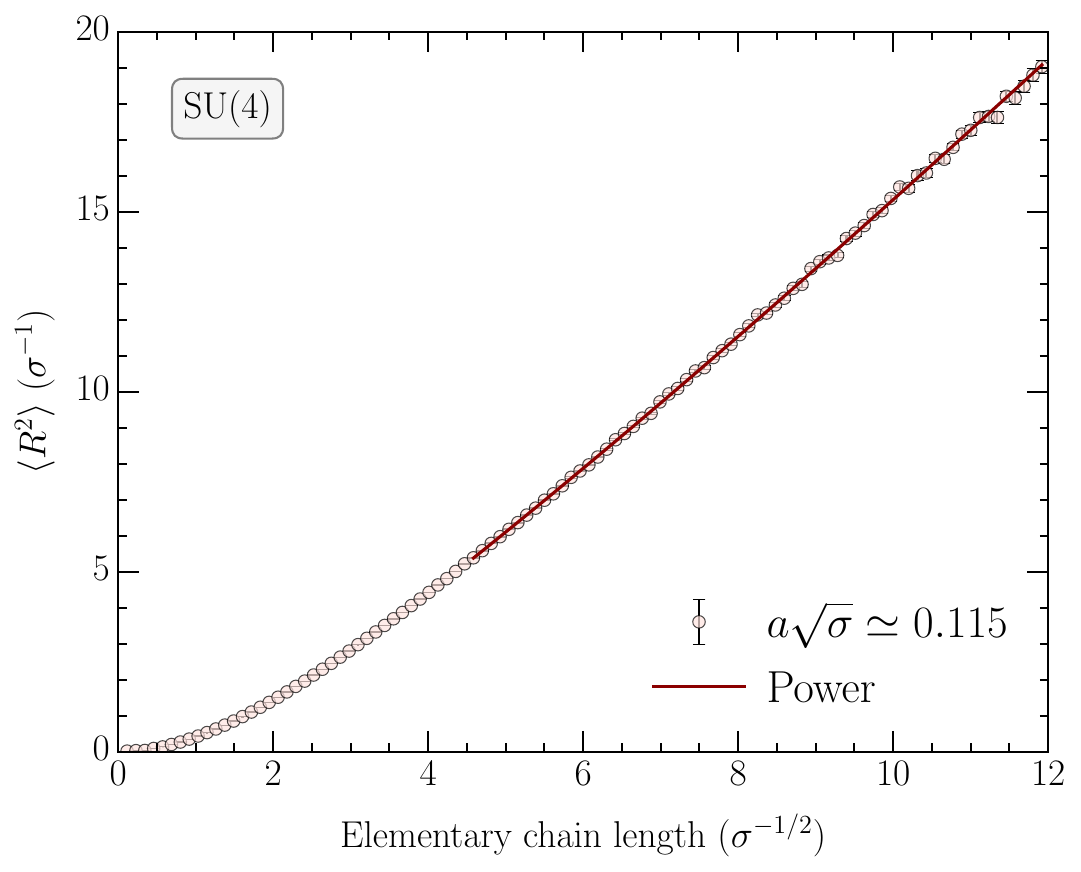}
	\caption{The mean-square end-to-end distance as a function of chain length for elementary chains in \(\SU(4)\) gauge theory, fit to the power-law ansatz of Eq.~\eqref{eq:powerlaw} for chain lengths \(L\sqrt{\sigma} \gtrsim 4.5\). Qualitatively similar behavior to elementary chains in \(\SU(3)\) is observed, though the end-to-end distances here increase more rapidly at long chain lengths with \(p \simeq 0.55\).}
	\label{fig:R2su4elementarypower}
	\vspace{-0.85em}
\end{figure}

With decay rates obtained for later use, we now turn to focus on the end-to-end distances for \(\SU(4)\), needed to fix the effective lattice spacing \(\tilde{a}\). It will be fascinating to see whether both elementary and doubly charged chains are described by a common effective spacing. Beginning with elementary chains, the mean-square end-to-end distances are shown as a function of chain length for each \(\SU(4)\) ensemble in Fig.~\ref{fig:R2su4elementarypower}. These are qualitatively similar to our \(\SU(3)\) example in Fig.~\ref{fig:R2su3}, appearing approximately linear at long chain lengths following some nontrivial short-distance behavior. Interestingly, the end-to-end distances for elementary chains increase more rapidly here than in \(\SU(3)\).

We have again fit the generalized WLC ansatz in Eq.~\eqref{eq:wlcmodified} to the \(\SU(4)\) elementary chain data, and found similar results to \(\SU(3)\). That is, the short-distance behavior is close to \(\braket{R^2} \approx \frac{1}{2} L^2\), but the ansatz cannot simultaneously describe both short- and long-distance behavior. Instead, we again model the data with the power law in Eq.~\eqref{eq:powerlaw}. We hold the minimum chain length considered for the fits constant across our lattice simulation spacings, tuned to be the smallest possible length while retaining \(\chi^2/\mathrm{dof} \simeq 1\) simultaneously across all \(\SU(4)\) ensembles. Notably, this is satisfied by \(L_\mathrm{min}\sqrt{\sigma} \simeq 4.5\), the same as \(\SU(3)\). These fits are overlaid on the data in Fig.~\ref{fig:R2su4elementarypower}, and the fit parameters and \(\chi^2\) values are summarized in Table~\ref{tab:powerlawfitselementary}.

\begin{table}
	\caption{The details of our power-law fits to the end-to-end distances for elementary chain lengths \(L\sqrt{\sigma} \gtrsim 4.5\) in \(\SU(4)\), including the fit parameters \(p\), \(A\) and \(L_0\) in units of the string tension, and the \(\chi^2/\mathrm{dof}\) values.}
	\label{tab:powerlawfitselementary}
	\begin{ruledtabular}
		\begin{tabular}{@{}ccccc@{}}
			\(a\sqrt{\sigma}\) & \(p\) & \(A \, \sigma^{1-\nu}\) & \(L_0 \sqrt{\sigma}\) & \(\chi^2/\mathrm{dof}\) \\
			\hline
			0.239 & 0.549(7) & 0.65(2) & 1.07(6) & 1.00 \\
			0.191 & 0.544(6) & 0.69(2) & 1.17(6) & 0.75 \\
			0.159 & 0.545(6) & 0.70(2) & 1.19(6) & 1.00 \\
			0.115 & 0.549(7) & 0.74(3) & 1.23(6) & 1.04 \\
		\end{tabular}
	\end{ruledtabular}
\end{table}

We find that the power-law exponent \(2p\) is constant within statistical uncertainty across all lattice simulation spacings, with \(p \simeq 0.55\). This is slightly larger than our \(\SU(3)\) value (\(p \simeq 0.53\)), reflecting that the \(\SU(4)\) end-to-end distance grows faster with chain length, and is correspondingly closer to the self-avoiding-walk value \(p \simeq 0.588\).

\begin{figure}
	\includegraphics[width=0.47\textwidth]{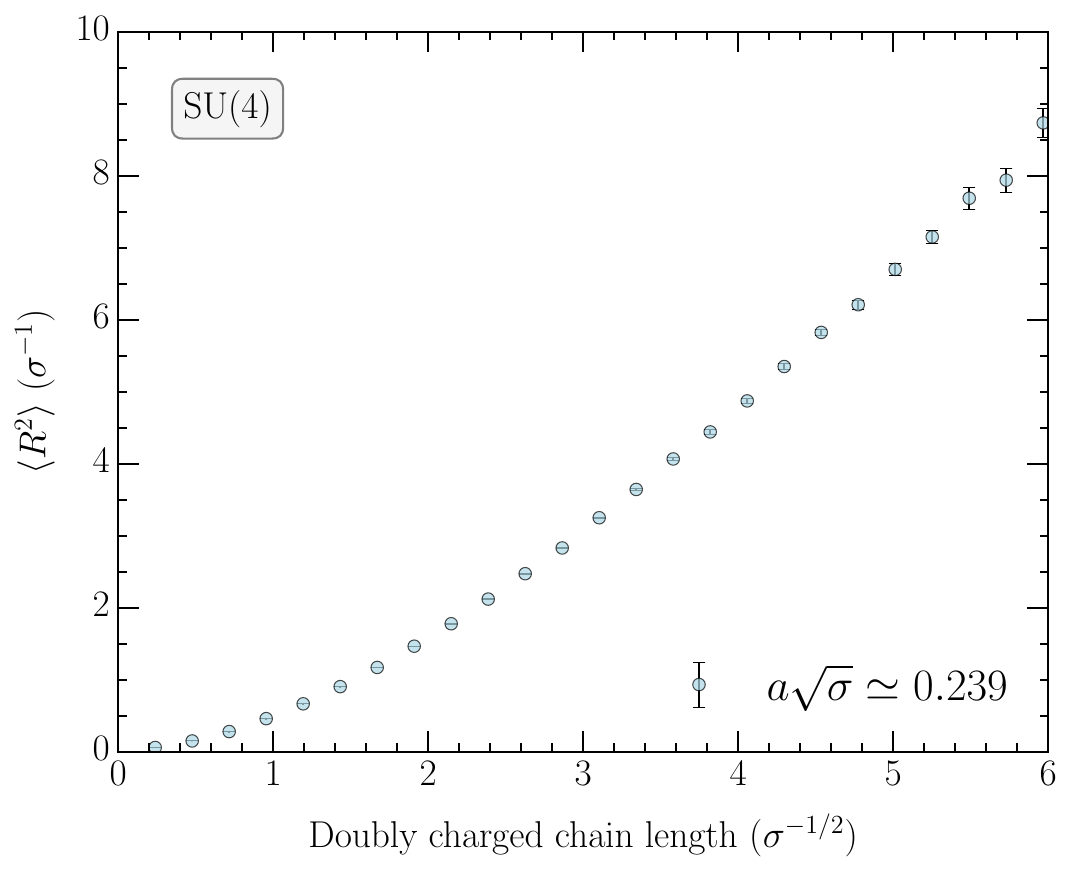}
	\hfill
	\includegraphics[width=0.47\textwidth]{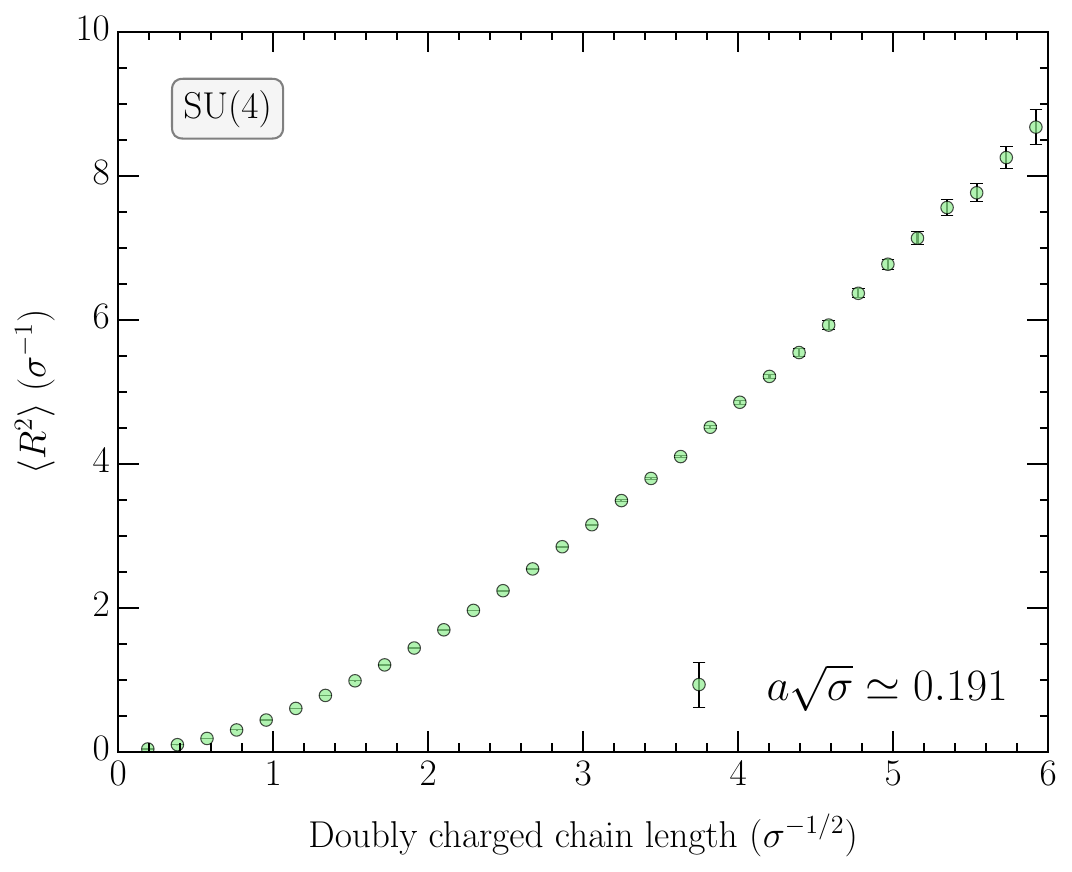}
	\includegraphics[width=0.47\textwidth]{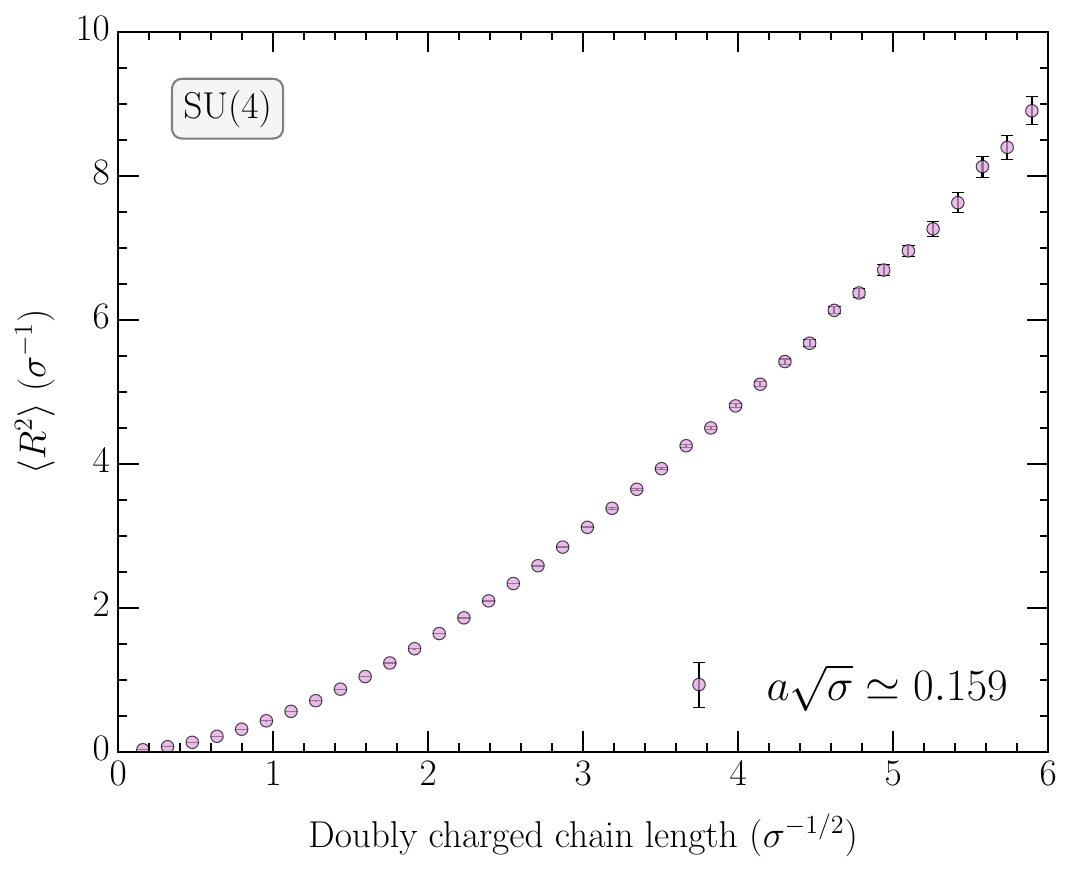}
	\hfill
	\includegraphics[width=0.47\textwidth]{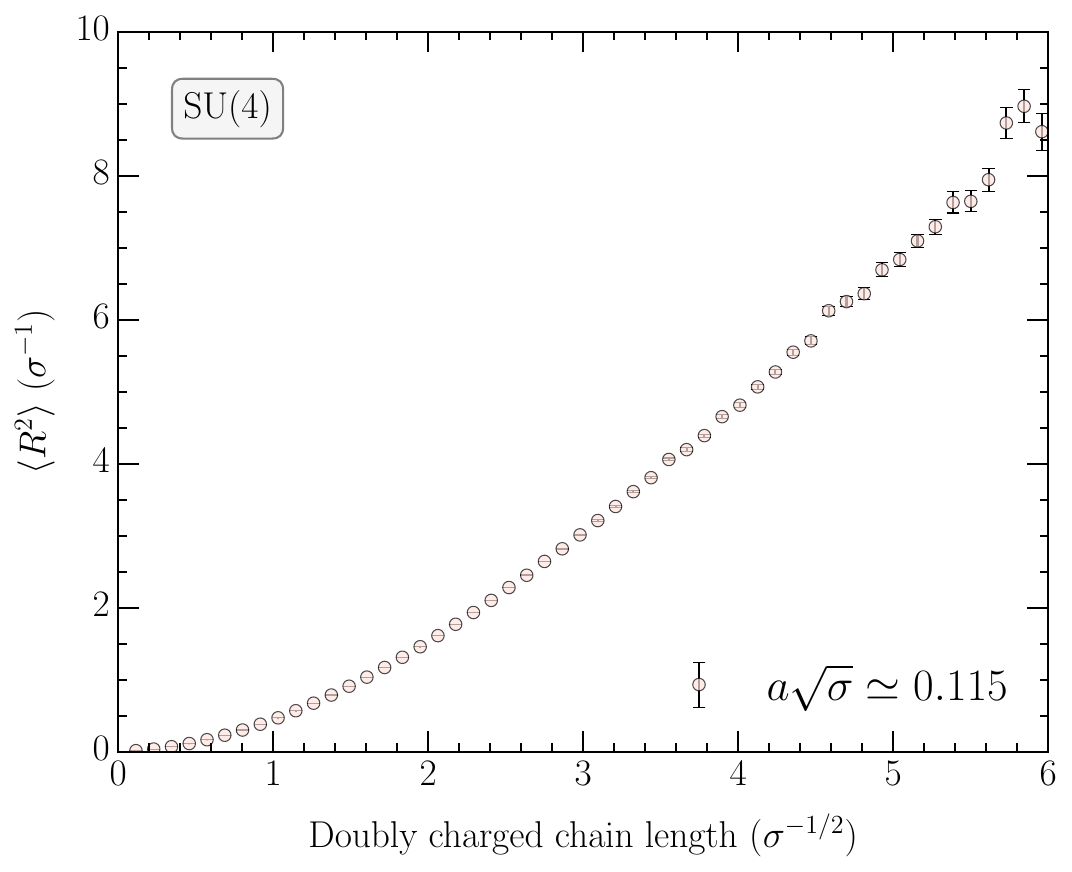}
	\caption{The mean-square end-to-end distance as a function of chain length for doubly charged chains in \(\SU(4)\). A power law is still an accurate description of the long-distance data, however the fit parameters cannot be precisely determined due to larger statistical uncertainties and a narrow fit range.}
	\label{fig:R2su4doublycharged}
\end{figure}

Looking now at doubly charged chains, their mean-square end-to-end distances as a function of chain length are presented in Fig.~\ref{fig:R2su4doublycharged}. Since there are significantly fewer doubly charged chains in the vortex network relative to elementary chains, we have considerably less statistics here and can only reach a maximum chain length half of that for elementary chains. This makes it difficult to pinpoint the long-distance behavior. For instance, although the data for doubly charged chains is still accurately modeled by the power law in Eq.~\eqref{eq:powerlaw} above a minimum chain length \(L_\mathrm{min} \sqrt{\sigma} \simeq 3\), the resulting statistical uncertainties on the fit parameters (\(p, A, L_0\)) are uncomfortably large. This is due to the larger errors on the end-to-end distances and the narrower fit range. This admits a large range of fit parameters that describe the data equally well. For this reason, we forgo exploring the power-law fits in depth for doubly charged chains.

Still, with the objective of fixing \(\tilde{a}\), we follow the same procedure used for \(\SU(3)\) and perform linear fits [Eq.~\eqref{eq:linear}] at the commencement of the power-law scaling regime, with the upper bound varied to fit over as large a window as tolerated by the \(\chi^2/\mathrm{dof}\). This is fixed to be \(L\sqrt{\sigma} \simeq [4.5,7.0]\) for elementary chains in \(\SU(4)\), and \(L\sqrt{\sigma} \simeq [3.0,4.2]\) for doubly charged chains. These fits are included in Appendix~\ref{app:linearfits}, with the fit details supplied in Table~\ref{tab:linearfitselementary}.

\begin{table}
	\caption{The details of our linear fits to the end-to-end distances for elementary chain lengths \(L\sqrt{\sigma} \in [4.5,7.0]\) and doubly charged chain lengths \(L\sqrt{\sigma} \in [3.0,4.2]\) in \(\SU(4)\), including the fit parameters \(\xi_0\) and \(L_0\) in units of the string tension, and the \(\chi^2/\mathrm{dof}\) values. The persistence lengths at \(a = 0\), obtained via continuum extrapolations linear in \(a^2\), are also provided.}
	\label{tab:linearfitselementary}
	\begin{ruledtabular}
		\begin{tabular}{@{}ccccS[table-format=1.3(2)]cc@{}}
			\multirow{2}{*}{\(a\sqrt{\sigma}\)} & \multicolumn{3}{c}{Elementary} & \multicolumn{3}{c}{Doubly charged} \\
			& \(\xi_0 \sqrt{\sigma}\) & \(L_0 \sqrt{\sigma}\) & \(\chi^2/\mathrm{dof}\) & \multicolumn{1}{c}{\(\xi_0 \sqrt{\sigma}\)} & \(L_0 \sqrt{\sigma}\) & \(\chi^2/\mathrm{dof}\) \\
			\hline
			0.239 & 0.821(3) & 1.45(1) & 1.13 & 0.861(7)  & 1.22(2) & 0.93 \\
			0.191 & 0.853(3) & 1.51(1) & 0.69 & 0.882(12) & 1.27(3) & 1.59 \\
			0.159 & 0.872(3) & 1.54(2) & 1.07 & 0.898(11) & 1.31(3) & 1.15 \\
			0.115 & 0.883(3) & 1.55(1) & 1.14 & 0.889(8)  & 1.29(2) & 1.40 \\
			\hline
			0     & 0.905(4) &         &      & 0.903(10) &         &      \\
		\end{tabular}
	\end{ruledtabular}
\end{table}

\begin{figure}
	\includegraphics[width=0.49\textwidth]{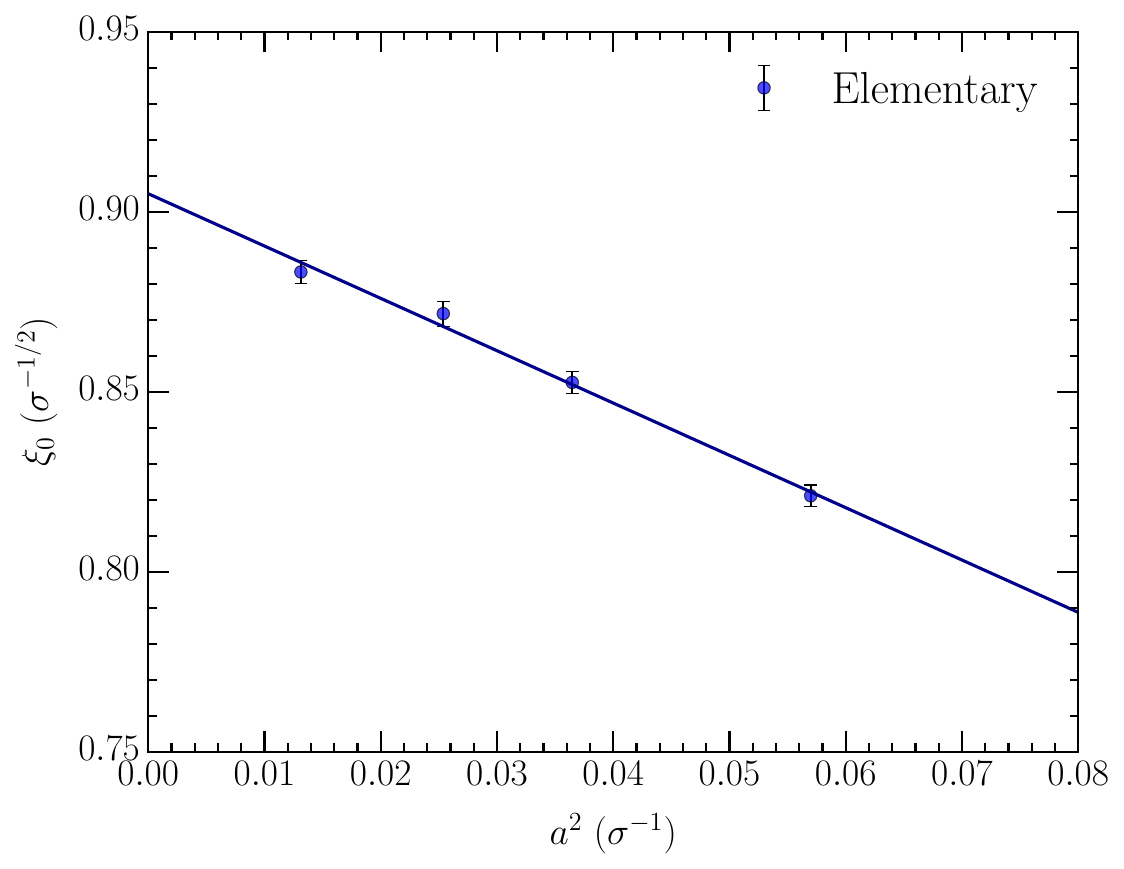}
	\hfill
	\includegraphics[width=0.49\textwidth]{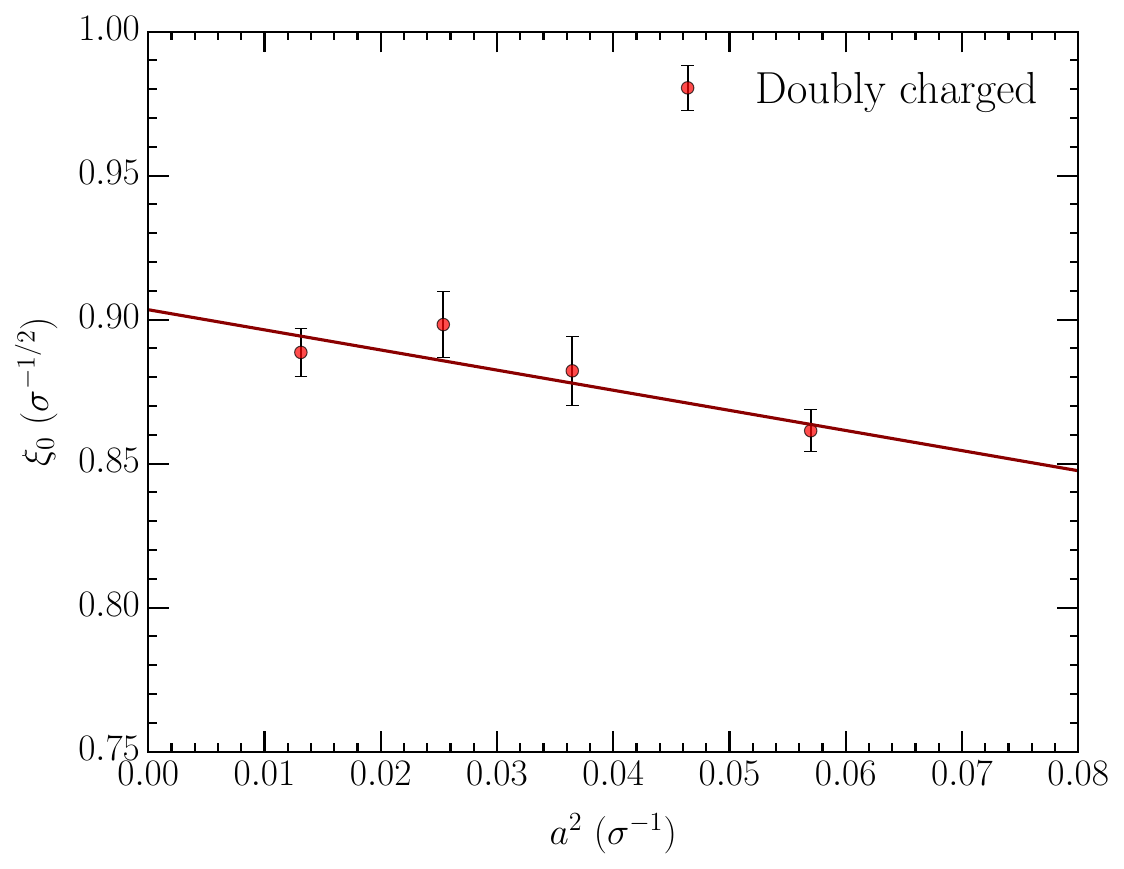}
	\caption{The persistence length \(\xi_0\) for elementary (\textbf{left}) and doubly charged (\textbf{right}) chains in \(\SU(4)\), extracted via the slope of a linear fit [Eq.~\eqref{eq:linear}] over intermediate chain lengths \(L\sqrt{\sigma} \in [4.5,7.0]\) and \(L\sqrt{\sigma} \in [3.0,4.2]\), respectively. In both cases, it is approximately linear as a function of \(a^2\), and extrapolations to the continuum limit are overlaid.}
	\label{fig:xi0elementary}
\end{figure}

For both types of center-vortex chain, the persistence length \(\xi_0\) mildly increases as the lattice simulation spacing decreases. As done for the stopping rates, this motivates exploring the possibility of extrapolating to the continuum limit via linear fits in \(a^2\). Thus, \(\xi_0\) is plotted as a function of \(a^2\) in Fig.~\ref{fig:xi0elementary}, and we see that both elementary (left panel) and doubly charged (right panel) chains reasonably follow a simple linear description. On that account, corresponding fits are overlaid, which capture the data well with \(\chi^2/\mathrm{dof} \simeq 0.95\) for elementary chains and \(\simeq 0.94\) for doubly charged chains.

The obtained continuum-extrapolated (\(a = 0\)) persistence lengths are additionally listed in Table~\ref{tab:linearfitselementary}. Remarkably, the elementary and doubly charged values are consistent with each other within statistical uncertainty, with \(\xi_0 \sqrt{\sigma} = 0.905(4)\) for the former and \(0.903(10)\) for the latter. This provides initial evidence that a single effective lattice spacing \(\tilde{a}\) can be used to describe elementary and doubly charged chains in \(\SU(4)\), and will be discussed further in the following section.

\section{Discussion} \label{sec:discussion}
\subsection{Large-distance power-law behavior} \label{subsec:saw}
Having presented our numerical results, we now proceed to interpret the findings in context of the effective model. At large length scales, the Yang-Mills data exhibits a nontrivial power-law behavior. The measured exponents, \(p = 0.528(6)\) for \(\SU(3)\) and \(p = 0.549(7)\) for \(\SU(4)\), are both below the value \(p_\mathrm{SAW} \simeq 0.588\) characterizing isolated self-avoiding walks in three dimensions. This comparison is particularly interesting because in the standard theory of polymer networks, where chains are allowed to meet at junctions, the large-distance growth of chains is still characterized by the universal self-avoiding-walk exponent~\cite{Duplantier:1989, Schafer:1992vsw, Duplantier:2019ega}. It is therefore natural to ask what may account for the exponents in the Yang-Mills data falling below the self-avoiding-walk value. In this regard, it is first important to note that the center-vortex network presents rare instances of self-intersection, so that the chains are not strictly self-avoiding. Accordingly, within the effective description, one possibility is that the suppression of multiple link occupation does not imply strict self-avoidance:\ a path may return to a previously visited lattice site through different links without occupying any link more than once.

The second possible ingredient arises from the underlying Cartan-flux structure proposed through the effective model. In that description, the dominant center-vortex condensate is governed by the quartic self-interactions of the diagonal fields \(\phi_i\) [see Eq.~\eqref{eq:3DWcv}]. This is the natural counterpart of the excluded-volume interactions responsible for self-avoiding behavior in polymer field theories. At this level, the elementary interactions distinguish the underlying Cartan fluxes:\ segments carrying the same elementary flux repel each other, whereas segments carrying different fluxes do not. However, we are yet to account for the nonoriented component in Eq.~\eqref{eq:3Dnonoriented}, parametrized by the mixing coefficient \(\zeta\). This represents a subleading interaction that allows Cartan monopoles to interpolate between different elementary fluxes along a single projected center-vortex chain. Consequently, these monopole-induced transitions may lead to deviations from strict self-avoiding behavior. This type of underlying transition could also be taking place in the Yang-Mills ensemble.

Interestingly, as proposed in Refs.~\cite{Engelhardt:1999xw, Junior:2025pwy}, if this underlying Cartan-flux picture indeed captures the microscopic organization of the Yang-Mills center-vortex network then the same hierarchy of interactions could also account for another property of the ensemble:\ topological charge density. In the thick center-vortex model, intersections between vortices carrying different elementary Cartan fluxes give rise to topological-charge lumps with fractional charge \(Q = \frac{1}{3}\), whereas intersections of equal Cartan fluxes produce \(Q = \frac{2}{3}\). Because vortices carrying the same elementary Cartan flux repel, whereas those carrying different fluxes do not, the former type of intersection is naturally favored. This could provide a simple explanation for the predominance of \(Q = \frac{1}{3}\) lumps observed in lattice Yang-Mills simulations~\cite{Mickley:2023exg}.

\subsection{Effective lattice spacing and center-vortex tension ratio} \label{subsec:atilde}
Our second consideration centers on using the measured persistence length \(\xi_0\), extracted from an initial linear regime in the numerical data, to set the effective lattice spacing \(\tilde{a}\) in the model. Before the onset of the asymptotic scaling regime, the effective model displays the non-backtracking-walk (NBW) behavior built into its local dynamics. As discussed previously, the value \(c = \frac{1}{5}\) [Eq.~\eqref{eq:xi0nbw}] is expected to provide a good representation of this regime in the percolating phase, at least over a certain range of length scales. Therefore, it is natural to perform this matching in the initial linear regime, where the slope provides a local measure of the persistence length, before the onset of power-law scaling observed in the Yang-Mills data at large \(L\). In future work, it would be interesting to determine whether the Yang-Mills ensemble and the effective model develop compatible long-distance scaling, which would require an independent numerical study of the effective model.

There are two distinct methods for fixing \(\tilde{a}\). The first possibility we consider is to directly match the persistence length from our numerical simulations, \(\xi_0\), to that of the model. This in turn sets the effective spacing \(\tilde{a}\) according to Eq.~\eqref{eq:xi0nbw} for a NBW, which leads to
\begin{equation}
	\tilde{a} = \frac{4}{5} \, \xi_0 \qquad \text{(persistence length)} \,.
\end{equation}
An alternative matching prescription is based on the mean-square displacement as a function of chain length. In the effective model, this is related to the persistence length through Eq.~\eqref{eq:R2xi},
\begin{equation}
	\braket{\tilde{R}^2} = 2 \left(\tilde{\xi}_\mathrm{p}(L) - \frac{\tilde{a}}{2}\right) L \approx \frac{1+c}{1-c} \, \tilde{a} \, L \,.
\end{equation}
Under the NBW value \(c = \frac{1}{5}\), equating this to Eq.~\eqref{eq:R2continuum} for the continuum Yang-Mills relation yields the matching prescription
\begin{equation}
	\tilde{a} = \frac{4}{3} \, \xi_0 \qquad \text{(end-to-end distance)} \,.
\end{equation}
Using the continuum-extrapolated values of \(\xi_0\) from Sec.~\ref{sec:simulations}, the corresponding \(\SU(3)\) and \(\SU(4)\) effective lattice spacings under both prescriptions are listed in Table~\ref{tab:matching}.

Our measurements of \(\xi_0\), and hence \(\tilde{a}\), for charge-one and charge-two center vortices are very similar and agree well within statistical uncertainty. This supports the description in terms of a single effective lattice spacing, and the \(\SU(4)\) values quoted in Table~\ref{tab:matching} are calculated using an inverse-variance-weighted average of the two. Previously, the use of a single spacing was assumed in the phenomenological work of Ref.~\cite{Engelhardt:2005qu}, though we now have numerical evidence validating this assumption. There, the obtained value of \(\sigma\tilde{a}^2 = 0.68 \!\implies\! \tilde{a} = 0.83/\sqrt{\sigma}\) for \(\SU(4)\) corresponds to a typical center-vortex thickness, rather than to an effective monomer size associated with the local stiffness. Nevertheless, it lies between our two determinations discussed above.

\begin{table}
	\caption{The effective lattice spacing \(\tilde a\) obtained under two different matching prescriptions, based on persistence lengths or end-to-end distances, between the numerical data and the effective NBW model with \(c = \frac{1}{5}\).}
	\label{tab:matching}
	\begin{ruledtabular}
		\begin{tabular}{@{}ccc@{}}
			Gauge group & Persistence length & End-to-end distances \\
			\hline
			\(\SU(3)\) & \(0.506(2)/\sqrt{\sigma}\) & \(0.844(4)/\sqrt{\sigma}\) \\
			\(\SU(4)\) & \(0.724(3)/\sqrt{\sigma}\) & \(1.206(5)/\sqrt{\sigma}\) \\
		\end{tabular}
	\end{ruledtabular}
\end{table}

The distinction between matching the numerical data to the model based on persistence length and end-to-end distance raises the question as to which quantity should be used to determine \(\tilde a\). The usual strategy in polymer physics, where isolated chains themselves are the fundamental objects, the end-to-end properties are commonly used to fix the model parameters. In contrast, for the present effective model, the physical spacing \(\tilde a\) encoding the local stiffness of center-vortex lines is the fundamental parameter. Consequently, matching the local persistence lengths appears to be the most natural prescription, while the end-to-end distances between junctions should emerge as a prediction of the effective theory rather than as an input quantity.

With the effective spacing fixed by the persistence length, we now evaluate the center-vortex tension ratio via Eq.~\eqref{eq:muratiorates}, using Table~\ref{tab:matching} together with the continuum stopping rates from Table~\ref{tab:stoppingrates}. We find the doubly charged to elementary center-vortex tension ratio in \(\SU(4)\) Yang-Mills theory to be
\begin{equation} \label{eq:stringtensionratio}
	\frac{\mu_2}{\mu_1} = \frac{\lambda_{(2)} \, \tilde{a} + \ln 5}{\lambda_{(1)} \, \tilde{a} + \ln 5} =  1.366(2) \,.
\end{equation}
This value is quite close to the ratio \(\mu_2/\mu_1 = 4/3 \simeq 1.333\)  [Eq.~\eqref{eq:muratiomodel}] obtained in the simplest effective model, where nonoriented components are neglected. By contrast, end-to-end matching would give the slightly larger value \(\mu_2/\mu_1 \simeq 1.563\).

Thus, the Yang-Mills value can be viewed as a relatively small deviation from the prediction of the minimal effective model. At this first step of the hierarchy, the ratio \(\mu_2/\mu_1 = 4/3\) follows directly from its Cartan structure:\ the dominant center-charge sectors are associated with magnetic weights of minimum squared norm, whose ratios determine the vortex-network tensions [cf.\ Eqs.~\eqref{eq:networktensions} and \eqref{eq:dominanttensions}].

As discussed above, the next step in the effective description introduces a milder component associated with Cartan monopoles, allowing transitions between different Cartan fluxes along nonoriented vortex components and giving rise to confining flux tubes. In this case, the Cartan structure of \(\mathfrak{su}(N)\) also leads to the asymptotic Casimir law for the \(k\)-string tensions~\cite{Junior:2023tjg, Junior:2024urr},
\begin{equation}
	\frac{\sigma_k}{\sigma_1} = \frac{k(N-k)}{N-1}\,.
\end{equation}
For \(\SU(4)\), this gives \(\sigma_2/\sigma_1 = 4/3\). Lattice simulations provide the Yang-Mills result \(\sigma_2/\sigma_1 = 1.375\)~\cite{Lucini:2004my, Athenodorou:2021qvs}. The proximity of this value to the center-vortex tension ratio suggests that both observables reflect a common Cartan-flux origin that persists beyond the minimal effective description. Such deviations can naturally arise from additional processes within the effective hierarchy. For example, monopole-fusion events involving different adjoint magnetic charges provide one such mechanism~\cite{Junior:2024urr}.

\section{Conclusions} \label{sec:conclusion}
In this work, we have investigated the implications of a previously proposed confinement mechanism, formulated in terms of the \(N\) physically equivalent Cartan fluxes and the elementary \(N\)-matching process, for the properties of the center-vortex network in the percolating phase. By carrying out a detailed lattice characterization of the polymer-like mechanical properties of center vortices in \(\SU(3)\) and \(\SU(4)\) Yang-Mills theory, this allows for quantitative connections to be established between the effective model and lattice observables. This provides the reference description for the inclusion of nonoriented (Cartan monopole) components, which generate the infrared scales underlying confinement observables.

Initially, we showed that applying a Villain approximation to this reference description in \(\SU(N)\), when the ensemble enters the percolating regime, leads to the dynamical emergence of higher center-charge vortices through the local coalescence of elementary Cartan fluxes. This qualitatively agrees with numerical Yang-Mills simulations, which in \(\SU(4)\) reveal a percolating network of charge-one center-vortex paths connected by typically shorter charge-two objects. For \(\SU(4)\), this reference description yields \(\mu_2/\mu_1 = 4/3\) for the ratio of charge-one and charge-two network tensions, reflecting the squared norms of the dominant magnetic weights in the corresponding center-charge sectors.

To confront this reference description with the Yang-Mills ensemble, we performed the first lattice study of the polymer physics of center-vortex chains using their mean-square end-to-end distance. This analysis reveals a coherent picture for the geometric and mechanical properties of center-vortex chains, exhibiting polymer-like behavior with asymptotic power-law scaling, \(\braket{R^2} \propto L^{2p}\) with \(p \simeq 0.528(6)\) in \(\SU(3)\) and \(p \simeq 0.549(7)\) in \(\SU(4)\). These exponents are both systematically below the self-avoiding-walk value \(p_\mathrm{SAW} \simeq 0.588\), possibly reflecting monopole-induced transitions between Cartan fluxes that arise from inclusion of the nonoriented vortex component.

In addition, elementary and doubly charged vortices display statistically compatible local persistence lengths, \(\xi_0\sqrt{\sigma} = 0.905(4)\) and \(0.903(10)\) respectively, supporting the description in terms of a single effective lattice spacing. In combination with the measured chain-length distributions, which confer exponential decay rates, this allowed us to extract the ratio of network tensions directly from the lattice data. The result, \(\mu_2/\mu_1 \simeq 1.366\), lies remarkably close to the reference value of \(4/3\), providing strong quantitative support for the description based on elementary Cartan fluxes.

An equally striking correspondence emerges when this result is compared with explicit confinement observables, in particular with the asymptotic \(\SU(4)\) \(k\)-string tension ratio measured from Wilson loops. Even though the center-vortex chains and confining flux tubes are distinct objects---the former strongly fluctuating and percolating, the latter smooth and massive---their corresponding simplest descriptions both yield the characteristic value \(4/3\) as a consequence of the underlying Cartan structure of \(\mathfrak{su}(N)\). Moreover, lattice simulations yield \(\sigma_2/\sigma_1 \simeq 1.375\), while the present determination gives \(\mu_2/\mu_1 \simeq 1.366\). This may reflect corrections in a common effective description of the underlying physics.

Overall, the present results provide strong quantitative support for the leading component of the hierarchical infrared description in \(\SU(3)\) and \(\SU(4)\) Yang-Mills theories considered here. This, in turn, strengthens the broader hierarchical confinement mechanism developed in previous works. The inclusion of Cartan monopole defects on top of this reference induces confining flux tubes with \(N\)-ality, which can exhibit Nielsen-Olesen profiles and Casimir scaling, as well as corrections to these behaviors.

In future work, one could explore the correlation between center vortices and the topological charge density~\cite{Biddle:2019gke} in greater depth to see whether the fractional-charge \(Q \simeq \frac{1}{3}\) objects preferentially form near transverse intersections of center-vortex sheets. Of particular interest are the three vortex surfaces emerging from an \(\SU(3)\) center-monopole worldline, which, in the effective Cartan-flux model, carry different elementary fluxes. As these surfaces bend through spacetime, they may intersect transversely away from the junction, potentially producing (\(Q \simeq \frac{1}{3}\)) topological-charge lumps. Observing such correlations in Yang–Mills configurations would provide further support for this picture. Similar studies could also be carried out in \(\SU(4)\), where fractional charges of magnitude \(\frac{1}{4}\) are expected.

In doing so, it would also be beneficial to compute the end-to-end distances for doubly charged chains in \(\SU(4)\) to higher precision. This will allow the power-law properties for doubly charged chains to be precisely pinpointed, informing whether a similar exponent as for elementary chains (\(p \simeq 0.55\)) also applies for doubly charged chains and building on the observed deviation from self-avoiding-walk behavior.

\section*{Acknowledgments}
This work was supported with supercomputing resources provided by the Phoenix High-Performance Computing (HPC) service at Adelaide University. This research was undertaken with the assistance of resources and services from the National Computational Infrastructure (NCI), which is supported by the Australian Government. This work was supported by the Pawsey Supercomputing Centre through the Pawsey Centre for Extreme Scale Readiness (PaCER) program. This research was supported by the Australian Research Council through Grant No.~DP210103706. This work was funded in part by the Simons Foundation's Simons Collaboration on Color Confinement and QCD Strings award \# 994314. Financial support from the Brazilian agency CNPq under Contract No.~309971/2021-7 is also gratefully acknowledged.

\appendix

\section{Defining and adjoint weights of \texorpdfstring{\(\mathfrak{su}(N)\)}{su(N)}} \label{app:suNweights}
The Cartan (diagonal) generators \(T_q\) of \(\SU(N)\) in the defining representation are
\begin{equation}
	T_q = \frac{1}{\sqrt{2q(q+1)N}} \operatorname{diag} [\underbrace{1, \ldots, 1}_q, -q, 0, \ldots, 0] \,,
\end{equation}
where the initial \(q\) elements are equal to \(1\). The \(N\) weights \(\omega_i\), \(i = 1, \ldots, N\), of the defining representation are then given by the \((N-1)\)-tuples 
\begin{equation} \label{eq:weights}
	\omega_i = (T_1|_{ii}, T_2|_{ii},\ldots, T_{N-1}|_{ii}) \,.
\end{equation}
Equivalently, the tuple \(\omega_i\) is formed by the common eigenvalues of the defining weight vector \(\ket{\omega_i}\), which is a column matrix with element \(1\) at the \(i\)-th row and zero otherwise,
\begin{equation}
	T_q \ket{\omega_i} = T_q|_{ii} \ket{\omega_i} = \omega_i|_q \ket{\omega_i} \;.
\end{equation}
Since the generators are traceless, it follows that
\begin{equation} \label{eq:Nmatching}
	\omega_1 + \ldots + \omega_N = 0 \,.
\end{equation}
The defining weights also satisfy (no implicit sum over \(i\))
\begin{align} \label{eq:definingweights}
	\omega_i \cdot \omega_i &= \frac{N-1}{2N^2} \,, & \omega_i \cdot \omega_j &= -\frac{1}{2N^2} \quad (i\neq j) \,, & \sum_{i=1}^N \omega_i|_q \, \omega_i|_p &= \frac{\delta_{qp}}{2N} \,.
\end{align}

With these conventions, the \(N\) elementary magnetic weights \(\beta_i\), defined in Eq.~\eqref{eq:magneticweights}, are given by 
\begin{gather} \label{eq:betai}
	\beta_i = 2N \omega_i \,.
\end{gather}
A general magnetic weight can then be written as
\begin{align}
	\lambda &= \sum_{i=1}^N m_i \, \beta_i \,, & m_i\in\mathbb Z \,.
\end{align}
Its center charge is \(\sum m_i \pmod N\). Due to Eqs.~\eqref{eq:Nmatching} and \eqref{eq:betai}, a common integer shift of all coefficients \(m_i\) leaves \(\lambda\) unchanged. Thus, for any weight in the charge-\(q\) sector, \(0 \leq q \leq N-1\), we may choose a representation such that \(\sum m_i = q\). From Eq.~\eqref{eq:definingweights}, it follows that
\begin{equation}
	\lambda^2 = 2N \sum_{i=1}^N m_i^2 - 2q^2 \,.
\end{equation}
Since \(m_i^2 \geq m_i\) for every integer \(m_i\), we have \(\sum m_i^2 \geq q\), with equality if and only if \(q\) coefficients are equal to one and the remaining \(N-q\) vanish. Therefore, the minimum-norm weights in the charge-\(q\) sector are
\begin{align}
	\lambda_\mathrm{min} &= \beta_{i_1} + \cdots + \beta_{i_q} \,, & i_1, \ldots, i_q &\ \mathrm{distinct} \,,
\end{align}
and have squared norm
\begin{equation}
	\lambda_\mathrm{min}^2 = 2q(N-q) \,.
\end{equation}
All such weights are related by permutations of the \(\beta_i\) (Weyl transformations).

The \(N(N-1)\) roots of \(\mathfrak{su}(N)\), which are also the weights of the adjoint representation, are
\begin{equation} \label{eq:roots}
	\alpha_{ij} = \omega_i - \omega_j \quad (i \neq j) \,,
\end{equation}
and satisfy
\begin{align} \label{eq:rootsproperties}
	\alpha^2 &= \alpha \cdot \alpha = 1/N \,, & \omega_i \cdot \alpha_{ij} &= \frac{1}{2N} \,, & \omega_i \cdot \alpha_{jk} &= 0 \quad (j, k \neq i) \,.
\end{align}
With these conventions, the coroot lattice, which is defined by \(\nu\cdot\omega_i \in \mathbb Z\) for every defining weight \(\omega_i\), is generated by \(\nu_{ij} = 2N\alpha_{ij}\).

\subsection{\texorpdfstring{\(\mathfrak{su}(2)\)}{su(2)}} \label{subapp:su2weights}
The Cartan subalgebra of \(\mathfrak{su}(2)\) is one-dimensional:
\begin{gather}
	T_1 = \frac{1}{2\sqrt{2}} \left(\begin{array}{cc}
		1 &  0 \\
		0 & -1
	\end{array}\right) \,, \\[0.5em]
	\beta_1 = \sqrt{2} \,, \qquad\qquad \beta_2 = -\sqrt{2} \,.
\end{gather}

\subsection{\texorpdfstring{\(\mathfrak{su}(3)\)}{su(3)}} \label{subapp:su3weights}
The Cartan subalgebra of \(\mathfrak{su}(3)\) is two-dimensional:
\begin{gather}
	T_1 = \frac{1}{2\sqrt{3}} \left(\begin{array}{ccc}
		1 &  0 & 0 \\
		0 & -1 & 0 \\
		0 &  0 & 0
	\end{array}\right) \,, \qquad\qquad T_2 = \frac{1}{6} \left(\begin{array}{ccc}
		1 & 0 &  0 \\
		0 & 1 &  0 \\
		0 & 0 & -2
	\end{array}\right) \,, \\[0.5em]
	\beta_1 = (\sqrt{3},1) \,, \qquad\qquad \beta_2 = (-\sqrt{3},1) \,, \qquad\qquad \beta_3 = (0,-2) \,.
\end{gather}

\subsection{\texorpdfstring{\(\mathfrak{su}(4)\)}{su(4)}} \label{subapp:su4weights}
The Cartan subalgebra of \(\mathfrak{su}(4)\) is three-dimensional:
\begin{gather}
	T_1 = \frac{1}{4} \left(\begin{array}{cccc}
		1 &  0 & 0 & 0 \\
		0 & -1 & 0 & 0 \\
		0 &  0 & 0 & 0 \\
		0 &  0 & 0 & 0
	\end{array}\right) \,, \quad\quad T_2 = \frac{1}{4\sqrt{3}} \left(\begin{array}{cccc}
		1 & 0 &  0 & 0 \\
		0 & 1 &  0 & 0 \\
		0 & 0 & -2 & 0 \\
		0 & 0 &  0 & 0
	\end{array}\right) \,, \quad\quad T_3 = \frac{1}{4\sqrt{6}} \left(\begin{array}{cccc}
		1 & 0 & 0 &  0 \\
		0 & 1 & 0 &  0 \\
		0 & 0 & 1 &  0 \\
		0 & 0 & 0 & -3
	\end{array}\right) \,, \\[0.5em]
	\beta_1 = (2,2,2) \,, \qquad\qquad \beta_2 = (-2,2,2) \,, \qquad\qquad \beta_3 = (0,-4,2) \,, \qquad\qquad \beta_4 = (0,0,-6) \,.
\end{gather}

\section{Line tensions and physical decay rates} \label{app:tensiondecayrate}
In this appendix, we derive Eq.~\eqref{eq:singleoccupation} for the contribution of a single closed loop and discuss how the line tension is converted to a physical decay rate for simple noninteracting polymers.

\subsection{Single-loop contribution}
Consider expanding the 3D partition function \(\mathscr{Z}_\mathrm{c.v.}\) for \(\xi = 0\) in powers of \(V(l) = \tr\,\bigl(\Phi^\dagger(x)\Phi(y)\bigr)\),
\begin{align}
	\mathscr{Z}_\mathrm{c.v.} &= \sum_k \frac{(-1)^k}{k!} \sum_{l_1,\ldots,l_k} \frac{1}{\mathcal{N}} \int D\Phi \; V(l_1) \ldots V(l_k) \, e^{-Q[\Phi]} \,, & Q[\Phi] &= \sum_x \mathcal{Q}(\Phi(x)) \,.
\end{align}
We now focus on the contribution from links that form a closed non-self-intersecting loop \(\mathcal C\). For a loop of length \(k\), which is necessarily even, there are \(k!\) ways to form the same loop (\(k\) possibilities for the first monomer, \(k-1\) possibilities for the second, etc.) such that the contribution is
\begin{equation}
	\frac{1}{\mathcal{N}} \int D\Phi \; V(l_1) \ldots V(l_k) \, e^{-Q[\Phi]} \,,
\end{equation}
where the links \(l_1, \ldots, l_k\) correspond to a possible choice of ordering. Let us say that the end of \(l_1\) is the initial point of \(l_2\) (\(y_1 = x_2\)), the end of \(l_2\) is the initial point of \(l_3\) (\(y_2 = x_3\)), and so on. Recalling that \(V(l) = \sum_i \bar{\phi}_i(x) \phi_i(y)\), the nontrivial contributions are originated from loops where all the segments are of the same type. Let us separate the contribution of a given type \(i\) (we denote \(\phi_i = \phi\)),
\begin{equation}
	\int D\Phi \; \bar{\phi}(x_1) \phi(y_1) \, \bar{\phi}(x_2) \phi(y_2) \ldots \bar{\phi}(x_k) \phi(y_k) \, e^{-Q[\Phi]} = \int D\Phi \; |\phi(x_1)|^2 \, |\phi(x_2)|^2 \ldots |\phi(x_k)|^2 \, e^{-Q[\Phi]} \,.
\end{equation} 
Noting that \(Q\) is a sum over sites, the integral over variables at sites different from \(x_1, \ldots, x_k\) cancels with a similar factor from the normalization \(\mathcal{N}\). Thus, this contribution is
\begin{equation}
	\left(\frac{1}{z} \int d\phi \, |\phi|^2 \, e^{-\eta |\phi|^2 - \lambda |\phi|^4} \right)^k \equiv e^{-\mu k} \,,
\end{equation}  
which yields Eq.~\eqref{eq:singleoccupation}.

\subsection{Decay rate}
Now, in the absence of the quartic interaction, the partition function for all configurations of a given type is obtained by exponentiating the \(1\)-loop contribution \(\mathscr{Z}_1\)~\cite{Kawai:1980qq},
\begin{align}
	\mathscr{Z} &= \exp(\mathscr{Z}_1) \,, & \mathscr{Z}_1 &= \sum_{\mathcal C} e^{-\mu \, l(\mathcal C)} \,.
\end{align}
To analyze the loop-length distribution that would be obtained by sampling configurations from this ensemble, we organize the loops according to their length. Writing
\begin{align}
	\mathscr{Z}_1 &= \sum_k z_k \,, & z_k &= \mathcal N_k \, e^{-\mu k} \,,
\end{align}
where \(\mathcal N_k\) is the number of distinct closed paths containing \(k\) links, the full partition function becomes
\begin{equation}
	\mathscr{Z} = \prod_k \exp(z_k) = \prod_k \sum_{n_k=0}^{\infty} \frac{z_k^{n_k}}{n_k!} \,.
\end{equation}
Here, \(n_k\) represents the number of loops of length \(k\) in a given configuration. After normalization, \(n_k\) follows a Poisson distribution with parameter \(z_k\),
\begin{align}
	P(n_k) &= e^{-z_k} \frac{z_k^{n_k}}{n_k!} \,, & \braket{n_k} &= z_k \,.
\end{align}

For large \(k\), the number of \(k\)-step random walks returning to a given site behaves as \((2d)^k \, k^{-d/2}\), up to a \(k\)-independent factor~\cite{Kleinert:2009}. Summing over the lattice volume \(V\) and dividing by \(k\), since each closed loop can be rooted at any of its \(k\) links, gives, for large \(k\),
\begin{equation}
	\mathcal N_k \propto V \left(2d\right)^k k^{-d/2-1} \,.
\end{equation}
It follows that the loop-length distribution obtained by sampling the
ensemble behaves as
\begin{equation}
	\braket{n_k} \propto V \, k^{-d/2-1} \, e^{-(\mu - \ln 2d)k} \,.
\end{equation}
Thus, the exponential decay of \(\braket{n_k}\) is controlled not by the microscopic line cost \(\mu\) alone, but by the entropy-corrected combination \(\mu-\ln 2d\). In terms of the physical loop length \(L = k\tilde a\), the corresponding decay rate is therefore
\begin{equation} \label{eq:decayrate}
	\lambda = \frac{\mu-\ln 2d}{\tilde a} \,,
\end{equation}
which reproduces Eq.~\eqref{eq:decayratetension} in the main text for a simple random chain with \(g = 2d\).

In addition, evaluating the integral in Eq.~\eqref{eq:singleoccupation} with \(\lambda = 0\), we have
\begin{align}
	e^{-\mu} &= \frac{1}{\eta} \,, & \eta &= 2d + \rho = 2d + m^2 \tilde{a}^2 \,.
\end{align}
Combining this relation with Eq.~\eqref{eq:decayrate} gives
\begin{equation}
	\lambda = \frac{1}{\tilde a} \ln\left(1 + \frac{m^2\tilde a^2}{2d}\right) \,,
\end{equation}
which directly relates the mass parameter of the effective field theory to the decay rate extracted from the statistical distribution of loop lengths. For \(m\tilde{a} \ll 1\), it reduces to
\begin{equation}
	m^2 \simeq \frac{2d}{\tilde a} \, \lambda \,.
\end{equation}
In this regime, where the length scale \(1/m\) is much larger than \(\tilde a\), the lattice field reproduces the continuum free-field behavior at distances large compared with \(\tilde a\), with \(-m^2\) identified as the pole of the corresponding propagator.

\section{Villain approximation in the Cartan subalgebra} \label{app:villainapprox}
For a diagonal matrix \(D\), we can directly apply the Villain approximation with the usual optimized parameters for each factor,
\begin{equation}
	e^{\beta \, \tr\cos(D)} = \prod_i  e^{\beta \cos d_i} \approx (R_\mathrm{V})^N \sum_{\{n\}} e^{-\frac{\beta_\mathrm{V}}{2} \sum_{i=1}^N (d_i - 2\pi n_i)^2} \,,
\end{equation}
where \(\{n\} = \{n_1, \ldots, n_N\}\). This is a good approximation for any diagonal \(D\). Now, the diagonal elements \(d_i\) and the set of independent integers \(n_i\) can be rewritten as
\begin{align}
	d_i &= \bar{d} + d^\circ_i \,, & \bar{d} &= \frac{1}{N} \sum_i d_i \,, & \sum_i d^\circ_i &= 0 \,, \\
	n_i &= \bar{n} + n^\circ_i \,, & \bar{n} &= \frac{1}{N} \sum_i n_i \,, & \sum_i n^\circ_i &= 0 \,.
\end{align}
Thus,
\begin{equation}
	\sum_{i=1}^N (d_i - 2\pi n_i)^2 = N (\bar{d} - 2\pi \bar{n})^2 + \sum_{i=1}^N (d^\circ_i - 2\pi n^\circ_i)^2 \,.
\end{equation}
In particular, if \(D\) is traceless, we can write \(D = \Delta \cdot T \implies d_i^\circ = \Delta \cdot \omega_i\). Additionally, for a given set of integers \(n_i^\circ\), which sum to zero, there is a unique weight \(\nu\) in the coroot lattice such that
\begin{equation}
	n_i^\circ = \nu \cdot \omega_i \,.
\end{equation}
Therefore, the sum over \(\{n\}\) can be replaced as a sum over \(\{\nu\}\),
\begin{equation} \label{eq:genvillain}
	e^{\beta \tr \cos(D)} \approx \mathrm{const.} \sum_{\{\nu\}} e^{-\frac{\beta_\mathrm{V}}{2} \sum_{i=1}^N ((\Delta - 2\pi \nu) \cdot \omega_i)^2} = \mathrm{const.} \sum_{\{\nu\}} e^{-\frac{\beta_\mathrm{V}}{4N} (\Delta - 2\pi \nu)^2} \,,
\end{equation} 
where we used Eq.~\eqref{eq:definingweights} to write
\begin{gather}
	\sum_{i=1}^N ((\Delta - 2\pi\nu) \cdot \omega_i)^2 = \frac{1}{2N} (\Delta - 2\pi\nu)^2 \,.
\end{gather}
This is the Villain approximation in the Cartan subalgebra used in
Eq.~\eqref{eq:generalizedvillain}.

From here, Eq.~\eqref{eq:genvillain} can be linearized with
\begin{align}
	\sum_{\{\nu\}} e^{-\frac{\beta_\mathrm{V}}{4N} (\Delta - 2\pi \nu)^2} &\propto \sum_{\{\nu\}} \int dJ \; e^{-\frac{N}{\beta_\mathrm{V}} J^2} e^{i J \cdot (\Delta - 2\pi\nu)} \\
	&= \int dJ \; e^{-\frac{N}{\beta_\mathrm{V}} J^2} e^{i J \cdot \Delta} \, \sum_{\{\nu\}} e^{-i2\pi J \cdot \nu} \,.
\end{align}
This can then be summed over the coroot lattice using the Poisson formula
\begin{equation}
	\sum_{\{\nu\}} e^{-i2\pi J \cdot \nu} = \sum_{\{w\}} \delta(J - w) \,.
\end{equation}
Here, \(\{w\}\) sums over the weight lattice, \(w \cdot \nu \in \mathbb{Z}\), generated by the integer combination of the defining weights \(\omega_i\). This generalizes the usual integer-summation formula
\begin{equation}
	\sum_{\nu \in \mathbb{Z}} e^{-i 2\pi j \nu} = \sum_{w \in \mathbb{Z}} \delta(j - w)
\end{equation}
to the periodic lattice of coroots. In this manner, we get
\begin{equation}
	e^{\beta \tr \cos(\Delta \cdot T)} \approx \mathrm{const.} \sum_{\{w\}} e^{-\frac{N}{\beta_\mathrm{V}} w^2} e^{i w \cdot \Delta} \,.
\end{equation}

\section{End-to-end chain properties} \label{app:endtoend}
Let us discuss, in a general context, how the persistence properties of chains can be characterized by their end-to-end behavior. There is a large amount of classical well-established results, especially in the context of polymers~\cite{deGennes:1979, Rubinstein:2003}. Let \(\mathbf{r}_i\) be the position of a path starting at \(\mathbf{r}_1\) and ending at \(\mathbf{r}_{n+1}\) formed by \(n\) monomers \(\mathbf{s}_i\), \(i = 1, \ldots, n\) of length \(b\). The \(i\)th position \(\mathbf{r}_i\) can then be expressed as
\begin{align}
	\mathbf{r}_i &= \mathbf{r}_1 + \sum_{j=1}^{i-1} \mathbf{s}_j \,, & |\mathbf{s}_i| &= b \,.
\end{align}
This setting is quite general:\ it can be applied to paths growing in Euclidean space with a given angular distribution when adding monomers, or to paths in a lattice with a given rule at each step. For example, in Yang-Mills simulations, the monomer length \(b\) is equal to the lattice simulation spacing \(a\), to be taken as small as possible. Meanwhile, in context of the effective model, the monomer length instead corresponds with the finite effective spacing \(\tilde{a}\).

The end-to-end displacement vector is
\begin{equation}
	\mathbf{R} = \mathbf{r}_{n+1} - \mathbf{r}_1 = \sum_{i=1}^{n} \mathbf{s}_i \,,
\end{equation}
such that the mean-square displacement is given by
\begin{align}
	\braket{R^2} &= \sum_{i,j=1}^{n} \braket{\mathbf{s}_i \cdot \mathbf{s}_j} = n b^2 + 2 \sum_{i < j} \braket{\mathbf{s}_i \cdot \mathbf{s}_j} \,, & R &= |\mathbf{R}|\;.
\end{align}
Introducing the tangent vectors \(\mathbf{t}_i = \mathbf{s}_i / b\) at each step, and recalling our expression for the global persistence length \(\xi_\mathrm{p}(L)\) in Eq.~\eqref{eq:globalpersistence}, it is clear that \(\braket{R^2}\) and \(\xi_\mathrm{p}(L)\) are exactly related by
\begin{equation} \label{eq:R2xiapp}
\begin{aligned}
	\braket{R^2} &= nb \, \biggl(b + \frac{2b}{n} \sum_{i < j} \braket{\mathbf{t}_i \cdot \mathbf{t}_j} \biggr) \\[0.4em]
	&= 2 \left(\xi_\mathrm{p}(L) - \frac{b}{2}\right) L \,, & L &= nb \,.
\end{aligned}
\end{equation}
This recovers Eq.~\eqref{eq:R2xi} used in the main text.

In regard to Yang-Mills simulations, we are interested in taking the continuum limit,
\begin{align}
	b &= a \to 0 \,, & n &\to \infty \,, & L &= na \ \mathrm{fixed} \,.
\end{align}
For a wormlike chain (WLC) [Eq.~\eqref{eq:wlcpersistence}], well-defined limiting behavior is obtained by holding the dominant persistence length term \(\xi_0 = a/(1-c)\) fixed and rearranging for \(c\),
\begin{equation}
	c(a) = 1 - \frac{a}{\xi_0} = 1 - \frac{L}{n \xi_0} \,.
\end{equation}
As a consequence, the persistence length simplifies in the continuum limit to 
\begin{equation}
\begin{aligned}[b]
	\xi_\mathrm{p}(L) &= \lim_{n\to\infty} \xi_0 \left[1 - \frac{\xi_0}{L} \left(1 - \Bigl(1 - \frac{L}{n\xi_0} \Bigr)^n \right) \right] \\[0.4em]
	&= \xi_0 \left[1 - \frac{\xi_0}{L} \left(1 - e^{-\frac{L}{\xi_0}}\right) \right] \,.
\end{aligned}
\end{equation}
Finally, applying Eq.~\eqref{eq:R2xiapp} with \(b = a \to 0\), we arrive at Eq.~\eqref{eq:wlc} for the mean-square displacement of a continuum WLC,
\begin{equation} \label{eq:R2wlc}
	\braket{R^2} = 2 \xi_0 L \left[1 - \frac{\xi_0}{L} \left(1 - e^{-\frac{L}{\xi_0}} \right) \right] \,.
\end{equation}

\section{Linear fits} \label{app:linearfits}
In this appendix, we display the linear fits of Sec.~\ref{sec:simulations} in \(\SU(4)\) to the mean-square end-to-end distance data over intermediate length scales. Solid lines are used to indicate the fit window, while dashed lines show to what extent the linear fits deviate from the data at short and long chain lengths.

\begin{figure}[H]
	\includegraphics[width=0.49\textwidth]{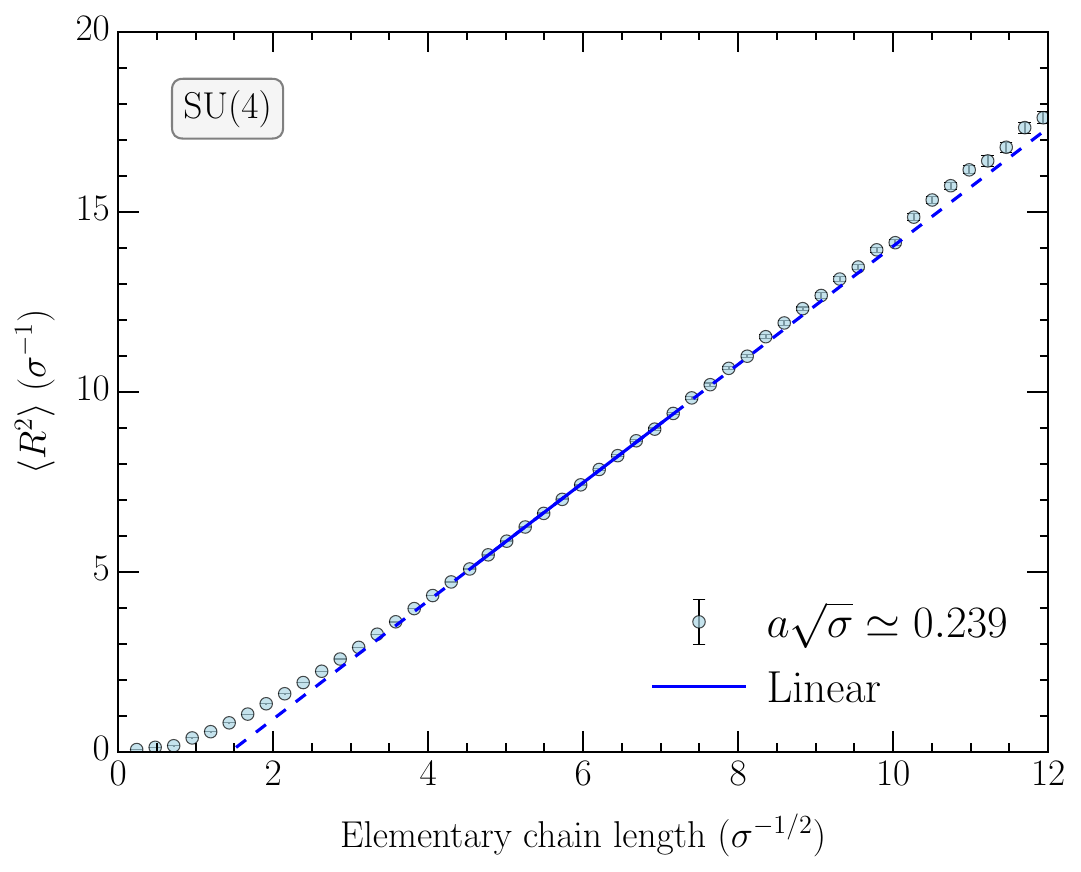}
	\hfill
	\includegraphics[width=0.49\textwidth]{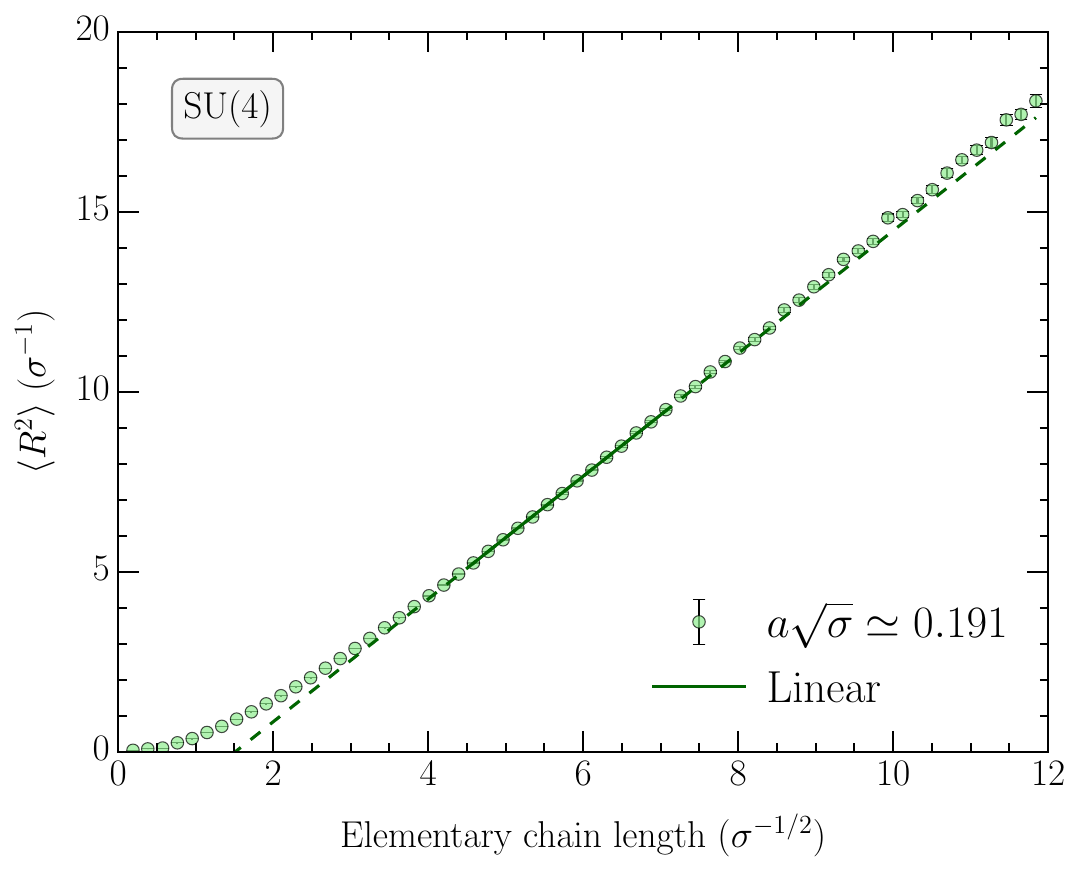}%
	\vspace{0.5em}
	\includegraphics[width=0.49\textwidth]{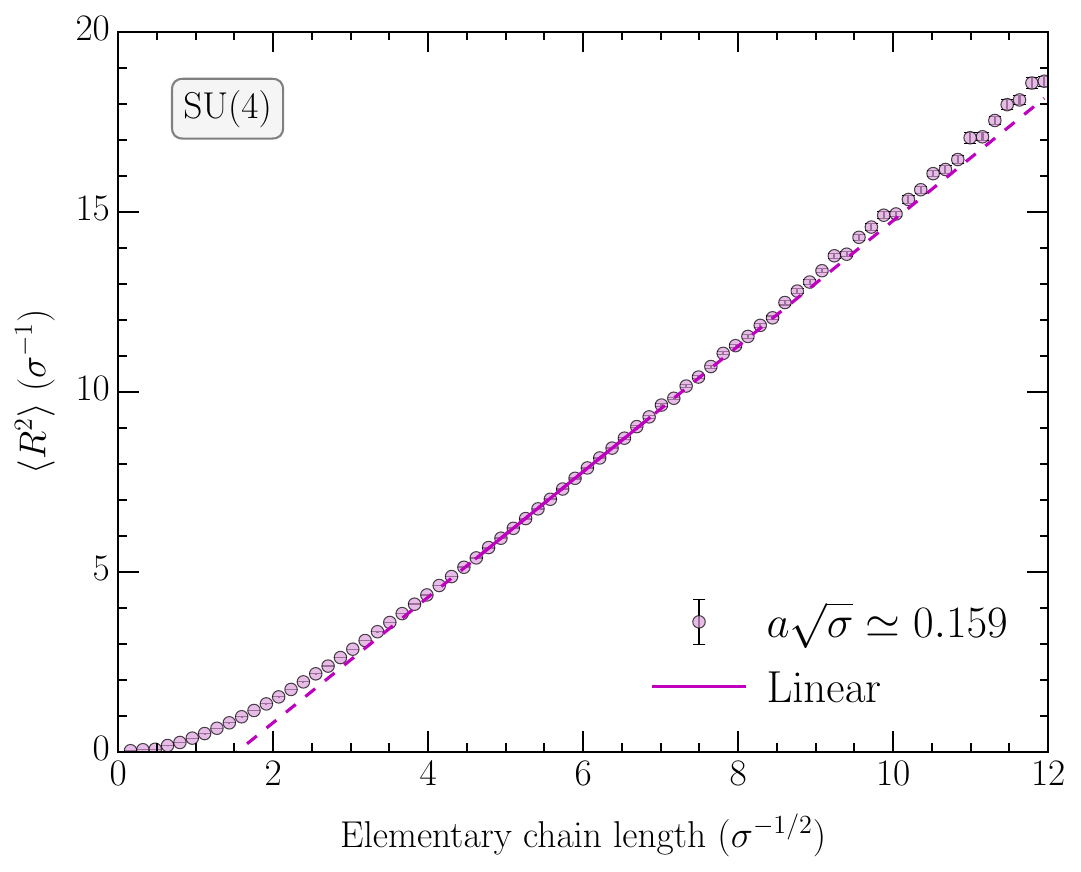}
	\hfill
	\includegraphics[width=0.49\textwidth]{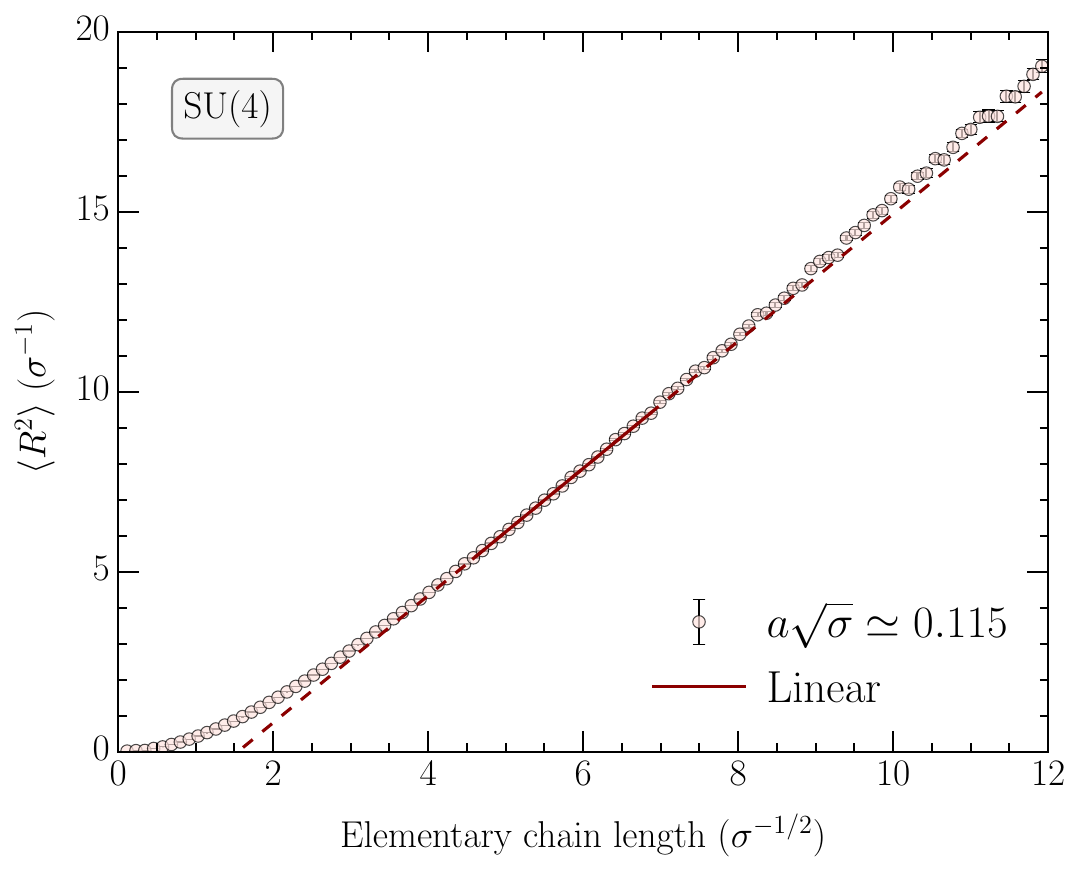}
	\caption{The linear ansatz of Eq.~\eqref{eq:wlcmodified} fit to the mean-square end-to-end distances for intermediate elementary chain lengths \(L\sqrt{\sigma} \in [4.5,7.0]\) in \(\SU(4)\).}
	\label{fig:R2su4elementarylinear}
\end{figure}

\begin{figure}[ht]
	\includegraphics[width=0.49\textwidth]{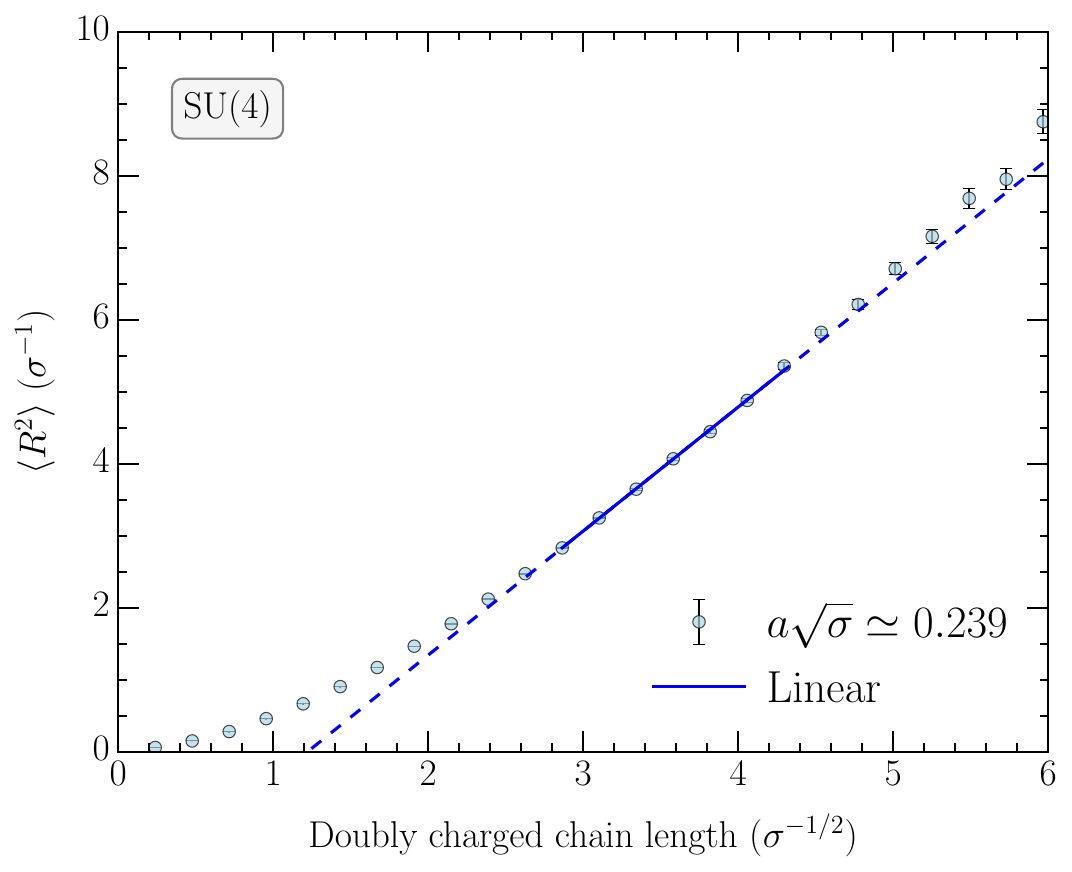}
	\hfill
	\includegraphics[width=0.49\textwidth]{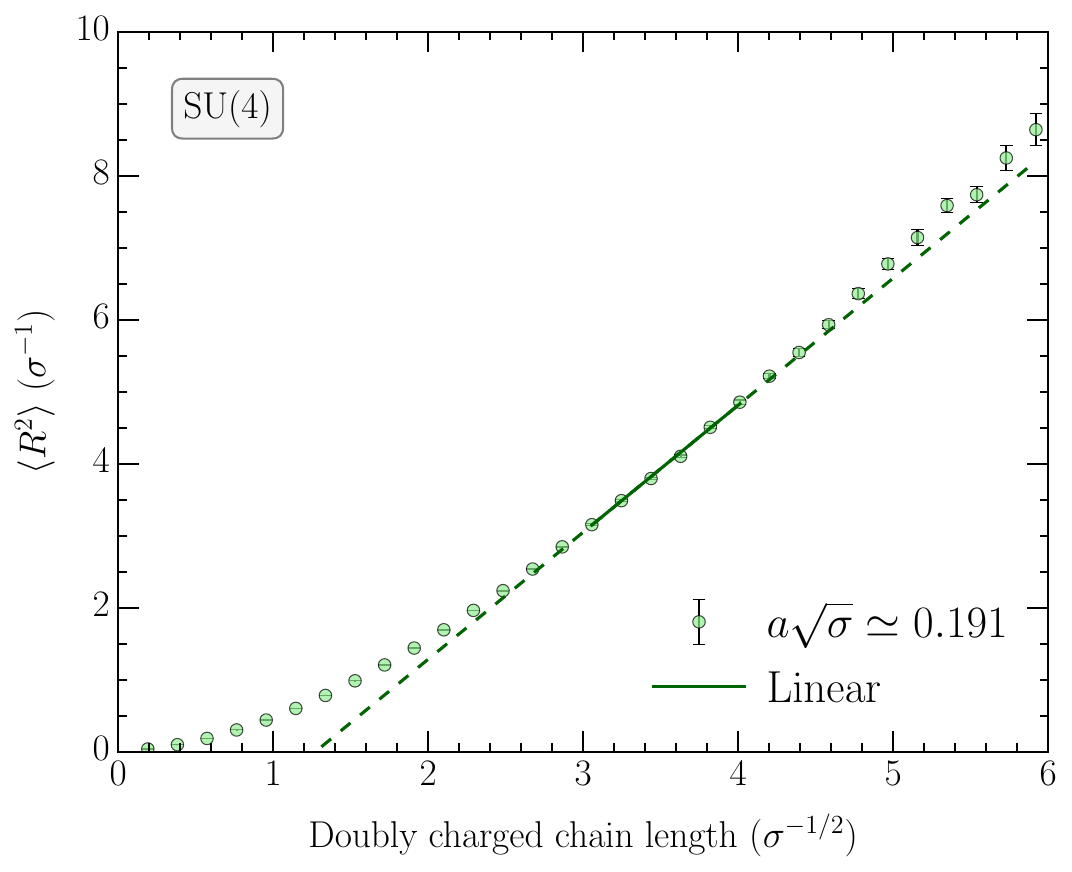}%
	\vspace{0.5em}
	\includegraphics[width=0.49\textwidth]{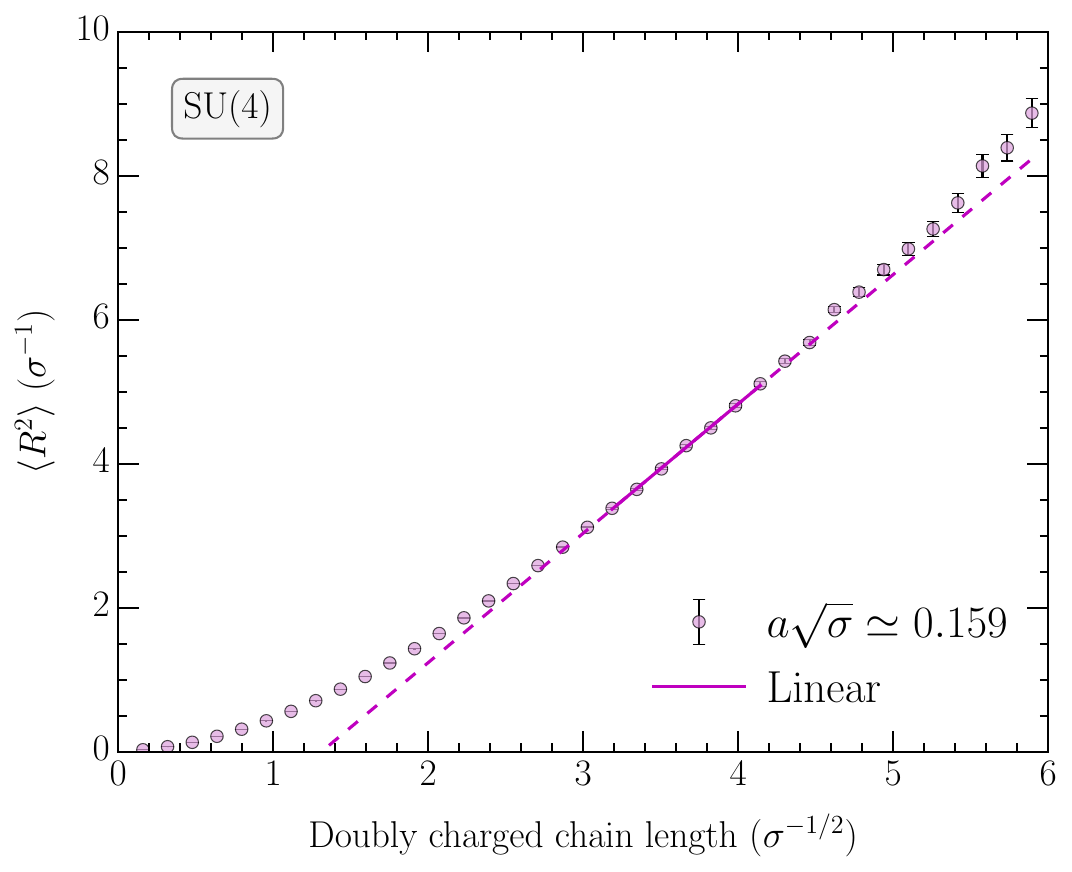}
	\hfill
	\includegraphics[width=0.49\textwidth]{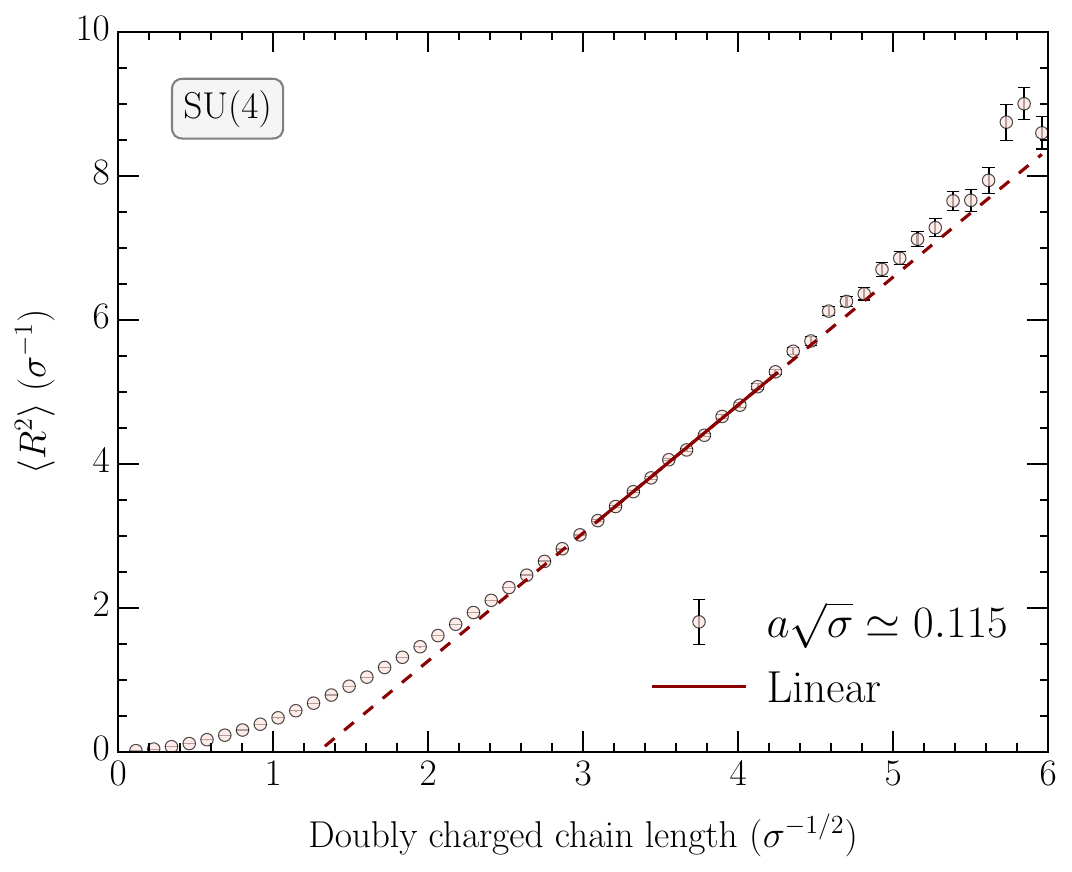}
	\caption{Linear fits for intermediate chain lengths \(L\sqrt{\sigma} \in [3.0,4.2]\) to the squared end-to-end distances for doubly charged chains in \(\SU(4)\).}
	\label{fig:R2su4doublychargedlinear}
\end{figure}

\bibliography{main}

\end{document}